\documentclass[12pt]{article}
\usepackage[
 letterpaper,
 left=2.5cm,
 right=2.5cm,
 top=4cm,
 bottom=2.5cm
]{geometry}
\usepackage[T1]{fontenc}
\usepackage{mathpazo}
\usepackage{amsmath}
\usepackage{graphicx}
\usepackage{tikz-cd}
\usepackage{multicol}
\usepackage{booktabs}
\usepackage{threeparttable}
\usepackage{subcaption}
\usepackage{xcolor}
\usepackage{adjustbox}
\usepackage{changepage}
\usepackage{array}
\usepackage{tabularx}
\usepackage[authoryear,round]{natbib}
\usepackage{setspace}
\usepackage{makecell}
\renewcommand{\thefootnote}{\arabic{footnote}}
\usepackage[hidelinks]{hyperref}

\begin{document}

\title{When Less Is More: Managing AI Adoption with Adaptive Incentive Design}
\author{
 Jie Gong\footnotemark[1]
 \hspace{2em} Jiayi Hou\footnotemark[1]
 \hspace{2em} Jin Li\footnotemark[1]
 \hspace{2em} Fei Pu\footnotemark[1]
 \hspace{2em} Xinjue Yao\footnotemark[1]
}
\date{\today}
\maketitle
\begin{abstract}
\noindent
% This paper studies gaming and adaptive incentive design in AI adoption. In our setting, a medical-device company mandated 200 queries a month from each employee to encourage AI use. The mandate led more employees to use the AI tool for the first time, but usage patterns showed signs of gaming: query counts bunched at the threshold, and 31 percent of queries were unrelated to work or exact duplicates. Following employee feedback that the mandate was too demanding, the firm adapted the incentive by lowering the target to 100 in a staggered rollout across branches. The adjustment led employees to cut unrelated and duplicate queries while maintaining work-related ones. Among sales employees, monthly sales rose 7 percent. The findings demonstrate the gains from adaptive incentive design: retaining a mandate while adjusting its intensity can improve both the quality of AI use and employee performance.

We study gaming and adaptive incentive design in AI adoption. A large medical-device company required roughly 5,000 employees to submit at least 200 queries per month to an internal AI assistant. First-time use increased significantly after the mandate, but usage patterns suggested gaming: query counts bunched at the threshold, and 31 percent of queries were repeated or off-task. The firm subsequently revised incentive design and lowered the target to 100. Using staggered implementation across branches, we estimate that the adjustment reduced query volume by 30 percent, with repeated or off-task queries accounting for about 90 percent of the decline. The estimated change in non-repeated, work-related queries was small and statistically insignificant. Among sales employees, monthly sales increased by 7 percent. The findings identify incentive adaptation as a consequential margin of technology adoption: recalibrating usage requirements can reduce gaming and improve employee performance.

\vspace{0.5em}

\vspace{0.25em}

\end{abstract}
\renewcommand{\thefootnote}{\fnsymbol{footnote}}
\footnotetext[1]{Faculty of Business and Economics, The University of Hong Kong; HKU Centre for AI, Management and Organization.}
\renewcommand{\thefootnote}{\arabic{footnote}}
\setcounter{footnote}{0}

\clearpage

\section{Introduction}
% Generative AI can improve the productivity of knowledge workers \citep{noy2023experimental, brynjolfsson2025generative,cui2026effects, dellacqua2026navigating}. Yet making AI available does not ensure that employees will use it \textcolor{red}{\citep{dillon2025shifting,hartley2025labor,humlum2025unequal,bick2026rapid}}, much less use it productively. Workers face fixed costs of learning and, therefore, are reluctant to experiment with AI and integrate it into their workflow. For firms, this creates an adoption problem \textcolor{red}{\citep{mcelheran2024ai,bonney2026microstructure,humlum2026still}}. 

Generative AI can raise the productivity of knowledge workers \citep{noy2023experimental, brynjolfsson2025generative,cui2026effects, dellacqua2026navigating}. Yet access does not ensure adoption. Workers may prefer established routines, be reluctant to spend time and effort learning a new technology, or fear that demonstrating AI's capabilities will threaten their job security \citep{dillon2025shifting,hartley2025labor,humlum2025unequal, li2025overcoming,bick2026rapid}. Firms therefore increasingly track AI use, incorporate it into performance evaluations, and sometimes impose minimum-use mandates.

A mandate may push workers past initial learning costs, but it also turns AI usage into a performance measure that workers can game. Workers can inflate measured use by repeating prompts, asking trivial questions, or submitting queries unrelated to their jobs, the classic problem of rewarding the measured action while hoping for a different outcome \citep{kerr1975folly, holmstrom1991multitask}. Such behavior has earned the name ``tokenmaxxing'' in the technology sector \citep{bousquette2026tokenmaxxing}. 

How much this distortion matters remains an empirical question. Press accounts document that gaming occurs but cannot reveal how prevalent it is. Limited gaming may be a tolerable cost of overcoming adoption inertia; extensive gaming may impose substantial costs on employees and give firms an inflated picture of AI adoption. Our first question is therefore: how do workers meet an AI-usage target, and how much of measured use appears to reflect gaming?

Because firms do not know ex ante how workers will respond to a given target, managing an AI-usage mandate is a dynamic problem.\footnote{Static models characterize incentive intensity given a specified relationship between measured performance and the principal's objective \citep{holmstrom1991multitask, baker1992incentive}. Dynamic models emphasize both the commitment costs of revision \citep{laffont1988dynamics} and the opportunity to learn a measure's susceptibility to gaming and revise it \citep{courty2003dynamics}.} If gaming proves substantial, a second question follows: how should a firm respond?  We study one adaptive response: retaining the mandate while lowering its intensity after an initial period. Its consequences depend on what the higher target had sustained. If it sustained work-related use that was still taking root, lowering the target may slow further experimentation. If it mainly sustained queries submitted to meet the target, lowering the target may reduce gaming. Adjustment therefore provides a diagnostic of the activity sustained by the stronger pressure.  We ask: what disappears when a firm relaxes an initially demanding AI-use target?

Answering these questions requires observing the activities behind usage counts and the variation in mandate intensity. Our setting provides both.

We study a large medical-device company that introduced an internal AI assistant to roughly 5,000 non-production employees. The tool was designed to help employees retrieve and synthesize internal information, including technical documents, customer records, and regulatory material. As voluntary take-up remained low, the firm required employees to submit at least 200 AI queries per month, with monetary penalties for non-compliance. Three months later, following concerns that the  requirement was too demanding, management lowered the target to 100 queries. The adjustment was implemented at different times across branches, creating within-firm variation in mandate intensity.

We observe both usage counts and the content of each query, allowing us to distinguish non-repeated, work-related use from repeated or off-task activity \citep{chatterji2025people,handa2025economic,counts2026enterprise}, a margin that standard measures of technology adoption do not capture \citep{brynjolfsson1996paradox,forman2005location,bloom2012americans,hosseinimaasoum2025generative,schubert2025organizational}. Moreover, the staggered branch-level rollout of the lower target creates within-firm variation in mandate intensity, which we use to estimate how adjusting the target affects the quantity and composition of AI use.

Our first result is that the initial 200-query mandate had mixed consequences. Before the mandate, only 14.8\% of employees had ever submitted a query. The 200-query requirement was followed by a large wave of first-time users: 31.2\% of employees used the tool for the first time in April, followed by another 9.1\% in May. 

At the same time, usage was strongly organized around the requirement. Employees bunch at the threshold: among employees with any usage, 12\% submit exactly 200 queries, and 95\% submit no more than 220. The content of the queries shows that 31\% of the queries are repeated or personal or off-task. Usage is also heavily back-loaded within the month. Query activity is low and stable during the first half of the month, but increases sharply as the deadline approaches. The mandate brought many employees to the tool, but it also generated substantial gaming.

Our second result is that lowering the target reduced usage volume and changed the composition of AI use. Descriptively, when the target falls from 200 to 100, the spike in usage shifts to the new threshold, but the behaviors associated with quota filling become less pronounced. The share of low-quality queries falls from 31\% to 18\%, and the end-of-month surge attenuates. 

Using the staggered branch-level rollout of the lower target, we estimate that reducing the target lowers the monthly query volume by 29 queries, a 30\% decline relative to the sample mean. Repeated or off-task queries account for about 90\% of the total decline, while the change in the number of non-repeated, work-related queries is small and statistically insignificant. The number of the most work-relevant queries increases and improvements in work relevance are concentrated in the lower part of the distribution. The adjustment therefore removed mainly the queries generated to satisfy the higher target.

Our third result is that this shift in usage composition was accompanied by improved output performance. Among sales employees, the only group for whom we observe individual performance, lowering the target increased monthly sales by 7\%. Event-study estimates show flat pre-trends and persistent post-treatment effects. This result rejects the simple view that more AI usage is always better. In this setting, a lower target produced less measured usage, a better composition of use, and higher sales.

The pattern is consistent with a shift away from gaming and toward more valuable use. Under the higher target, employees spend time producing queries to satisfy the quota. When the target is lowered, they cut back disproportionately on repeated or off-task activity. Heterogeneity results point in the same direction. Employees whose prior usage clustered near the 200-query threshold respond most strongly to the lower target. Among employees whose usage was not anchored near 200, total usage changes little: the decline in low-quality queries is offset by an increase in high-quality queries. Even employees who had used the tool before the mandate reduce total and low-quality usage after the adjustment, suggesting that the original target induced gaming even among employees with prior interest in the tool.

These findings show that AI mandates have both benefits and costs. They can initiate experimentation when voluntary take-up is slow, but they can also induce employees to manufacture usage. The managerial problem is therefore not simply whether mandates work or fail; they can do both. Because firms do not know ex ante how workers will respond, the relevant design question is how to learn from early implementation and recalibrate adoption incentives over time.

This study contributes to two strands of the literature. The first is the empirical literature on how workers game performance metrics \citep{oyer1998fiscal, jacob2003rotten, courty2004empirical, larkin2014cost, benson2015agents}. Prior studies typically detect these responses from patterns in reported outcomes, such as bunching, timing shifts around fiscal deadlines, quota adjustments, or suspicious strings of test answers. We provide direct, input-level evidence on how workers respond to a target-based performance metric. The full text and timestamp of every query allow us to observe how workers meet the target, distinguishing gaming activities from productive, work-related use.

Second, we provide causal evidence on the effects of adjusting a performance metric. Existing work shows why performance measures may need to change over time as their weaknesses become apparent and workers learn how to game them \citep{courty2003dynamics,li2021learning}. Related work on workplace technology adoption views implementation as a process of organizational co-invention rather than a one-time decision \citep{bresnahan2017adoption}. We contribute to these literatures by providing causal evidence on the effects of a particular adjustment: retaining the usage mandate while lowering its intensity after an initial period. Our findings suggest that the relevant design problem is not whether to use a mandate, but how to revise it as the organization learns.

% Third, the paper speaks to how information technology changes what firms can measure and contract on. In standard multitask models, firms want to reward valuable actions but observe only imperfect proxies \citep{holmstrom1991multitask, baker1992incentive}. Monitoring technologies relax this constraint by making worker inputs more observable and therefore more contractible \citep{hubbard2000demand, baker2004contractibility}. AI tools produce detailed logs of worker interactions, making one input—usage—easy to observe and verify. But observability does not solve the alignment problem. The firm can count queries precisely, yet counted usage can still diverge from productive usage, and the size of that gap depends on how the target is set. Technology therefore expands what the firm can measure, but does not determine how the measure should enter incentives. Aligning the quantity of use with its value remains the central design problem.

\section{Institutional Background}
\label{sec:background}

\subsection{Company background}

We study MedTech Corp, a large medical-device company with operations across China.\footnote{``MedTech Corp'' is a pseudonym used to protect the firm's confidentiality.} The firm employs around 9,000 workers and is organized into 46 local branches. These branches are important for our empirical design because implementation of the revised AI mandate later occurred at the branch level. Our main empirical analysis uses 42 branches after excluding four very small branches.

The firm's workforce consists of frontline production workers and
non-production employees. Non-production employees include sales, R\&D staff, technical support staff, and administrative staff. They regularly use internal product, customer, and regulatory information, making them natural potential users of the firm's AI assistant.

\subsection{The AI assistant}

The firm built an internal AI assistant and opened it to all non-production employees on March 1, 2025. The AI assistant is built on DeepSeek large language models and fine-tuned on the firm's internal knowledge base, including technical documentation, customer records, and regulatory archives. The tool gives employees a single interface for retrieving and synthesizing information that was previously scattered across internal systems.

Employees use the AI assistant through a chat-based interface, submitting natural-language queries and receiving generated responses. This design is particularly useful for tasks that require quick access to firm-specific knowledge.

Two examples illustrate the types of tasks the AI assistant could support. A salesperson preparing for a hospital visit could use the assistant to create an account profile and develop a tailored sales pitch in about 10 minutes, compared with 40--60 minutes using manual search. An R\&D engineer could look up a competitor's product attributes or diagnose abnormal test results without searching through scattered documents.

% Jiayi Rely: In sales, sales representatives use it to generate pre‑visit account snapshots, role-specific opening scripts, and structured needs‑discovery checklists, as well as to convert tender technical requirements into parameter alignment tables and standardize post‑visit memos and follow‑up actions (e.g., cutting pre‑visit prep from ~40–60 minutes to ~10 minutes). In R\&D, engineers use it for competitor benchmarking, structured troubleshooting of abnormal experimental results (hypotheses + verification plans), and supplier question-list generation to reduce search and rework.

\subsection{The Initial Usage Mandate and Its Adjustment}

\paragraph{The initial mandate.}
Despite these potential uses, employees did not adopt the tool organically. In the first month after the AI assistant was released, only 14.8\% of the employees had submitted at least one query. 

In response, the company decided to make usage mandatory. On March 31, 2025, the CEO's office issued a formal mandate requiring all non-production employees, about 5,000 workers, to submit at least 200 queries per month. Employees who failed to meet this requirement for the first time would receive a formal warning, and those who repeatedly failed to comply would be fined RMB 200. This fine was about 2\% of average monthly compensation, or 7\% of average monthly performance-based pay.

\paragraph{Employee feedback and policy adjustment.}
Three months after the initial mandate, the CEO's office received feedback about the 200-query requirement. Branches reported that employees considered producing 200 queries per month time-consuming and difficult to complete. Headquarters concluded that the original target was too demanding and decided to reduce its intensity. Headquarters did not review usage or query content before making this decision. In July 2025, the monthly requirement was lowered from 200 to 100.

\paragraph{Branch implementation.}
The decision to reduce the target was made at the headquarters, but it was not immediately communicated to employees in every branch. The CEO's office communicated the change through a memo instead of a formal notice. Some branches interpreted it as a guideline and waited for written instructions or further confirmation, while others treated it as a new binding threshold. 

Branches that adopted the change informed employees through internal notices or meetings. Branches that did not adopt the change made no such announcement. The AI interface remained unchanged. It only showed each employee's cumulative monthly query count without the applicable target.

As a result, local HR teams communicated the adjustment to employees at different times. Nineteen branches switched to the 100-query threshold immediately in July. Other branches switched later, and 15 branches continued enforcing the 200-query requirement through December 2025. Appendix Figure~\ref{fig:appendix_policy_timeline} summarizes the rollout.

\section{Data and Empirical Strategy}
\label{sec:data}
We first describe the query logs, personnel records, and sales data used to measure employee responses. We then explain how the staggered branch-level implementation of the revised target identifies the effects of the policy adjustment.
\subsection{Data}
\paragraph{Data sources and sample.}
The AI platform logs the full text and timestamp of every query each employee submits. We link these logs to personnel records and sales data to build an employee-month panel running from March through December 2025, beginning with the first month in which the tool was available company-wide. Personnel records provide branch assignment, job role, tenure, and demographics.

For the main empirical analysis, we exclude employees at headquarters and in the AI department, restrict the sample to non-production employees, and exclude four very small branches with fewer than 30 non-production employees.\footnote{All remaining branches have at least 80 non-production employees; this restriction removes very small branches whose organizational structure differs from the rest of the branch network; the main findings are robust to including all 46 branches.} The baseline sample covers about 5,000 non-production employees across 42 branches, or roughly 44,000 employee-month observations. For a subsample of more than 500 sales employees, we also observe individual-level monthly sales.

\paragraph{Usage quantity and first use.}
We construct measures of both usage quantity and usage quality. For quantity, we use the platform's compliance-counted query measure, which is the same measure employees saw on the interface and the firm used to enforce the mandate. We also construct a measure of first use. An employee records first use in the first month in which they submit at least one compliance-counted query. This measure captures initial experimentation with the tool.

\paragraph{Query content and quality.}
For usage quality, we construct two text-based measures. The first is repetition. At the query level, a query is classified as repeated if it exactly duplicates one
of the employee's own earlier queries in the same month. This conservative
measure captures the most direct form of low-effort, box-checking behavior.\footnote{Employees may repeat a query when they try to obtain a better answer. In our setting, however, repetition is concentrated near the monthly deadline and falls when the target is lowered, even though the underlying AI tool does not change.} 

The second measure captures whether a query is related to work. Following
\citet{chatterji2025people}, we use an LLM-based classifier to assign each query
to one of three work-relevance categories: \emph{clearly work-related}, \emph{generic or weakly work-related}, or \emph{personal or off-task}. Clearly work-related queries are directly connected to the employee's responsibilities or the firm's operations. Generic or weakly work-related queries may support work but are not specific to the employee's role, task, or organizational context. Personal or off-task queries are unlikely to support work. The classifier uses the query text and job-context information, including the employee's position, branch, department, and the firm's line of business, but not treatment status, query date, or performance outcomes.

We validate the classification against human labels and through
an alternative embedding-based measure. Agreement between the model and the human-majority label is high.\footnote{In a validation sample of 1,000 randomly selected queries, three human annotators classify each query into the same three work-relevance
categories. Cohen's $\kappa$ between the model and the human-majority label is 0.815.} We also construct an embedding-based
work-relevance index that compares each query with both the firm's business
context and the employee's role context. The results are similar to those using the main classification. Appendix~\ref{app:classification} reports the full taxonomy, classifier prompt, additional agreement measures, examples, and the embedding-based construction.

We combine repetition and work relevance to construct the main query-quality outcomes. A query is classified as 
\emph{high quality} if it is not repeated and is either clearly work-related or generic or weakly work-related. A query is classified as
\emph{low quality} if it is repeated or personal/off-task.\footnote{Generic or weakly work-related queries account for about 3\% of all queries, and the results are similar under a more conservative definition that also classifies them as low quality.}

% For one supporting analysis, we also use a finer work-relevance score returned by
% the same classifier. The score ranges from 1 to 10 and captures how closely the
% query matches the employee's routine job content. We do not use this score to
% define the main high- and low-quality measures. Instead, we use it only to
% identify the clearest cases of work use: non-repeated queries that receive the
% maximum score of 10. The scoring anchors are reported in Appendix~\ref{app:classification}.

\paragraph{Employee-month outcomes.}
We aggregate these query-level classifications to the employee-month level. The main count outcomes include the total number of queries, the number of
repeated or off-task queries and the number of non-repeated, work-related queries. These
outcomes are defined for all employee-months; employees who submit no queries
in a month receive zeros for each count outcome. Keeping
inactive employees in the sample allows us to decompose changes in total
usage without conditioning on continued use.

We also construct composition outcomes among active users, including the share of repeated or off-task queries and average work relevance score. In supporting analyses, we also use a finer 1--10 work-relevance score, a 0--1 prompt-clarity score, and query-intent categories that distinguish \emph{Asking}, \emph{Doing}, and \emph{Expressing}. These outcomes are defined only for employee-months with at least one
query. Appendix~\ref{app:classification} reports their anchors, examples, and validation. In robustness checks, we verify that the results are not driven by changes
in the composition of active users.

\subsection{Empirical strategy}
\label{sec:empirical_strategy}

Our analysis proceeds in three steps. We first document whether previously inactive employees began to use the AI assistant after the 200-query mandate. We then describe employee behavior under the 200-query target, focusing on the quantity, content, and timing of their queries. Because the 200-query mandate was implemented company-wide, we treat
the first two steps as descriptive. Finally, we use the staggered rollout of the lower target across branches to causally estimate the effects of reducing the target from 200 to 100 queries.

% Our analysis proceeds in two steps. We first document employee AI usage under the 200-query target, examining both the quantity and the composition of queries. We then study the effects of lowering the target from 200 to 100 queries. To do so, we exploit the staggered rollout of the lower target across branches.

The key source of identification is the variation in when each branch implemented the lower target. The firm-level decision to reduce the target from 200 to 100 was common to all branches and is not the variation we exploit. Because implementation was delegated to local HR teams, branches switched to the adjusted regime at different times. This branch-level timing variation forms the basis for our difference-in-differences design.

Intuitively, the design compares employees in branches that have already switched to the 100-query regime with employees in branches that are still operating under the 200-query target in the same month. The comparison group consists of both later-switching and never-switching branches. 

The query logs provide a useful check that this branch-level treatment variable
captures the target employees faced in practice. Appendix
Figure~\ref{fig:branch_month_bunching} plots, for each branch-month, the share of
active employees whose monthly query counts fall near the 200- and 100-query
thresholds. The figure shows that bunching remains centered near 200 until a
branch switches and then shifts toward 100.
This pattern confirms that the revised target was enforced through branch-level communication and implementation.

We estimate the following specification: 
\begin{equation}
Y_{ijt} = \alpha_i + \lambda_t + \beta \, LowerTarget_{jt} + \varepsilon_{ijt},
\label{eq:did}
\end{equation}
where \(Y_{ijt}\) is an outcome for employee \(i\) in branch \(j\) in month \(t\). \(LowerTarget_{jt}=1\) if branch \(j\) operates under the 100-query target in month \(t\), and zero otherwise. \(\alpha_i\) are employee fixed effects and absorb stable differences across workers, such as job role or baseline propensity to use AI. \(\lambda_t\) are month fixed effects and absorb firm-wide shocks, such as common changes in demand, AI awareness, or seasonality. The coefficient \(\beta\) captures the average effect of a branch switching to the lower target on AI usage and, in the sales subsample, employee sales performance.

We also estimate an event-study specification by replacing \(LowerTarget_{jt}\) with indicators for event time relative to the month in which branch \(j\) first implemented the 100-query target, omitting the month before implementation. The event-study estimates serve two purposes. First, they show how quickly employees adjusted their usage behavior after the lower target was implemented. Second, they allow us to examine whether switching and comparison branches followed similar trends before the policy adjustment. 

The identifying assumption is that, absent the local switch to the 100-query regime, employees in switching branches would have followed similar trends in AI usage and performance as those in branches that had not yet switched or never switched. The relevant concern is that branches switching at different times may have followed different underlying trajectories.

We assess this concern in three ways. First, branches that ever switched and branches that remained under the 200-query target are similar in employee characteristics and baseline AI usage (Appendix Table~\ref{tab:balance_branch}). Second, baseline branch characteristics and AI usage do not individually or jointly predict the timing of lower-target implementation (Appendix Table~\ref{tab:switching_timing}).\footnote{A joint F-test of baseline differences between switching and never-switching branches yields \(p=0.49\). A joint Wald test in the implementation-timing model yields \(p=0.670\).}  Third, the event-study estimates show no systematic differential pre-trends across the main outcomes.

A related concern is spillovers across branches. If employees in not-yet-treated branches learned about the lower target from already-treated branches, they might adjust their behavior before their own branch formally switched. Two facts suggest that this concern is limited. First, branches are geographically dispersed and HR enforcement is branch-specific. Second, Appendix Figure~\ref{fig:branch_month_bunching} shows that many employees in late- and never-switching branches continue to submit about 200 queries per month (near the old threshold) after other branches had switched, indicating that they still viewed the 200-query target as binding until their own branch adopted the lower target. 

Because treatment timing is staggered, standard two-way fixed effects estimates may be sensitive to treatment-effect heterogeneity across cohorts. We therefore verify robustness using the interaction-weighted estimator of \citet{sun2021estimating}, which avoids using already-treated cohorts as controls for later-treated cohorts. These estimates are reported in Section~\ref{sec:robustness}.

For the usage outcomes, we cluster standard errors at the branch level across the 42-branch sample. Sales performance is observed for employees in eight branches. We also cluster standard errors at the branch level in the baseline sales specification. Section~\ref{sec:robustness} reports wild-cluster-bootstrap inference, leave-one-sales-branch-out estimates, and additional checks addressing the implementation month, employee turnover, alternative measures of work relevance, and count-data functional form.

\section{Employee Responses to the 200-Query Mandate}
\label{sec:initial_mandate}

This section documents employee responses to the firm's initial 200-query mandate. We examine two margins. The first is whether previously inactive employees began to use the AI assistant after usage became mandatory. This captures first use and initial experimentation with the tool. The second is how employees met the requirement, as reflected in the quantity, content, and timing of their queries. Because the initial mandate applied to all branches, the evidence in this section is descriptive.

\subsection{First Use of the AI Assistant}

\begin{figure}[!htbp]
  \centering
  \includegraphics[width=0.85\textwidth]{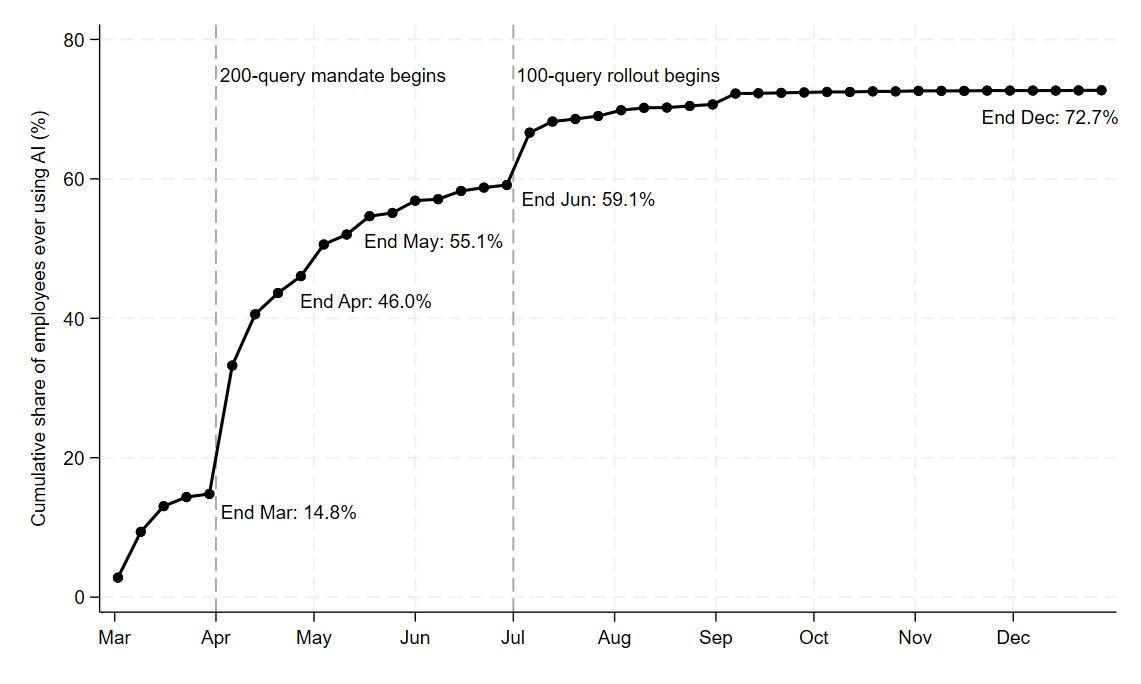}
  \caption{Cumulative Share of Employees Who Used the AI Assistant}
  \label{fig:cum_adoption_weekly}
  \captionsetup{font=footnotesize,justification=justified,singlelinecheck=false}
  \caption*{\footnotesize \textit{Notes:} This figure plots the cumulative share of employees who had submitted at least one AI query by the end of each week, from March through December 2025. Data labels report end-of-month values on the weekly series. The AI assistant was opened company-wide on March 1, 2025. The first dashed vertical line marks the beginning of the 200-query mandate, which started on March 31 and made April the first full compliance month. The second dashed vertical line marks the beginning of the 100-query target rollout in July. The sample includes non-production employees covered by the mandate and excludes headquarters, the AI department, and frontline production workers.}
\end{figure}

We first examine whether the mandate brought previously inactive employees to the tool. 
Figure~\ref{fig:cum_adoption_weekly} plots the cumulative share of employees who had submitted at least one query by the end of each week. The AI assistant was opened company-wide on March 1, 2025, but voluntary use was limited. By the end of March, only 14.8\% of employees had submitted at least one query.

First use increased sharply after the 200-query mandate began on March 31. By the end of April, the cumulative share of employees who had used the assistant reached 46\%. By the end of June, before the rollout of the lower target, it reached 59.1\%.
We interpret first use as initial experimentation with the tool, not necessarily integration into regular work.

This pattern shows that the mandate overcame substantial adoption
inertia. Making the tool available did not generate broad experimentation on its own. Once usage became a formal requirement, however, a large share of previously inactive employees started using the tool for the first time. 
% \footnote{The mandate did not move everyone. Employees who never submitted a query
% during the sample period were, on average, older, more experienced, and more likely to
% hold senior positions than employees who activated the tool.}

This extensive-margin success raises the natural question: how did employees meet the target? Do they integrate the tool into their daily work, or do they generate usage just to satisfy the metric? We next examine employee-level usage patterns.

\subsection{How Employees Met the 200-Query Requirement}

We examine usage under the 200-query target along three dimensions: the distribution of monthly query counts, the content of the queries, and the timing of usage within the month.

\begin{figure}[!htbp]
  \centering
  \begin{subfigure}[t]{0.48\textwidth}
    \centering
    \includegraphics[width=\linewidth]{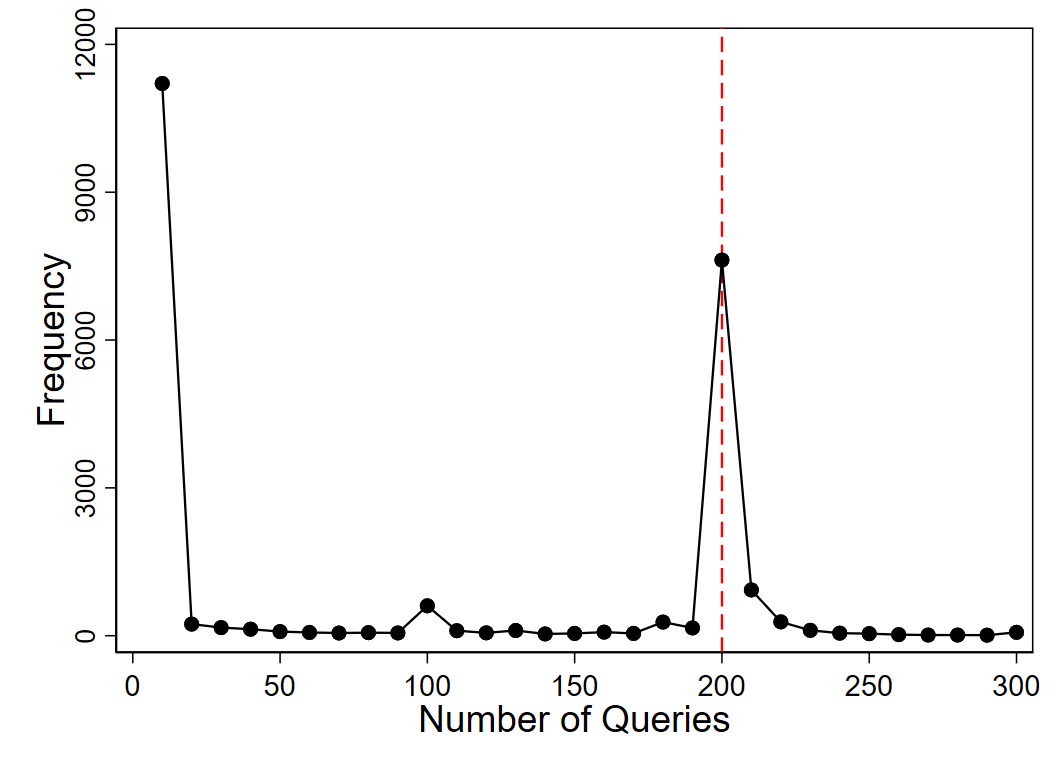}
    \caption{Distribution of Number of Queries}
    \label{fig:bunching_200}
  \end{subfigure}\hfill
  \begin{subfigure}[t]{0.48\textwidth}
    \centering
    \includegraphics[width=\linewidth]{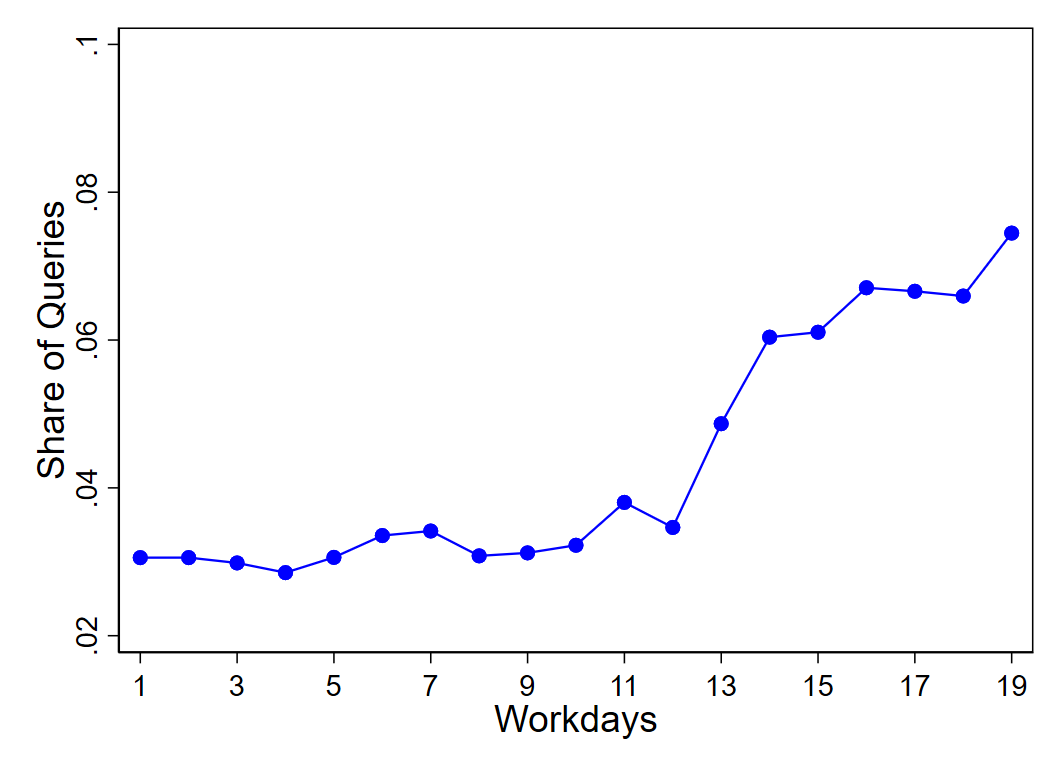}
    \caption{Share of Monthly Queries by Workday}
    \label{fig:timing_gaming_200}
  \end{subfigure}
  \caption{Distribution of Query Counts and Timing within the Month (Target = 200)}
  \label{fig:query_patterns_200}
  \captionsetup{font=footnotesize,justification=justified,singlelinecheck=false}
  \caption*{\footnotesize \textit{Notes:} This figure shows AI query usage under a target of 200 queries. Panel (a) plots the distribution of query counts. Dots plot the frequency of employee-month observations in 10-query bins, where the x-axis label denotes the lower bound of the bin. For example, the dot at 200 corresponds to employee-month observations with 200--209 monthly queries. Panel (b) plots the within-month distribution of query usage across workdays. The y-axis shows the share of queries, and the x-axis shows workdays. The sample excludes weekends and public holidays. Panel (b) is restricted to the first 19 workdays, as May 2025 had only 19 workdays in total. All analyses exclude headquarters, the AI department, and frontline production workers.}
\end{figure}

Figure~\ref{fig:query_patterns_200}(a) plots the distribution of monthly query counts. The distribution has a sharp spike at 200 queries. Among active employee-month observations, 12\% record exactly 200 queries, and 95\% submit no more than 220. Thus, very few employees continue substantially beyond the required level. This concentration near the threshold is difficult to reconcile with organic work use. If employees were using the tool only when it was useful for their tasks, there is no reason for monthly query counts to bunch exactly at the threshold. The spike instead suggests that many employees treated 200 as the relevant target: they submitted enough queries to satisfy the requirement, but had little incentive to continue once the requirement was met.\footnote{Bunching at policy thresholds is well documented as a behavioral response to incentives; see, for example, \citet{saez2010taxpayers} and \citet{kleven2016bunching}.} 

The query text provides a second opportunity to assess how employees interact with the tool to meet the requirement. We classify a query as low quality if it repeats an earlier query by the same employee in the same month or is classified as personal or off-task. Under the 200-query target, 31\% of queries fall into this category. Thus, roughly one in three counted interactions appears unlikely to reflect substantive work use.

The timing of usage provides further evidence. Figure~\ref{fig:query_patterns_200}(b) plots the share of monthly queries submitted on each workday. Activity is low and relatively flat during the first half of the month, with each workday accounting for about 3\% of monthly queries. In the second half of the month, usage rises sharply. During the final workdays of the month, each workday accounts for 7--8\% of the month's queries. This pattern suggests that many employees rushed to meet the target near the deadline rather than using the assistant steadily as part of their daily workflow.

Taken together, the initial mandate had mixed consequences. First use increased sharply after the mandate, consistent with the policy bringing previously inactive employees to the tool and encouraging experimentation. At the same time, the quantity, content, and timing of use were strongly organized around the monthly requirement. The policy increased measured AI use, but it also generated substantial gaming behavior. This mixed result provides the context for the firm's subsequent adjustment.

\section{Adjusting the AI Usage Target}
We now examine the firm's adjustment of the mandate. After employees reported that the 200-query requirement imposed a substantial time burden, headquarters did not abandon the mandate. Instead, it reduced the monthly target to 100. Because branches implemented the revised policy at different times, this staggered implementation allows us to estimate the effects of this adjustment. We first compare behavior under the two targets descriptively and then estimate the causal effects of the adjustment on usage quantity, usage composition, and sales performance.

\subsection{Descriptive Patterns of Usage under the Revised Target}

We first compare the distribution of monthly query counts, query content, and timing within the month across the two target regimes.\footnote{Appendix Figure~\ref{fig:active_effects} shows a short-run increase in first use around branch implementation of the lower target. This pattern suggests that the adjustment did not discourage experimentation and may have brought some non-users to the tool.}

\begin{figure}[!htbp]
  \centering
  \begin{subfigure}[t]{0.48\textwidth}
    \centering
    \includegraphics[width=\linewidth]{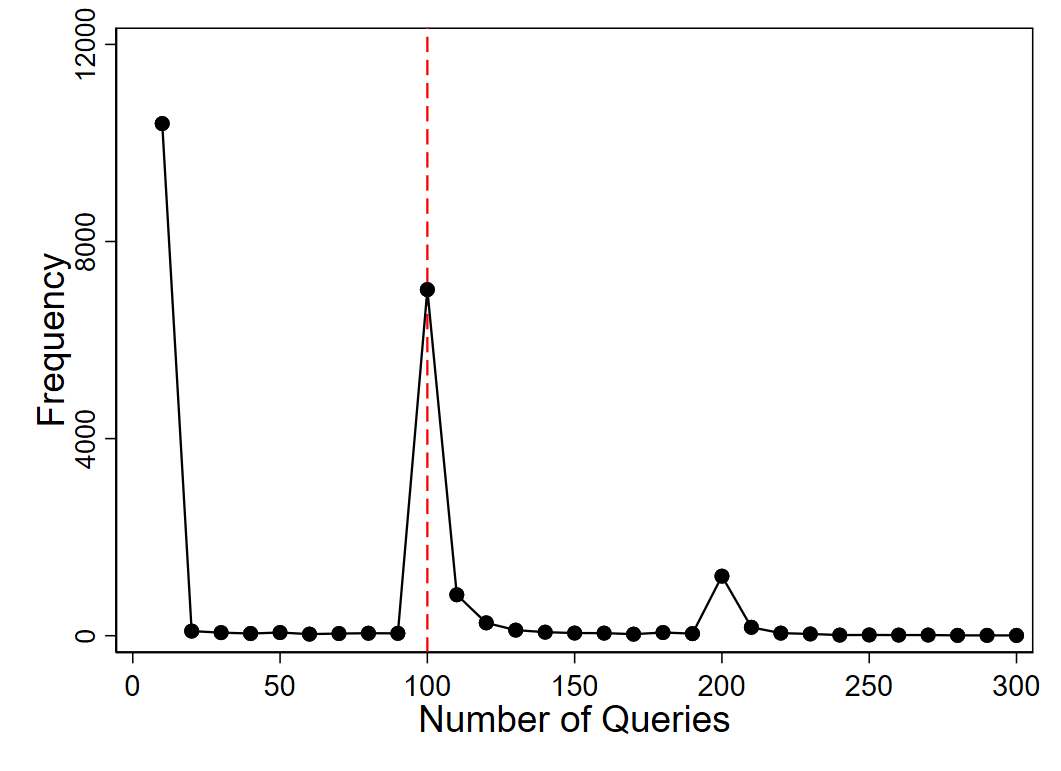}
    \caption{Distribution of Number of Queries}
    \label{fig:bunching_100}
  \end{subfigure}\hfill
  \begin{subfigure}[t]{0.48\textwidth}
    \centering
    \includegraphics[width=\linewidth]{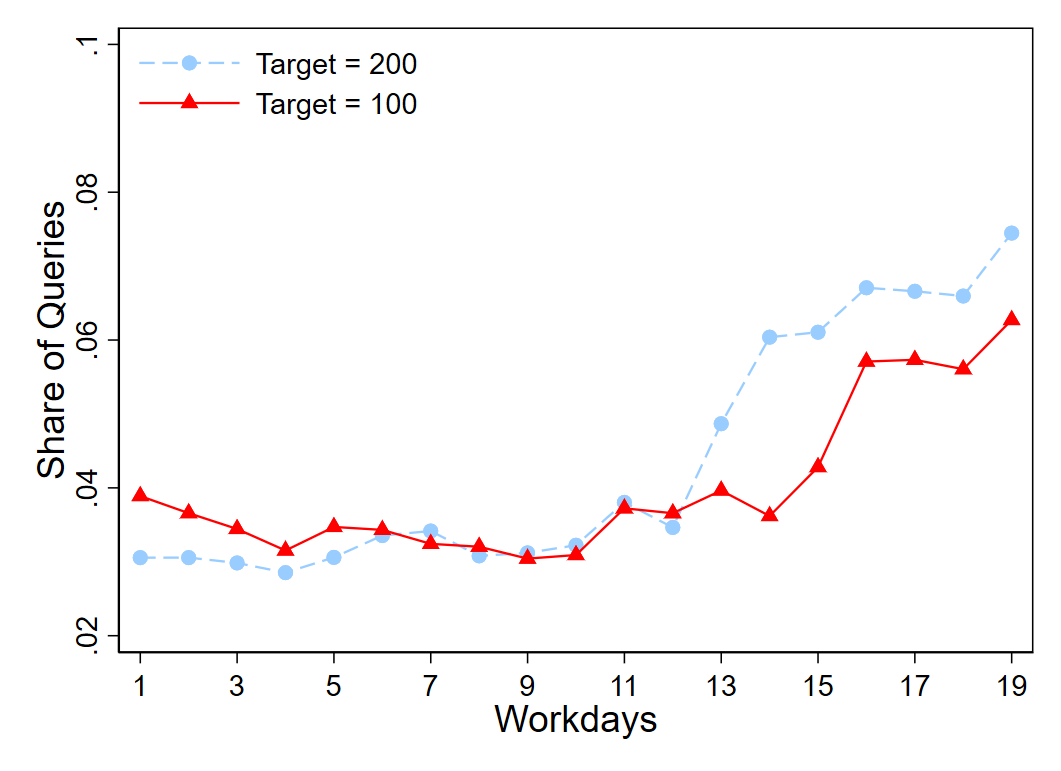}
    \caption{Share of Monthly Queries by Workday}
    \label{fig:timing_gaming_100}
  \end{subfigure}
  \caption{Distribution of Query Counts and Timing within the Month (Target = 100)}
  \label{fig:query_patterns_100}
\captionsetup{font=footnotesize,justification=justified,singlelinecheck=false}
\caption*{\footnotesize \textit{Notes:} This figure shows AI query usage under a target of 100 queries. Panel (a) plots the distribution of query counts. Dots plot the frequency of employee-month observations in 10-query bins, where the x-axis label denotes the lower bound of the bin. For example, the dot at 100 corresponds to employee-month observations with 100--109 monthly queries. Panel (b) plots the within-month distribution of query usage across workdays. The light blue dashed line corresponds to a 200-query target, and the red line to a 100-query target. The y-axis shows the share of queries, and the x-axis shows workdays. The sample excludes weekends and October because of the extended  holiday. Panel (b) is restricted to the first 19 workdays for comparison with the 200-query target. All analyses exclude headquarters, the AI department, and frontline production workers.}
\end{figure}

Figure~\ref{fig:query_patterns_100}(a) shows that the distribution of monthly query counts continues to have a sharp spike, but the spike shifts from 200 to 100, the new threshold. The share hitting the exact threshold is similar across the two regimes: among employees with positive usage, 11\% submit exactly 100 queries under the revised target, compared with 12\% who submit exactly 200 under the initial target. At the same time, more employees continue beyond the target: 25\% submit more than 110 queries, compared with 5\% who submit more than 220 under the 200-query target. The revised target therefore remained salient but was less binding for a larger share of active users.\footnote{A smaller bump remains near 200, possibly reflecting persistence of the old benchmark.}

The content of usage also improves. Employees still submit repeated or personal/off-task queries, but their share falls from 31\% under the 200-query target to 18\% under the 100-query target. This decline is consistent with less quota-filling activity under the lower target.

The end-of-month surge also attenuates. Figure~\ref{fig:query_patterns_100}(b) shows that usage still increases towards the end of the month, but the gap between early-month and late-month usage narrows, suggesting that with fewer queries required, employees do less last-minute catch-up before the deadline.

These comparisons show that employees continued to treat the
target as a threshold, but behaviors most suggestive of quota filling were less pronounced under the adjusted target. Measured usage is lower, but the share of non-repeated, work-related queries is higher. We next use the staggered branch-level implementation to estimate which marginal uses disappeared.

\subsection{Effects on Quantity and Composition of AI Use}

We now use the staggered branch-level rollout of the reduced target to estimate these effects more formally. We focus on three variables related to usage: total monthly number of queries, number of low-quality (repeated or off-task) queries, and number of high-quality (non-repeated and work-related) queries. We also examine the effects on employee performance among sales personnel. Table~\ref{tab:effects_baseline} reports the pooled
difference-in-differences estimates.

\begin{table}[!ht]
  \centering
  \begin{threeparttable}
    \captionsetup{font=small,justification=centering,singlelinecheck=false}
    \caption{Difference-in-Differences Estimates of AI Usage Quantity, Quality, and Performance}
    \label{tab:effects_baseline}

    \newcolumntype{Y}{>{\centering\arraybackslash}X}
    \setlength{\tabcolsep}{2.5pt}

    \begin{tabularx}{\textwidth}{l
      >{\hsize=1.05\hsize\centering\arraybackslash}X
      >{\hsize=1.15\hsize\centering\arraybackslash}X
      >{\hsize=1.25\hsize\centering\arraybackslash}X
      >{\hsize=.55\hsize\centering\arraybackslash}X}
      \toprule
      & (1) & (2) & (3) & (4) \\
      \midrule
      & {\small\shortstack[c]{Monthly Number of\\Queries\\\phantom{Queries}}}
      & {\small\shortstack[c]{Monthly Number\\of Repeated or\\Off-Task Queries}}
      & {\small\shortstack[c]{Monthly Number\\of Non-Repeated\\Work-Related Queries}}
      & {\small\shortstack[c]{Log\\(Sales)\\\phantom{Queries}}} \\
      \midrule
      $\textit{LowerTarget}_{jt}$ 
      & $-28.69^{***}$ 
      & $-25.97^{***}$ 
      & $-2.72$ 
      & $0.07^{**}$ \\
      & $(5.81)$ 
      & $(3.05)$ 
      & $(4.62)$ 
      & $(0.03)$ \\
      \midrule
      Month FE & Yes & Yes & Yes & Yes \\
      Individual FE & Yes & Yes & Yes & Yes \\
      Observations & 43,678 & 43,678 & 43,678 & 6,064 \\
      \bottomrule
    \end{tabularx}

    \begin{tablenotes}[flushleft]
      \footnotesize
      \item \textit{Notes:} This table reports the estimated effects of the AI usage target adjustment on employees' query behavior and sales performance using employee-month panel data. The dependent variable in column (1) is the monthly number of queries submitted by an employee in a given month. The dependent variable in column (2) is the monthly number of queries that are repeated or classified as personal or off-task. The dependent variable in column (3) is the monthly number of non-repeated queries classified as clearly, generically, or weakly work-related. The dependent variable in column (4) is the logarithm of individual monthly sales. All specifications are estimated using a two-way fixed effects model with individual and calendar-month fixed effects. Standard errors clustered at the branch level are in parentheses. * $p<0.10$, ** $p<0.05$, *** $p<0.01$.
    \end{tablenotes}
  \end{threeparttable}
\end{table}

We find that lowering the target reduces measured AI use. Monthly
query volume falls by 28.69 queries per employee-month, a 30\% decline relative
to the sample mean (Table~\ref{tab:effects_baseline}, column (1)). By itself,
this reduction could be good or bad. It would be costly if employees were cutting
productive uses of the tool. It would be beneficial if the dropped queries were
mainly queries generated to meet the original requirement.

Estimates by query type show that the decline is concentrated almost entirely in repeated or off-task use. These queries fall by 25.97 per employee-month, accounting for about 90\% of the total reduction. Non-repeated, work-related queries fall by only 2.72; the point estimate is much smaller and statistically insignificant.

\begingroup
\newcommand{\queryoutcomeplot}[2]{%
  \begin{tikzpicture}
    \node[inner sep=0,outer sep=0] (plotimage) {\includegraphics[width=\linewidth]{#1}};
    % The outcome title is embedded in the PNG. Cover only that title and
    % replace it here, while preserving the y-axis tick labels and estimates.
    \begin{scope}[overlay]
      \fill[white]
        ([xshift=0pt,yshift=2pt]plotimage.south west)
        rectangle
        ([xshift=11.5pt,yshift=-2pt]plotimage.north west);
      \node[
        rotate=90,
        align=center,
        font=\fontsize{6.0}{6.6}\selectfont
      ] at ([xshift=5.75pt]plotimage.west) {#2};
    \end{scope}
  \end{tikzpicture}%
}

\begin{figure}[!htbp]
  \centering
  \captionsetup[subfigure]{justification=centering}
  \begin{subfigure}[t]{0.48\textwidth}
    \centering
    \includegraphics[width=\linewidth]{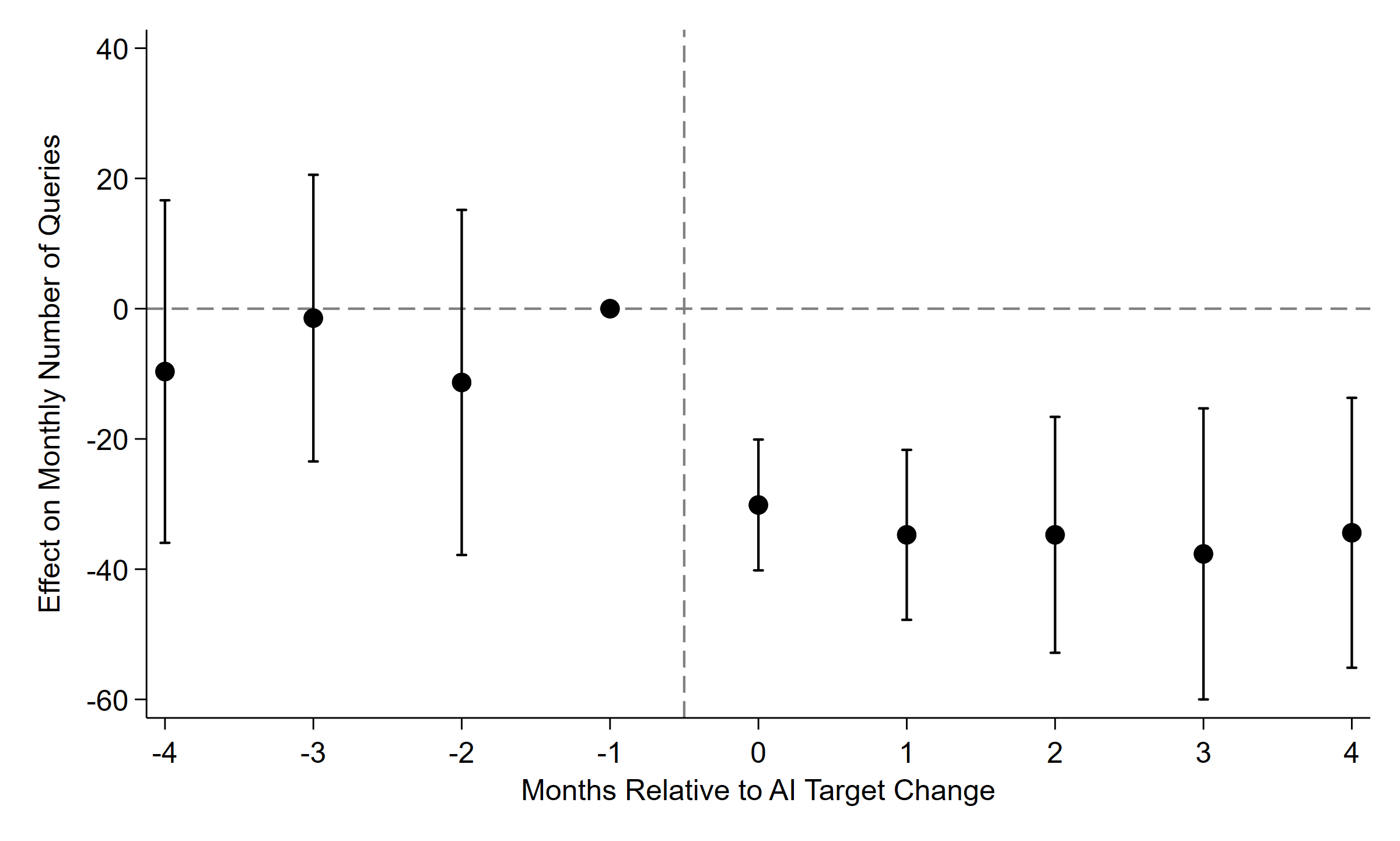}
    \caption{Monthly Number of Queries}
    \label{fig:event_study_num}
  \end{subfigure}\hfill
  \begin{subfigure}[t]{0.48\textwidth}
    \centering
    \queryoutcomeplot
      {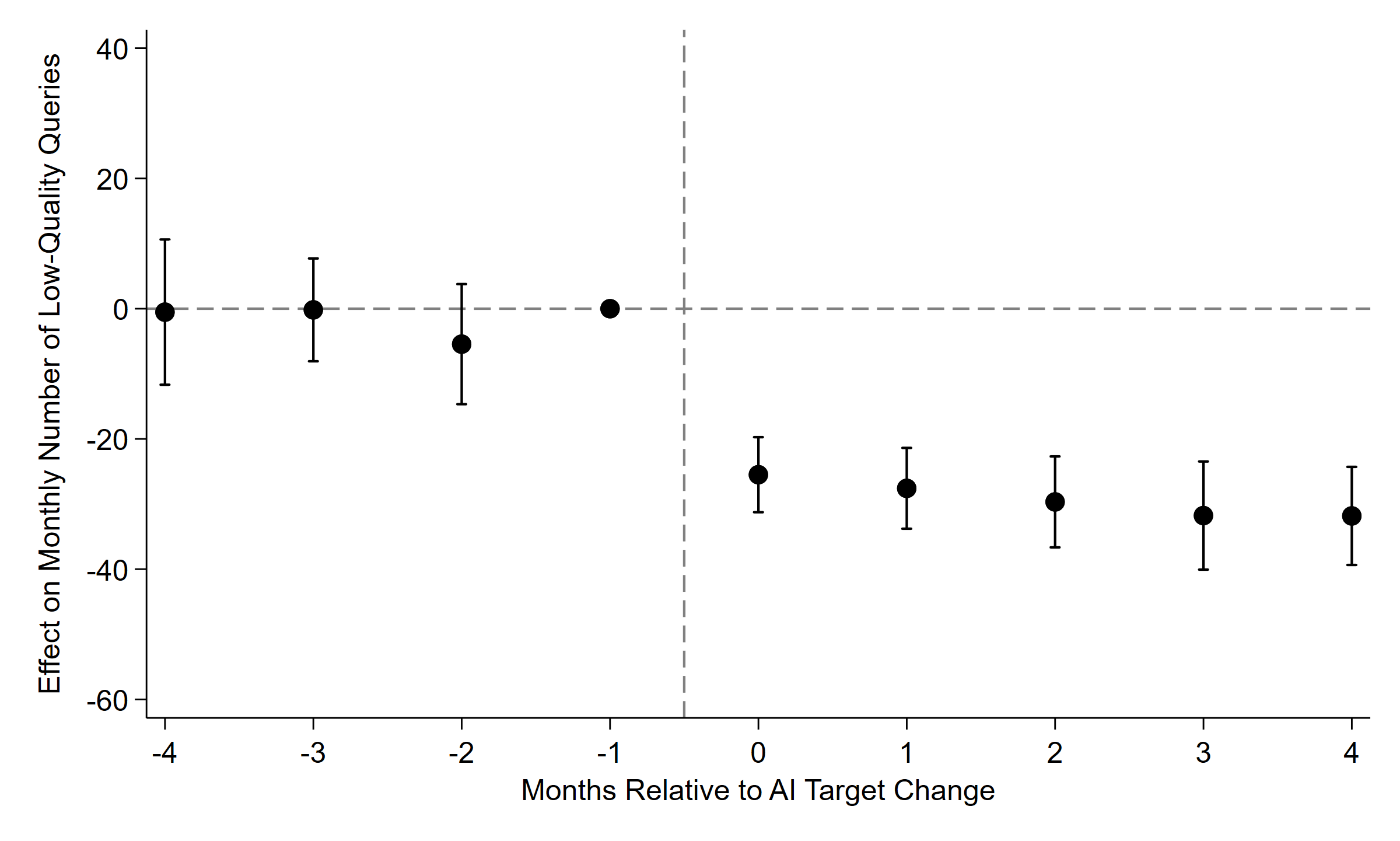}
      {Effect on Monthly Number of\\Repeated or Off-Task Queries}
    \caption{Monthly Number of Repeated or\\Off-Task Queries}
    \label{fig:event_study_low_quality}
  \end{subfigure}
  
  \vspace{1em}
  
  \begin{subfigure}[t]{0.48\textwidth}
    \centering
    \queryoutcomeplot
      {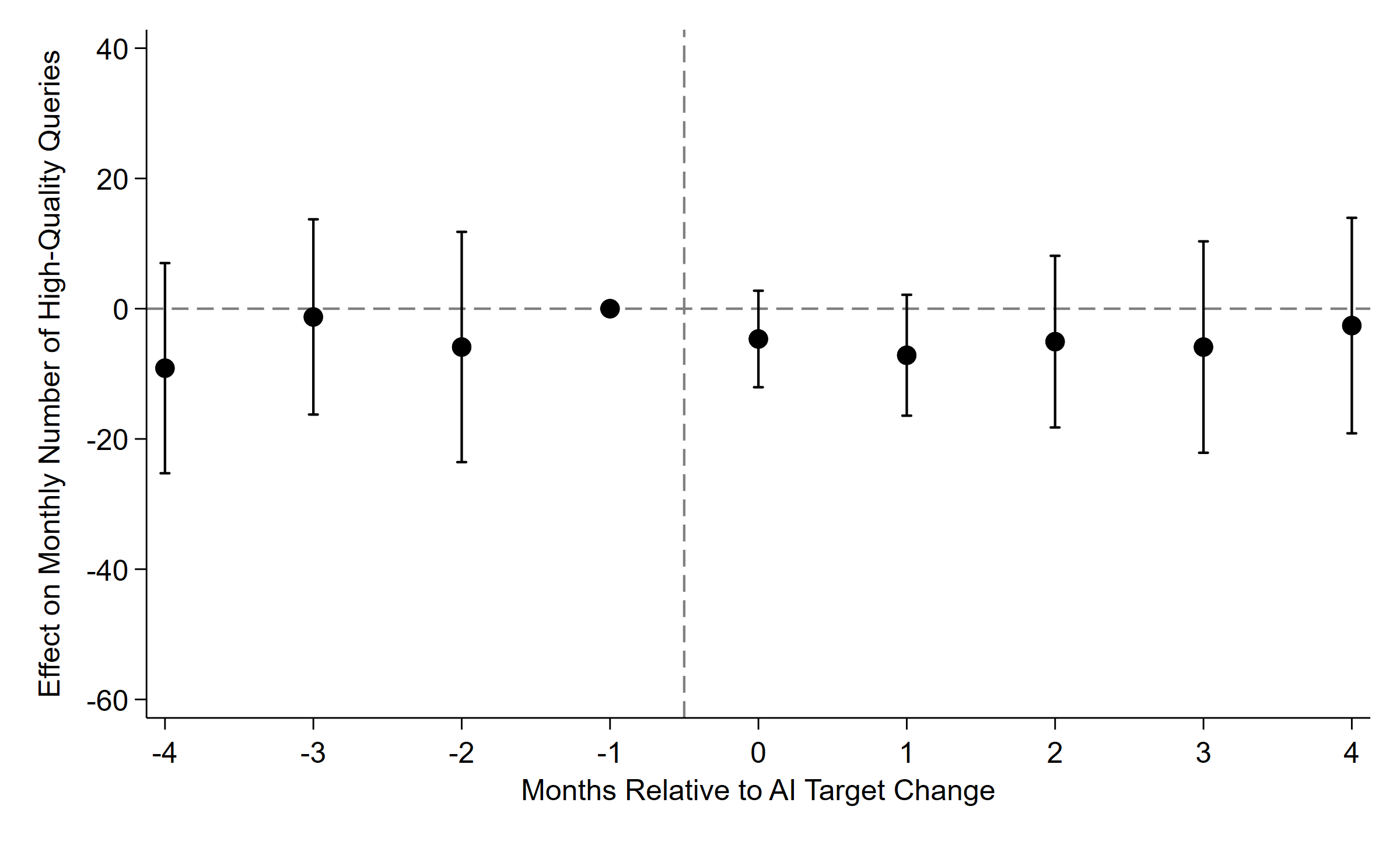}
      {Effect on Monthly Number of\\Non-Repeated Work-Related Queries}
    \caption{Monthly Number of Non-Repeated\\Work-Related Queries}
    \label{fig:event_study_high_quality}
  \end{subfigure}\hfill
  \begin{subfigure}[t]{0.48\textwidth}
    \centering
    \includegraphics[width=\linewidth]{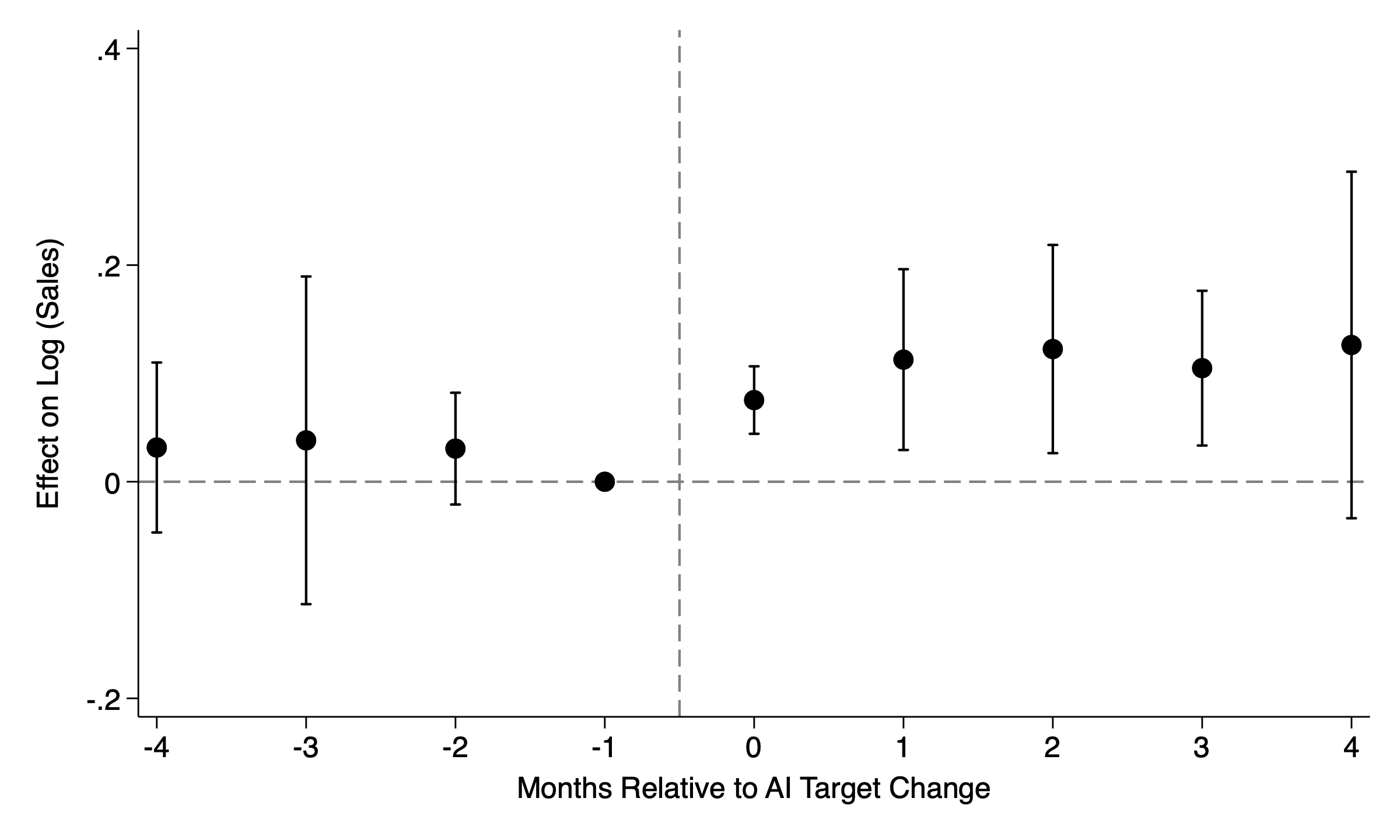}
    \caption{Log Monthly Sales}
    \label{fig:event_study_sales}
  \end{subfigure}
  
  \caption{Event-Study Estimates of AI Usage Quantity, Quality, and Performance}
  \label{fig:event_study_twfe}
  \captionsetup{font=footnotesize,justification=justified,singlelinecheck=false}
  \caption*{\footnotesize \textit{Notes:} This figure plots the dynamic effects of lowering the monthly query target, estimated using employee-month panel data from the event-study specification described in Section~\ref{sec:empirical_strategy}. The horizontal axis indexes event time in months relative to the policy implementation, with $k=-1$ (the month prior to the adjustment) as the omitted reference period. Panel (a) plots the effect on the total number of AI queries submitted by an employee in a given month. Panel (b) plots the effect on the monthly number of queries that are repeated or classified as personal or off-task. Panel (c) plots the effect on the monthly number of non-repeated queries classified as clearly, generically, or weakly work-related. Panel (d) plots the effect on the logarithm of individual monthly sales. All specifications are estimated using a two-way fixed effects model with individual and calendar-month fixed effects. Standard errors are clustered at the branch level. Markers denote the estimated coefficients and vertical bars denote 95\% confidence intervals.}
\end{figure}
\endgroup

The event-study estimates in Figure~\ref{fig:event_study_twfe} show the same
pattern dynamically. Total query volume and low-quality (repeated or off-task) query counts fall
immediately after a branch switches to the lower target and remain lower
throughout the post-treatment period. By contrast, the coefficients for high-quality (non-repeated and work-related) query counts remain close to zero after the switch. The pre-treatment coefficients are small
across the main usage outcomes, which is consistent with treated and not-yet-treated branches following similar trends before implementation.

The broad high- versus low-quality decomposition may mask changes within the work-related category. We therefore examine effects across the finer work-relevance distribution. Appendix Figure~\ref{fig:quality_distribution}, Panel (a), holds the pre-treatment relevance cutoffs fixed and estimates the effect of lowering the target on the monthly number of queries in each quintile. Query counts decline in each of the bottom four quintiles, with larger reductions toward the bottom, but increase in the top quintile. Consistent with this shift, the number of non-repeated queries receiving the maximum score on the 1--10 work-relevance scale increases by 11.5 per employee-month (Appendix Table~\ref{tab:effects_usage_content}, column (1)). Panel (b) examines a different margin: how the work-relevance score changes at different quantiles of the distribution. The improvement is concentrated in the lower tail. Work relevance increases substantially at the 10th and 25th percentiles and more modestly around the median, while the upper tail changes little. The two exercises show that lowering the target removes disproportionately more interactions from the bottom of the work-relevance distribution, improves the relevance of the remaining lower-tail queries, and increases the number of queries at the top.\footnote{Additional analyses show that prompts become more clearly specified after the target is lowered and that query intent shifts away from Asking and Expressing and toward Doing. Results are reported in Appendix Table~\ref{tab:effects_usage_content} and Appendix Figure~\ref{fig:event_study_usage_quality_interaction}.} 

Taken together, lowering the target reduced measured AI use, but the reduction was primarily compositional. Repeated or off-task queries accounted for about 90\% of the decline in usage. Non-repeated, work-related queries did not show a comparable decline, while the most task-specific queries increased. The adjustment therefore removed mainly the queries most likely to have been generated to satisfy the initial requirement. We next examine whether this shift in AI use was accompanied by changes in sales performance.

\subsection {Effects on Sales Performance}

We then examine the effect on employee performance. We observe individual output only for sales employees. In this subsample, lowering the target is
associated with a 7\% increase in monthly sales
(Table~\ref{tab:effects_baseline}, column (4)). The event-study estimates show
no visible pre-trend and indicate that sales rise after the policy change
(Figure~\ref{fig:event_study_twfe}, panel (d)). Because the sales sample spans eight branches, Section 5.5 reports wild-bootstrap inference and leave-one-branch-out estimates.\footnote{The corresponding AI-usage
estimates in the sales subsample show the same pattern as in the full sample:
the decline is concentrated in low-quality queries, with no statistically
significant decline in high-quality queries
(Appendix Table~\ref{tab:effects_baseline_sales_subsample}).}

We interpret the sales estimate as the reduced-form effect of the target
adjustment. The lower target may improve sales through several related channels:
employees may spend less time filling the quota, they may shift remaining AI use
toward more work-related tasks, or they may do less end-of-month catch-up that
crowds out regular sales work. The current data do not allow us to separate
these channels. However, the query decomposition is consistent with this
mechanism. In Appendix Table~\ref{tab:corr_ai_performance}, low-quality query
use is negatively associated with sales, while high-quality query use is
positively associated with sales, controlling for individual and month fixed
effects. These correlations are consistent with the
view that sales gains come with a shift away from low-value, target-driven usage.

The sales result shows that lower measured AI use was accompanied by higher performance. Together with the usage decomposition, this pattern suggests that the adjustment reduced the compliance burden without undermining productive use.

\subsection{Heterogeneous Responses to the Target Adjustment}
\label{sec:heterogeneity}

We next examine which employees were most affected by the firm's adjustment of the mandate. We estimate heterogeneous effects by interacting \(LowerTarget_{jt}\) in equation~\eqref{eq:did} with employee characteristics measured before the target adjustment. Appendix
Table~\ref{tab:heterogeneity_interaction} reports the results by prior usage, managerial status, work experience, and gender.

Panel A shows that the largest responses come from employees whose usage was previously anchored near the 200-query threshold. These employees reduced both the number of repeated or off-task queries and the number of non-repeated, work-related queries, and their monthly usage volume was more likely to cluster near the new threshold (Appendix Figure~\ref{fig:bunching_gamer}). By contrast, among employees whose prior use was not anchored near 200, total usage changed little; the adjustment mainly improved composition, as repeated or off-task use fell while non-repeated, work-related use rose, leaving total use almost unchanged.

Employees who had voluntarily tried the tool before the mandate also reduced total and repeated or off-task use after the target was lowered (Panel B). Thus, the original requirement appears to have been demanding even for employees with demonstrated prior interest in AI.

Differences by managerial status, experience, and gender are more modest. Managers show a larger reduction in total query volume and low-quality query counts, while differences by work experience and gender are small (Appendix Table~\ref{tab:heterogeneity_interaction}).

\subsection{Robustness}
\label{sec:robustness}

We report two sets of robustness checks. The first set addresses the research
design and inference.

% \paragraph{Treatment timing and implementation.}
% Because the timing variation in branch-level implementation is the key source of identification, we examine whether  this concern in three ways. First, branches that ever switched and branches that remained under the 200-query target are similar in baseline workforce characteristics and AI usage (Appendix Table~\ref{tab:balance_branch}; joint F-test, \(p=0.49\)). Moreover, baseline branch size, query volume, low-quality use, the share of employees near the 200-query threshold, and the share of active users do not individually or jointly predict implementation timing (Appendix Table~\ref{tab:switching_timing}; joint Wald test, \(p=0.670\)). Second, the branch-month usage patterns validate the implementation dates used in the analysis. Bunching remains centered near 200 until a branch switches and shifts toward 100 only after local implementation (Appendix Figure~\ref{fig:branch_month_bunching}). The continued bunching near 200 among not-yet-switching branches also suggests limited anticipation or spillovers. Third, the event-study estimates show no systematic differential pre-trends in the main usage and sales outcomes (Figure~\ref{fig:event_study_twfe}). Together, these checks provide no evidence that implementation timing was systematically related to pre-adjustment branch behavior or outcome trends.

\paragraph{Alternative estimators and timing specifications.}
Because treatment timing is staggered, standard two-way
fixed effects estimates can be biased when treatment effects vary across cohorts
or over event time. We therefore re-estimate the main specification using the
interaction-weighted estimator of \citet{sun2021estimating}. The estimates are
similar to the baseline results (Appendix Figure~\ref{fig:event_study_sa} and
Appendix Table~\ref{tab:effects_baseline_sa}), suggesting that the findings are
not driven by treatment-effect heterogeneity under staggered adoption.

We next check whether the estimates are sensitive to treatment of the implementation month or to differential branch trends. Dropping the month in which a branch switches
targets leaves the estimates stable, suggesting that short-run implementation
disruptions do not drive the results (Appendix
Table~\ref{tab:effects_monthly_queries_exclude_switch}). Adding
branch-specific linear time trends gives similar estimates for total queries,
low-quality queries, and sales (Appendix Table~\ref{tab:effects_monthly_queries_branch_trends}).\footnote{The high-quality-query estimate becomes more negative and marginally significant. However, the main decomposition remains: most of the decline in total query volume is still accounted for by the decline in low-quality queries.}

To assess whether the
results could arise from chance features of the rollout, we conduct a placebo test using 1,000 permutations of the observed lower-target implementation timing across branches. The resulting placebo estimates are centered around
zero, while the actual estimates for total queries, low-quality queries, and sales lie
far in the tails of the placebo distributions (Appendix
Figure~\ref{fig:placebo_tests}).\footnote{We do not report the corresponding
placebo test for high-quality query counts because the
corresponding baseline estimate is small and statistically insignificant.}

\paragraph{Inference.}
We also examine inference with a limited number of branch-level clusters. Wild
cluster bootstrap inference gives similar conclusions: the effects on total
queries, low-quality queries, and sales remain statistically significant, while
the effect on high-quality queries remains small and insignificant (Appendix
Table~\ref{tab:effects_monthly_queries_bootstrap}). Because sales are observed
for employees in eight branches, we also re-estimate the sales specification
after dropping one sales branch at a time. The leave-one-branch-out estimates
are close to the baseline estimate, suggesting that the performance result is not
driven by any particular branch (Appendix
Figure~\ref{fig:leave_one_out_sales}).

\bigskip

The second set of checks addresses sample composition, measurement, and functional
form. 
\paragraph{Sample composition, measurement, and functional form.}
Restricting the sample to employees who remain with the firm throughout
the sample period yields nearly unchanged estimates (Appendix
Table~\ref{tab:effects_monthly_queries_stayers}). The results are also similar
when we use the embedding-based work-relevance index rather than the LLM-based
classification. Under this alternative measure, lowering the target increases
measured work relevance (Appendix
Figure~\ref{fig:event_study_twfe_work_relevance_index} and Appendix
Table~\ref{tab:robust_work_relevance_index}). Finally, because query counts are
nonnegative and include many zeros, we re-estimate the quantity effects using a
Poisson model. The estimates are consistent with the baseline count results
(Appendix Table~\ref{tab:effects_queries_poisson}).

Across these robustness checks, we show that lowering the target consistently reduces total query volume, with the decline concentrated in repeated or off-task use. The estimate for work-related use remains substantially smaller than the decline in repeated or off-task queries. The positive sales estimate is also robust to wild-bootstrap inference and leave-one-branch-out estimation.

\section{Discussion}

This paper studies AI adoption as an incentive-design problem. Firms want employees to experiment with AI, but voluntary take-up is often slow. Rewarding productive use directly is difficult because the returns to individual uses are hard to observe. Firms may therefore rely on a readily measured input, such as the number of interactions with the tool. The challenge is that firms do not know ex ante how employees will respond to a given target. In our setting, the initial 200-query mandate was followed by a sharp increase in first use. At the same time, usage  showed substantial gaming. This is consistent with the goal-setting literature that goals can direct effort, but overly narrow or demanding goals can distort behavior \citep{locke2002building, schweitzer2004goal, ordonez2009goals}. 

The policy question is what firms should do once the effects of an initial policy become visible. One possible response to gaming is to eliminate the target. Our setting illustrates another: retain the mandate but reduce its intensity. Our results show that when the firm lowered the target to 100, monthly usage fell, but about 90\% of the decline came from repeated or off-task queries. Work-related use did not show a comparable decline, and sales increased among the employees for whom we observe individual output.

More broadly, the findings show why adoption incentives may need to be revised over time. The initial policy can generate useful information about how employees respond. In our setting, employee feedback revealed that the initial mandate imposed a substantial burden. Initial implementation therefore changed the value of the firm’s subsequent policy choices. An adjustment does not necessarily mean that the initial policy failed. It can be part of the process through which an organization learns how to manage a new technology.

% Measured adoption is therefore not the same as productive adoption \textcolor{red}{\citep{brynjolfsson2021jcurve,mcelheran2025rise,yotzov2026firm}}.

For firms considering incentives for AI adoption, the lesson is not that lower targets are always better. A target that is too low may fail to push employees past the initial costs of experimentation, while a target that is too high may induce them to just meet the requirement without productive engagement. Firms can monitor not only aggregate usage but also its content, and revise the policy as they learn how employees respond.

This study has several limitations. First, because the initial 200-query target was introduced to all branches at the same time, we cannot causally identify the effect of introducing the mandate. Our evidence concerns the effect of lowering an existing target. The results do not determine the optimal target level or the optimal sequence of targets.

Second, we observe employees' queries to the AI assistant, but we do not observe the entire work process. Specifically, we do not know whether employees use the AI responses or how those responses affect task-level output. Future work could better link AI usage to downstream decisions and task outcomes.

Third, the performance results pertain only to sales personnel. Although sales provides a straightforward individual-level performance measure and is important to the firm's actual business operations, AI may also affect other functions and business processes within the firm. For instance, in R\&D, technical support, and administrative units, the relevant performance outcomes may include faster project cycles, fewer errors, better documentation, or smoother internal coordination. Future work that measures these outcomes would be helpful for examining the broader effects of AI usage in organizations.

\clearpage

\bibliographystyle{aer}
\bibliography{references}

@article{holmstrom1991multitask,
  author  = {Holmstrom, Bengt and Milgrom, Paul},
  title   = {Multitask Principal-Agent Analyses: Incentive Contracts, Asset Ownership, and Job Design},
  journal = {Journal of Law, Economics, and Organization},
  year    = {1991},
  volume  = {7},
  number  = {Special Issue},
  pages   = {24--52}
}

@article{baker1992incentive,
  author  = {Baker, George},
  title   = {Incentive Contracts and Performance Measurement},
  journal = {Journal of Political Economy},
  year    = {1992},
  volume  = {100},
  number  = {3},
  pages   = {598--614}
}

@article{oyer1998fiscal,
  author  = {Oyer, Paul},
  title   = {Fiscal Year Ends and Nonlinear Incentive Contracts: The Effect on Business Seasonality},
  journal = {The Quarterly Journal of Economics},
  year    = {1998},
  volume  = {113},
  number  = {1},
  pages   = {149--185}
}

@article{larkin2014cost,
  author  = {Larkin, Ian},
  title   = {The Cost of High-Powered Incentives: Employee Gaming in Enterprise Software Sales},
  journal = {Journal of Labor Economics},
  year    = {2014},
  volume  = {32},
  number  = {2},
  pages   = {199--227}
}

@article{courty2004empirical,
  author  = {Courty, Pascal and Marschke, Gerald},
  title   = {An Empirical Investigation of Gaming Responses to Explicit Performance Incentives},
  journal = {Journal of Labor Economics},
  year    = {2004},
  volume  = {22},
  number  = {1},
  pages   = {23--56}
}

@article{jacob2003rotten,
  author  = {Jacob, Brian A. and Levitt, Steven D.},
  title   = {Rotten Apples: An Investigation of the Prevalence and Predictors of Teacher Cheating},
  journal = {The Quarterly Journal of Economics},
  year    = {2003},
  volume  = {118},
  number  = {3},
  pages   = {843--877}
}

@article{brynjolfsson1996paradox,
  author  = {Brynjolfsson, Erik and Hitt, Lorin M.},
  title   = {Paradox Lost? Firm-Level Evidence on the Returns to Information Systems Spending},
  journal = {Management Science},
  year    = {1996},
  volume  = {42},
  number  = {4},
  pages   = {541--558}
}

@article{bloom2012americans,
  author  = {Bloom, Nicholas and Sadun, Raffaella and Van Reenen, John},
  title   = {Americans Do IT Better: US Multinationals and the Productivity Miracle},
  journal = {American Economic Review},
  year    = {2012},
  volume  = {102},
  number  = {1},
  pages   = {167--201}
}

@article{forman2005location,
  author  = {Forman, Chris and Goldfarb, Avi and Greenstein, Shane},
  title   = {How Did Location Affect Adoption of the Commercial Internet? Global Village vs. Urban Leadership},
  journal = {Journal of Urban Economics},
  year    = {2005},
  volume  = {58},
  number  = {3},
  pages   = {389--420}
}

@article{sun2021estimating,
  author  = {Sun, Liyang and Abraham, Sarah},
  title   = {Estimating Dynamic Treatment Effects in Event Studies with Heterogeneous Treatment Effects},
  journal = {Journal of Econometrics},
  year    = {2021},
  volume  = {225},
  number  = {2},
  pages   = {175--199}
}

@article{brynjolfsson2025generative,
  author  = {Brynjolfsson, Erik and Li, Danielle and Raymond, Lindsey},
  title   = {Generative AI at Work},
  journal = {The Quarterly Journal of Economics},
  year    = {2025},
  volume  = {140},
  number  = {2},
  pages   = {889--942}
}

@article{noy2023experimental,
  author  = {Noy, Shakked and Zhang, Whitney},
  title   = {Experimental Evidence on the Productivity Effects of Generative Artificial Intelligence},
  journal = {Science},
  year    = {2023},
  volume  = {381},
  number  = {6654},
  pages   = {187--192}
}

@article{dellacqua2026navigating,
  author  = {Dell'Acqua, Fabrizio and McFowland III, Edward and Mollick, Ethan and Lifshitz, Hila and Kellogg, Katherine C. and Rajendran, Saran and Krayer, Lisa and Candelon, Fran\c{c}ois and Lakhani, Karim R.},
  title   = {Navigating the Jagged Technological Frontier: Field Experimental Evidence of the Effects of Artificial Intelligence on Knowledge Worker Productivity and Quality},
  journal = {Organization Science},
  year    = {2026},
  volume  = {37},
  number  = {2},
  pages   = {403--423},
  doi     = {10.1287/orsc.2025.21838}
}

@article{chatterji2025people,
  author      = {Chatterji, Aaron and Cunningham, Thomas and Deming, David J. and Hitzig, Zoe and Ong, Christopher and Shan, Carl Yan and Wadman, Kevin},
  title       = {How People Use ChatGPT},
  journal = {NBER Working Paper},
  number      = {34255},
  year        = {2025},
  doi         = {10.3386/w34255}
}

@article{handa2025economic,
  author  = {Handa, Kunal and Tamkin, Alex and McCain, Miles and Huang, Saffron and Durmus, Esin and Heck, Sarah and Mueller, Jared and Hong, Jerry and Ritchie, Stuart and Belonax, Tim and Troy, Kevin K. and Amodei, Dario and Kaplan, Jared and Clark, Jack and Ganguli, Deep},
  title   = {Which Economic Tasks Are Performed with AI? Evidence from Millions of Claude Conversations},
  journal = {arXiv Preprint arXiv:2503.04761},
  year    = {2025}
}

@article{kerr1975folly,
  title={On the Folly of Rewarding A, while Hoping for B},
  author={Kerr, Steven},
  journal={Academy of Management Journal},
  volume={18},
  number={4},
  pages={769--783},
  year={1975},
  publisher={Academy of Management Briarcliff Manor, NY 10510}
}

@article{cui2026effects,
  title={The Effects of Generative AI on High-Skilled Work: Evidence from Three Field Experiments with Software Developers},
  author={Cui, Kevin Zheyuan and Demirer, Mert and Jaffe, Sonia and Musolff, Leon and Peng, Sida and Salz, Tobias},
  journal={Management Science},
  year={2026},
  note={Articles in Advance},
  doi={10.1287/mnsc.2025.00535},
  publisher={INFORMS}
}

@article{ordonez2009goals,
  title={Goals Gone Wild: The Systematic Side Effects of Overprescribing Goal Setting},
  author={Ord{\'o}{\~n}ez, Lisa D and Schweitzer, Maurice E and Galinsky, Adam D and Bazerman, Max H},
  journal={Academy of Management Perspectives},
  volume={23},
  number={1},
  pages={6--16},
  year={2009},
  publisher={Academy of Management Briarcliff Manor, NY}
}

@article{schweitzer2004goal,
  title={Goal Setting as a Motivator of Unethical Behavior},
  author={Schweitzer, Maurice E and Ord{\'o}{\~n}ez, Lisa and Douma, Bambi},
  journal={Academy of Management Journal},
  volume={47},
  number={3},
  pages={422--432},
  year={2004},
  publisher={Academy of Management Briarcliff Manor, NY 10510}
}

@article{locke2002building,
  title={Building a Practically Useful Theory of Goal Setting and Task Motivation: A 35-year Odyssey},
  author={Locke, Edwin A and Latham, Gary P},
  journal={American Psychologist},
  volume={57},
  number={9},
  pages={705--717},
  year={2002},
  publisher={American Psychological Association}
}

@article{bousquette2026tokenmaxxing,
  author  = {Bousquette, Isabelle},
  title   = {Why Some Companies Say AI `Tokenmaxxing' Is Key to Survival},
  journal = {The Wall Street Journal},
  year    = {2026},
  month   = apr,
  day     = {14}
}

@article{saez2010taxpayers,
  title={Do Taxpayers Bunch at Kink Points?},
  author={Saez, Emmanuel},
  journal={American Economic Journal: Economic Policy},
  volume={2},
  number={3},
  pages={180--212},
  year={2010},
  doi={10.1257/pol.2.3.180},
  publisher={American Economic Association}
}

@article{kleven2016bunching,
  title={Bunching},
  author={Kleven, Henrik Jacobsen},
  journal={Annual Review of Economics},
  volume={8},
  number={1},
  pages={435--464},
  year={2016},
  publisher={Annual Reviews}
}

@article{bick2026rapid,
  author  = {Bick, Alexander and Blandin, Adam and Deming, David J.},
  title   = {The Rapid Adoption of Generative {AI}},
  journal = {Management Science},
  year    = {2026},
  note    = {Articles in Advance},
  doi     = {10.1287/mnsc.2025.02523}
}

@article{humlum2025unequal,
  author  = {Humlum, Anders and Vestergaard, Emilie},
  title   = {The Unequal Adoption of {ChatGPT} Exacerbates Existing Inequalities among Workers},
  journal = {Proceedings of the National Academy of Sciences},
  year    = {2025},
  volume  = {122},
  number  = {1},
  pages   = {e2414972121},
  doi     = {10.1073/pnas.2414972121}
}

@article{dillon2025shifting,
  author      = {Dillon, Eleanor W. and Jaffe, Sonia and Immorlica, Nicole and Stanton, Christopher T.},
  title       = {Shifting Work Patterns with Generative {AI}},
  journal = {NBER Working Paper},
  number      = {33795},
  year        = {2025},
  doi         = {10.3386/w33795}
}

@article{hartley2025labor,
  author      = {Hartley, Jonathan S. and Jolevski, Filip and Melo, Vitor and Moore, Brendan},
  title       = {The Labor Market Effects of Generative Artificial Intelligence},
  journal = {SSRN Working Paper},
  number      = {5136877},
  year        = {2025},
  note        = {Revised January 2026},
  doi         = {10.2139/ssrn.5136877}
}

@article{schubert2025organizational,
  author      = {Schubert, Gregor},
  title       = {Organizational Technology Ladders: Remote Work and Generative {AI} Adoption},
  journal = {SSRN Working Paper},
  number      = {5094265},
  year        = {2025},
  note        = {Revised January 2026},
  doi         = {10.2139/ssrn.5094265}
}

@article{hosseinimaasoum2025generative,
  author      = {{Hosseini Maasoum}, Seyed Mahdi and Lichtinger, Guy},
  title       = {Generative {AI} as Seniority-Biased Technological Change: Evidence from {U.S.} R\'esum\'e and Job Posting Data},
  journal = {SSRN Working Paper},
  number      = {5425555},
  year        = {2025},
  note        = {Revised November 2025},
  doi         = {10.2139/ssrn.5425555}
}

@article{counts2026enterprise,
  author  = {Counts, Scott and Chen, Yan and Dong, Jing and Sharma, Himanshu and Zaikin, Andrey and Hu, Rui and Kok, Alperen and Yilmaz, Gorkem Ozer and Suri, Siddharth and Tomlinson, Kiran and Jaffe, Sonia and Wang, Will},
  title   = {{AI} in the Enterprise: How People Use {M365 Copilot Chat}},
  journal = {arXiv Preprint arXiv:2605.23958},
  year    = {2026}
}

@article{li2025overcoming,
  title={Overcoming the Organizational Barriers to {AI} Adoption},
  author={Li, Jin and Zhu, Feng and Hua, Pascal},
  journal={Harvard Business Review},
  year={2025},
  month = nov,
  day={11},
  url={https://hbr.org/2025/11/overcoming-the-organizational-barriers-to-ai-adoption}
}

@article{laffont1988dynamics,
  title={The Dynamics of Incentive Contracts},
  author={Laffont, Jean-Jacques and Tirole, Jean},
  journal={Econometrica},
  volume={56},
  number={5},
  pages={1153--1175},
  year={1988},
  doi={10.2307/1911362},
  publisher={JSTOR}
}

@article{courty2003dynamics,
  title={Dynamics of Performance-Measurement Systems},
  author={Courty, Pascal and Marschke, Gerald},
  journal={Oxford Review of Economic Policy},
  volume={19},
  number={2},
  pages={268--284},
  year={2003},
  doi={10.1093/oxrep/19.2.268},
  publisher={Oxford University Press}
}

@article{bresnahan2017adoption,
  title={Adoption of New Information and Communications Technologies in the Workplace Today},
  author={Bresnahan, Timothy and Yin, Pai-Ling},
  journal={Innovation Policy and the Economy},
  volume={17},
  number={1},
  pages={95--124},
  year={2017},
  doi={10.1086/688846},
  publisher={University of Chicago Press Chicago, IL}
}

@article{li2021learning,
  title={Learning to Game the System},
  author={Li, Jin and Mukherjee, Arijit and Vasconcelos, Luis},
  journal={The Review of Economic Studies},
  volume={88},
  number={4},
  pages={2014--2041},
  year={2021},
  doi={10.1093/restud/rdaa065},
  publisher={Oxford University Press}
}

@article{cohen1960coefficient,
  author  = {Cohen, Jacob},
  title   = {A Coefficient of Agreement for Nominal Scales},
  journal = {Educational and Psychological Measurement},
  year    = {1960},
  volume  = {20},
  number  = {1},
  pages   = {37--46},
  doi     = {10.1177/001316446002000104}
}

@article{fleiss1971measuring,
  author  = {Fleiss, Joseph L.},
  title   = {Measuring Nominal Scale Agreement Among Many Raters},
  journal = {Psychological Bulletin},
  year    = {1971},
  volume  = {76},
  number  = {5},
  pages   = {378--382},
  doi     = {10.1037/h0031619}
}

@article{landis1977measurement,
  author  = {Landis, J. Richard and Koch, Gary G.},
  title   = {The Measurement of Observer Agreement for Categorical Data},
  journal = {Biometrics},
  year    = {1977},
  volume  = {33},
  number  = {1},
  pages   = {159--174},
  doi     = {10.2307/2529310}
}

@article{benson2015agents,
  title={Do Agents Game Their Agents’ Behavior? Evidence from Sales Managers},
  author={Benson, Alan},
  journal={Journal of Labor Economics},
  volume={33},
  number={4},
  pages={863--890},
  year={2015},
  publisher={University of Chicago Press Chicago, IL}
}

\clearpage % 在主文档中换页
\appendix

\clearpage
\refstepcounter{section}
\section*{Appendix \thesection: Figures and Tables}
\addcontentsline{toc}{section}{Appendix \thesection: Figures and Tables}
\label{app:figures_tables}

\renewcommand{\thepage}{A\arabic{page}}
\setcounter{page}{1}

% Use the current appendix letter, not hard-coded A
\renewcommand{\thefigure}{\thesection\arabic{figure}}
\renewcommand{\thetable}{\thesection\arabic{table}}

\setcounter{figure}{0}
\setcounter{table}{0}

%********** Appendix figures **********%
\begin{figure}[!htbp]
  \centering
  \caption{Policy Timeline and Staggered Branch Implementation}
  \label{fig:appendix_policy_timeline}
  \resizebox{0.95\textwidth}{!}{%
  \begin{tikzpicture}[x=1cm,y=1cm,>=stealth]
    \definecolor{timelineblue}{RGB}{58,67,185}
    \definecolor{timelinelight}{RGB}{120,120,120}

    \draw[line width=1pt, color=timelineblue] (0,0) -- (13.2,0);
    \draw[line width=1pt, color=timelinelight] (0.5,-4.15) -- (12.1,-4.15);
    \fill[timelinelight] (0.5,-4.15) circle (0.12);
    \fill[timelinelight] (12.1,-4.15) circle (0.12);

    \draw[line width=1.2pt, color=timelineblue] (0.45,0) circle (0.22);
    \node[align=center, text width=2.4cm, font=\small] at (0.45,1.05) {200-query mandate begins};
    \node[align=center, color=timelineblue, font=\small] at (0.45,-0.95) {Mar 31\\2025};

    \draw[line width=1.2pt, color=timelineblue] (6.6,0) circle (0.22);
    \node[align=center, text width=2.6cm, font=\small] at (6.6,1.05) {Target lowered\\$200 \rightarrow 100$};
    \node[align=center, color=timelineblue, font=\small] at (6.6,-0.8) {Jul};

    \foreach \x/\m in {7.7/Aug,8.8/Sep,9.9/Oct,11.0/Nov,12.1/Dec} {
      \draw[line width=0.8pt, color=timelinelight] (\x,-0.18) -- (\x,0.18);
      \node[color=gray!80, font=\small] at (\x,-0.8) {\m};
    }

    \node[anchor=west, color=timelineblue, font=\footnotesize] at (0.0,-2.95) {\textbf{Branch implementation}};

    \foreach \x/\n/\filled in {6.6/19/1,7.7/2/1,8.8/3/1,9.9/0/0,11.0/2/1,12.1/1/1} {
      \node[color=timelineblue] at (\x,-2.35) {\textbf{\n}};
      \ifnum\filled=1
        \fill[timelineblue] (\x,-2.95) circle (0.12);
      \else
        \draw[color=timelinelight, line width=0.8pt] (\x,-2.95) circle (0.12);
      \fi
    }

    \node[align=center, text width=8.5cm, color=timelinelight, font=\small] at (6.3,-4.7) {15 branches remain under the 200-query target through December 2025 (untreated)};
  \end{tikzpicture}%
  }
  \captionsetup{font=footnotesize,justification=justified,singlelinecheck=false}
  \caption*{\footnotesize \textit{Notes:} The figure summarizes the rollout of the mandatory AI-usage policy. The 200-query mandate for non-production employees began on March 31, 2025. Management lowered the target to 100 queries in July 2025. Branch-level implementation of the lower target was staggered: 19 branches switched in July, 2 in August, 3 in September, none in October, 2 in November, and 1 in December; 15 branches remained under the 200-query regime through December 2025.}
\end{figure}
\begin{figure}[!htbp]
  \centering

  \begin{subfigure}[t]{0.48\textwidth}
    \centering
    \includegraphics[width=\textwidth]{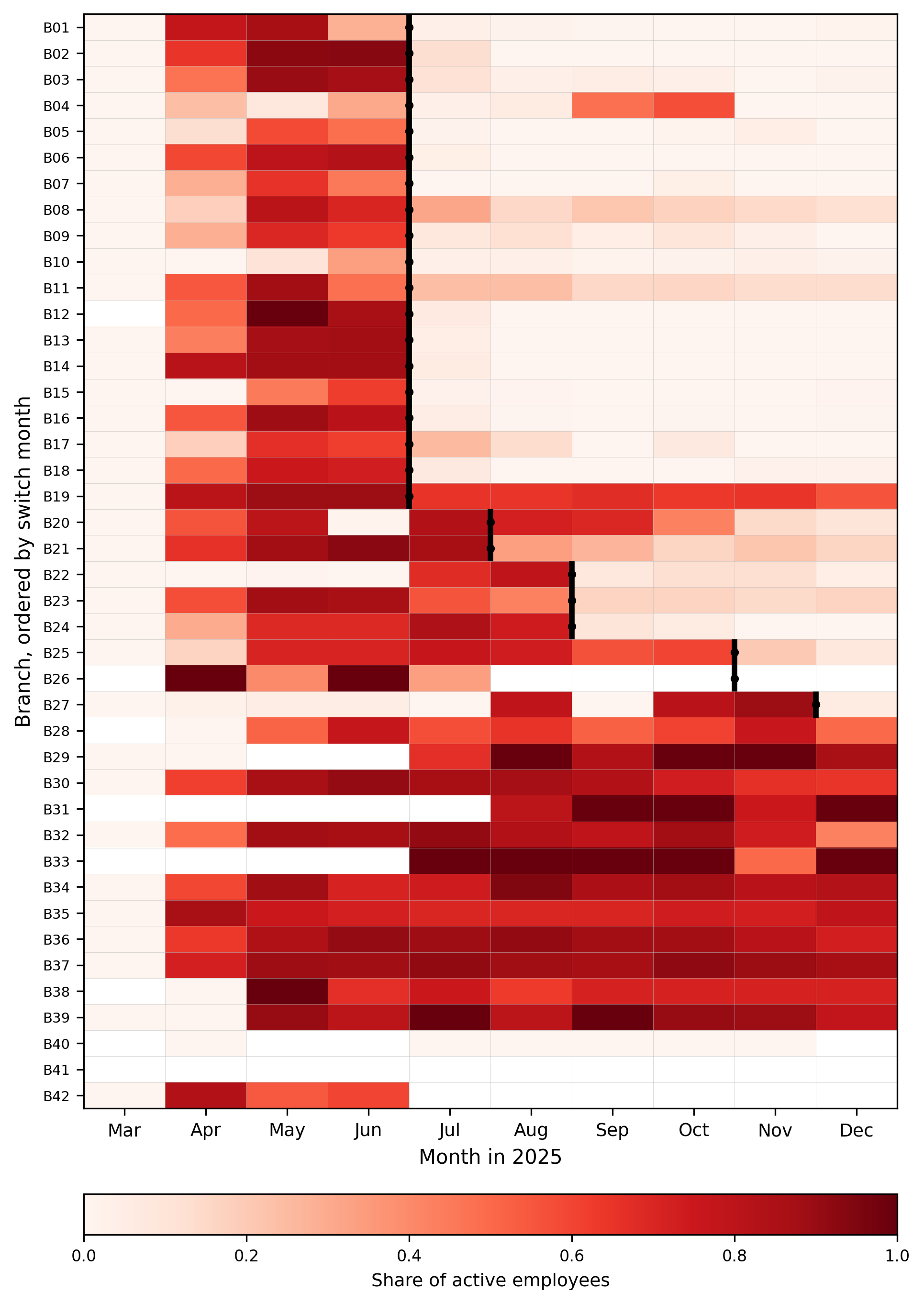}
    \caption{200-Query Target}
  \end{subfigure}
  \hfill
  \begin{subfigure}[t]{0.48\textwidth}
    \centering
    \includegraphics[width=\textwidth]{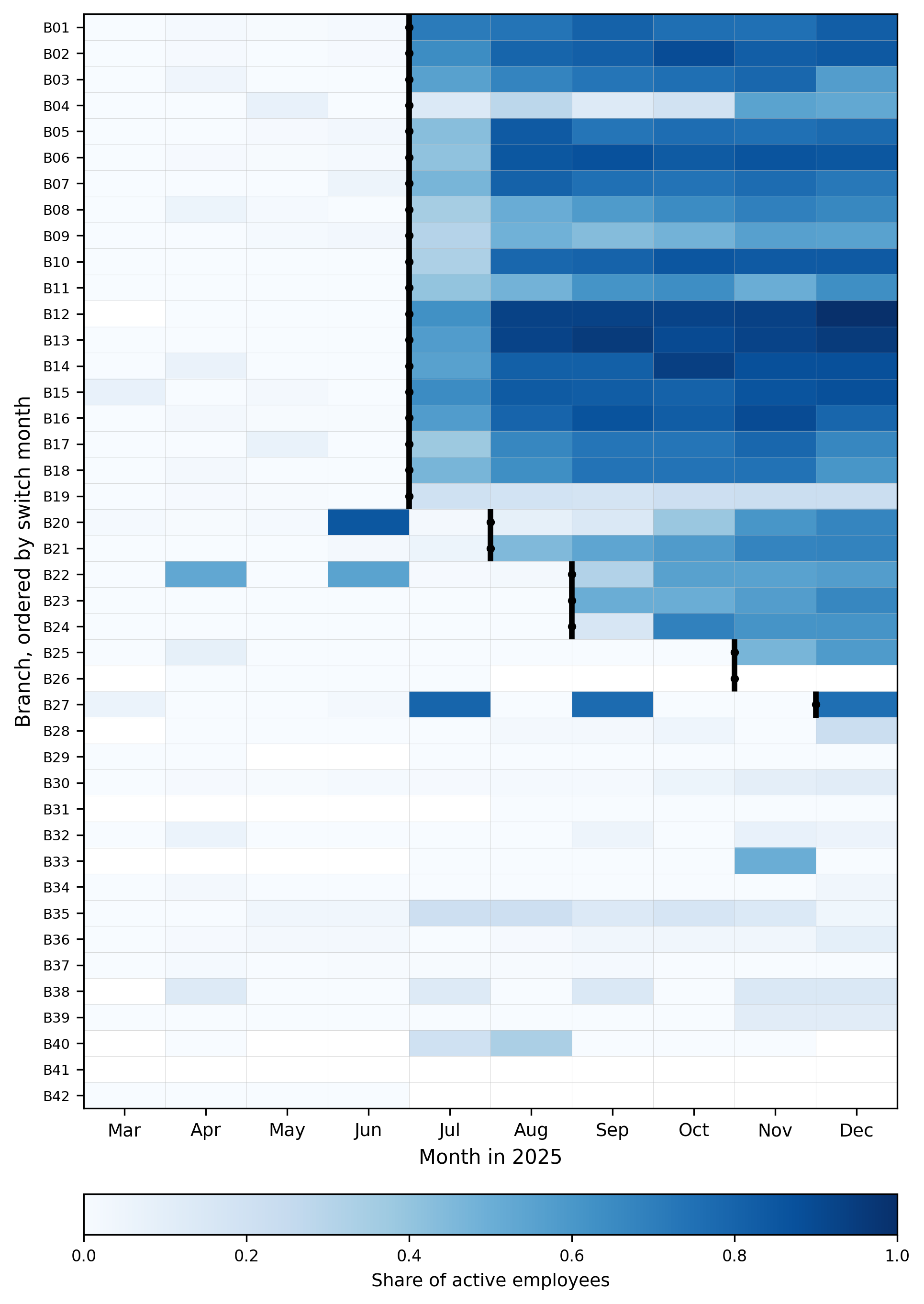}
    \caption{100-Query Target}
  \end{subfigure}

  \caption{Branch-Month Bunching Around the 200- and 100-Query Targets}
  \label{fig:branch_month_bunching}

  \captionsetup{font=footnotesize,justification=justified,singlelinecheck=false}
  \caption*{\footnotesize \textit{Notes:} This figure shows bunching at the branch-month level. Panel (a) reports the share of active employees in each branch-month whose monthly query count falls near the 200-query target, defined as 180--220 queries. Panel (b) reports the share whose monthly query count falls near the 100-query target, defined as 90--110 queries. Rows are branches ordered by the month in which they switched to the 100-query target, and columns are months from March to December 2025. The black vertical markers indicate the first month in which each branch switched from the 200-query target to the 100-query target. Cells with only one active employee are shown in white.}
\end{figure}

\begin{figure}[!htbp]
  \centering

  \begin{subfigure}[t]{\textwidth}
    \centering
    \includegraphics[width=0.7\linewidth]{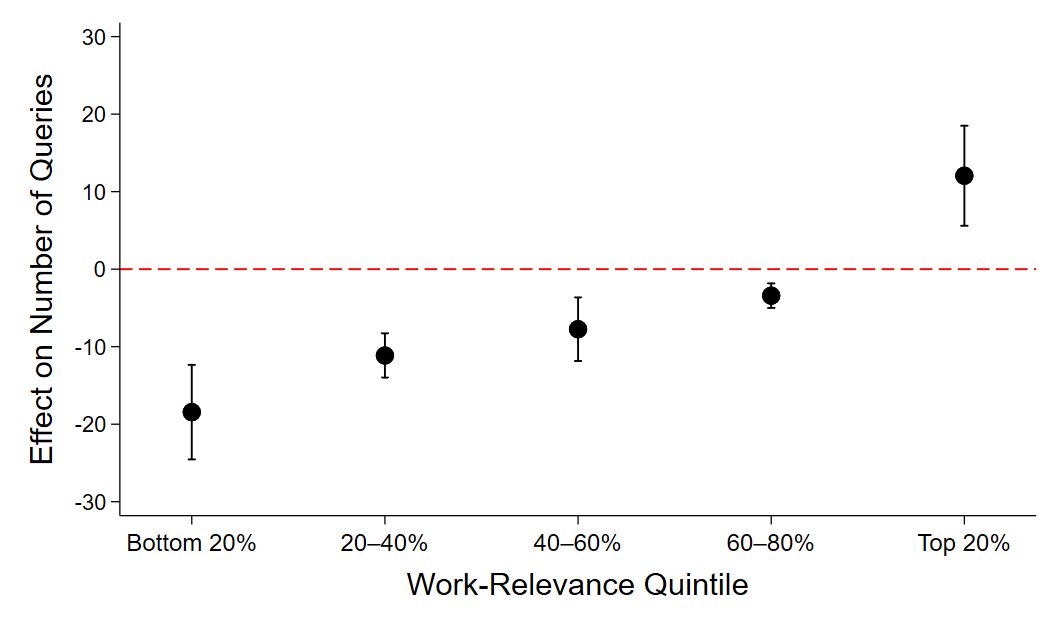}
    \caption{Effects on Query Counts by Work-Relevance Quintile}
    \label{fig:quintile_num}
  \end{subfigure}
  
  \vspace{1em}
  
  \begin{subfigure}[t]{\textwidth}
    \centering
    \includegraphics[width=0.7\linewidth]{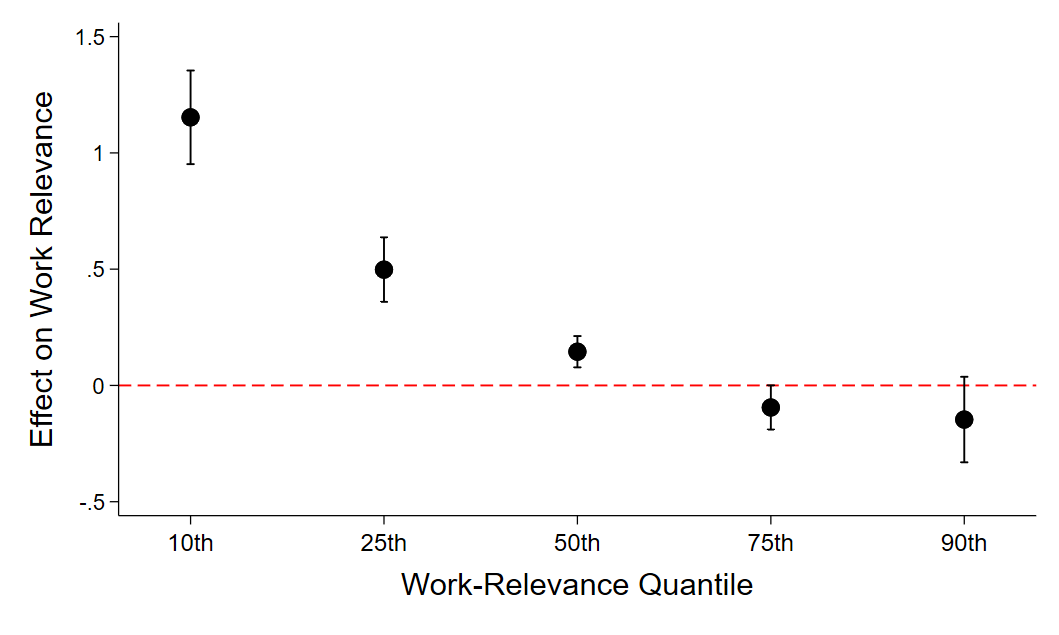}
    \caption{Quantile Treatment Effects on Work Relevance}
    \label{fig:quantile_reg}
  \end{subfigure}

  \caption{Effects of Lowering the AI Usage Target Across the Work-Relevance Distribution}
  \label{fig:quality_distribution}

\captionsetup{font=footnotesize,justification=justified,singlelinecheck=false}
\caption*{\footnotesize \textit{Notes:} 
This figure examines how the reduction in AI usage targets affects query counts across the work-relevance distribution.
Panel (a) plots the estimated treatment effect on the number of queries falling within the indicated quintile, defined by pre-treatment work-relevance thresholds. Larger negative values indicate a greater reduction in queries within that work-relevance quintile.
Panel (b) reports quantile treatment effects at the indicated percentiles of the work-relevance distribution. Positive values indicate that queries at those percentiles are more work-relevant after the target is lowered.
Markers denote point estimates and vertical bars denote 95\% confidence intervals.}
\end{figure}

\begingroup
\newcommand{\scoretenplot}[1]{%
  \begin{tikzpicture}
    \node[inner sep=0,outer sep=0] (plotimage) {\includegraphics[width=\linewidth]{#1}};
    % The old outcome title is embedded in the PNG. Cover only that title and
    % replace it here, while preserving the y-axis tick labels and estimates.
    \begin{scope}[overlay]
      \fill[white]
        ([xshift=0pt,yshift=2pt]plotimage.south west)
        rectangle
        ([xshift=11.5pt,yshift=-2pt]plotimage.north west);
      \node[
        rotate=90,
        align=center,
        font=\fontsize{5.6}{6.2}\selectfont
      ] at ([xshift=5.75pt]plotimage.west)
        {Effect on Monthly Number of Non-Repeated Queries\\with Work-Relevance Score $=10$};
    \end{scope}
  \end{tikzpicture}%
}

\begin{figure}[!htbp]
  \centering
  \captionsetup[subfigure]{justification=centering}
\begin{subfigure}[t]{0.48\textwidth}
  \centering
  \scoretenplot{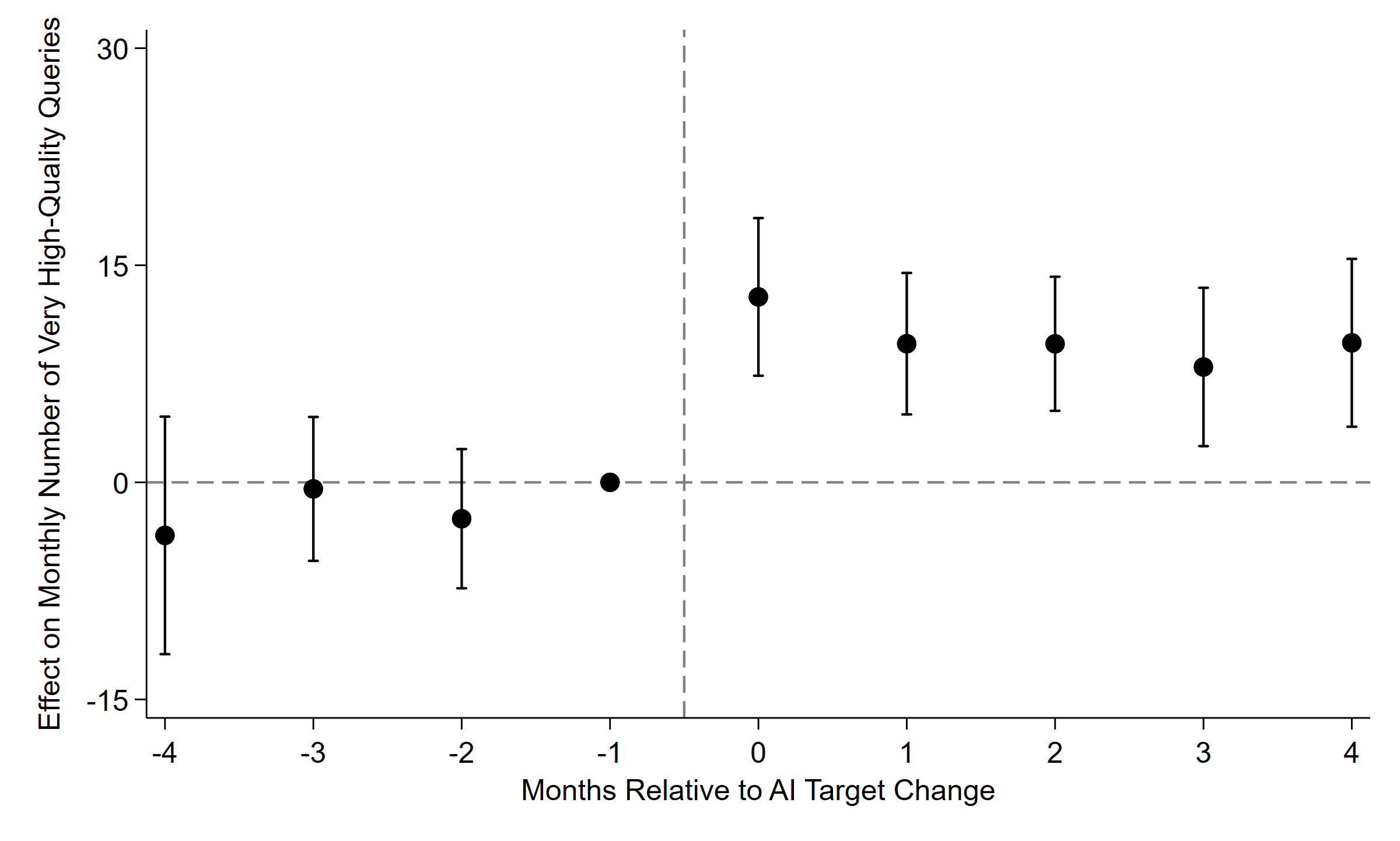}
  \caption{Non-Repeated Queries with\\Work-Relevance Score $=10$}
  \label{fig:event_study_very_high_quality}
\end{subfigure}\hfill
\begin{subfigure}[t]{0.48\textwidth}
  \centering
  \includegraphics[width=\linewidth]{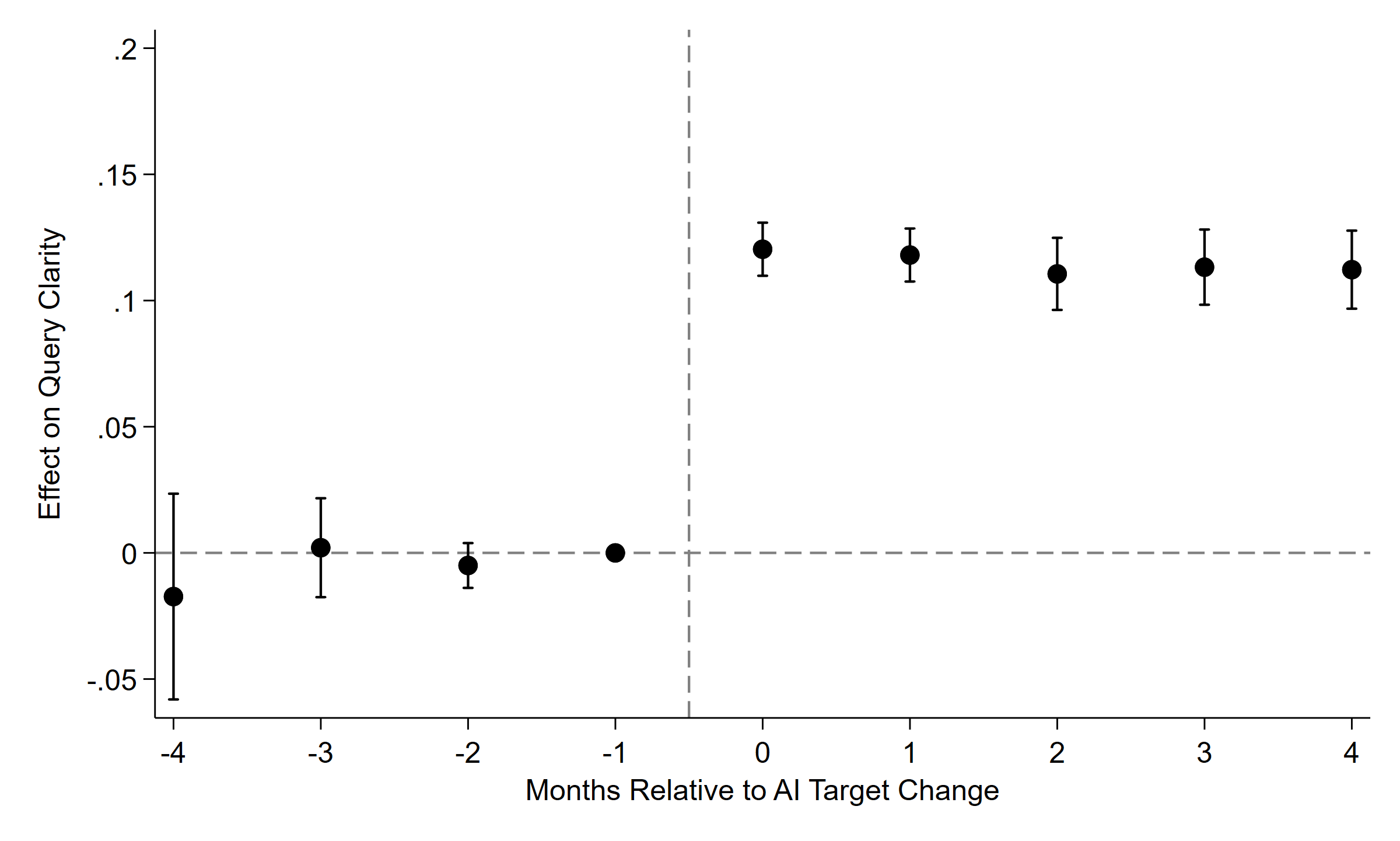}
  \caption{Prompt Clarity}
  \label{fig:event_study_query_clarity}
\end{subfigure}

\vspace{1em}

\begin{subfigure}[t]{0.31\textwidth}
  \centering
  \includegraphics[width=\linewidth]{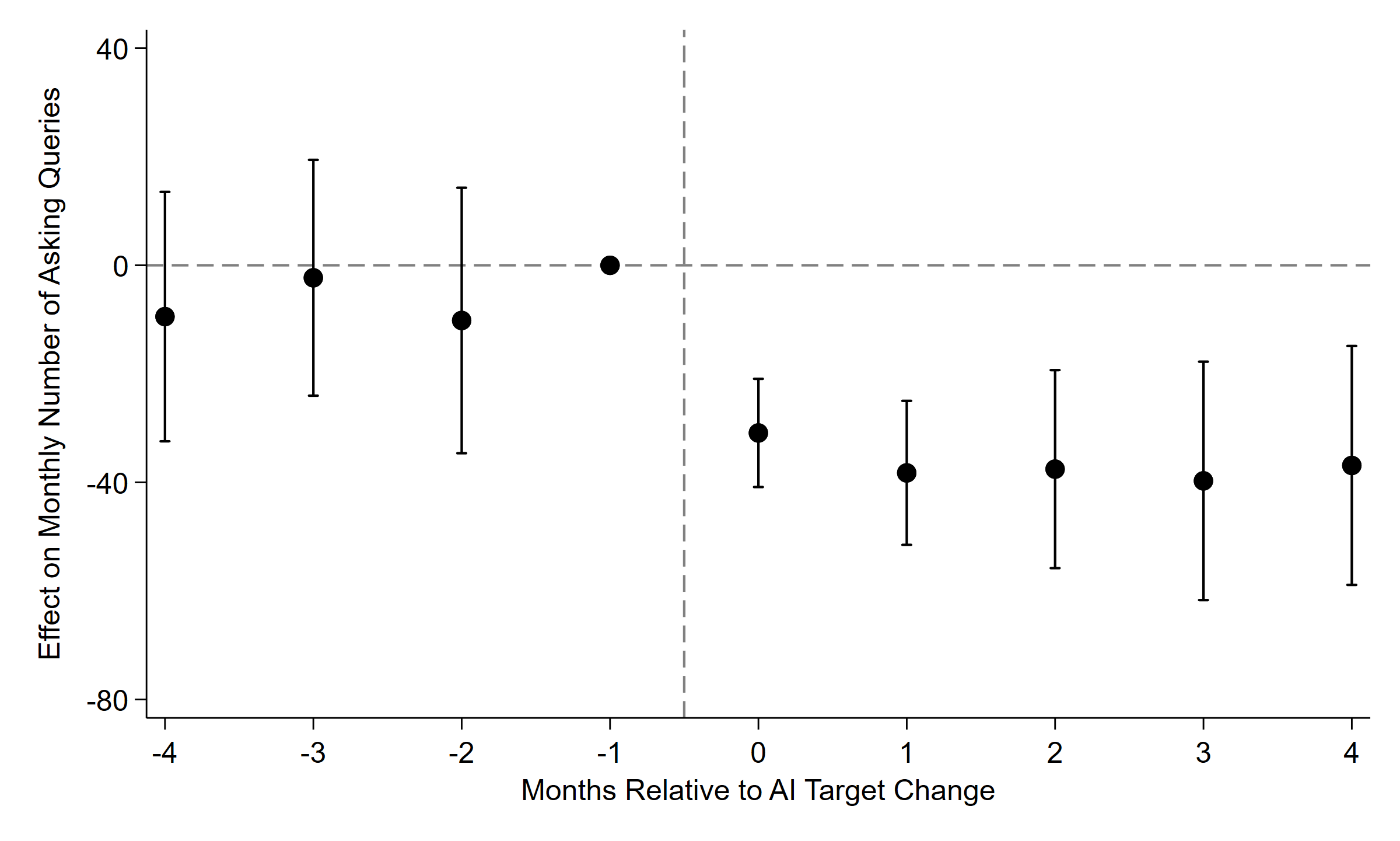}
  \caption{Asking Queries}
  \label{fig:event_study_asking}
\end{subfigure}\hfill
\begin{subfigure}[t]{0.31\textwidth}
  \centering
  \includegraphics[width=\linewidth]{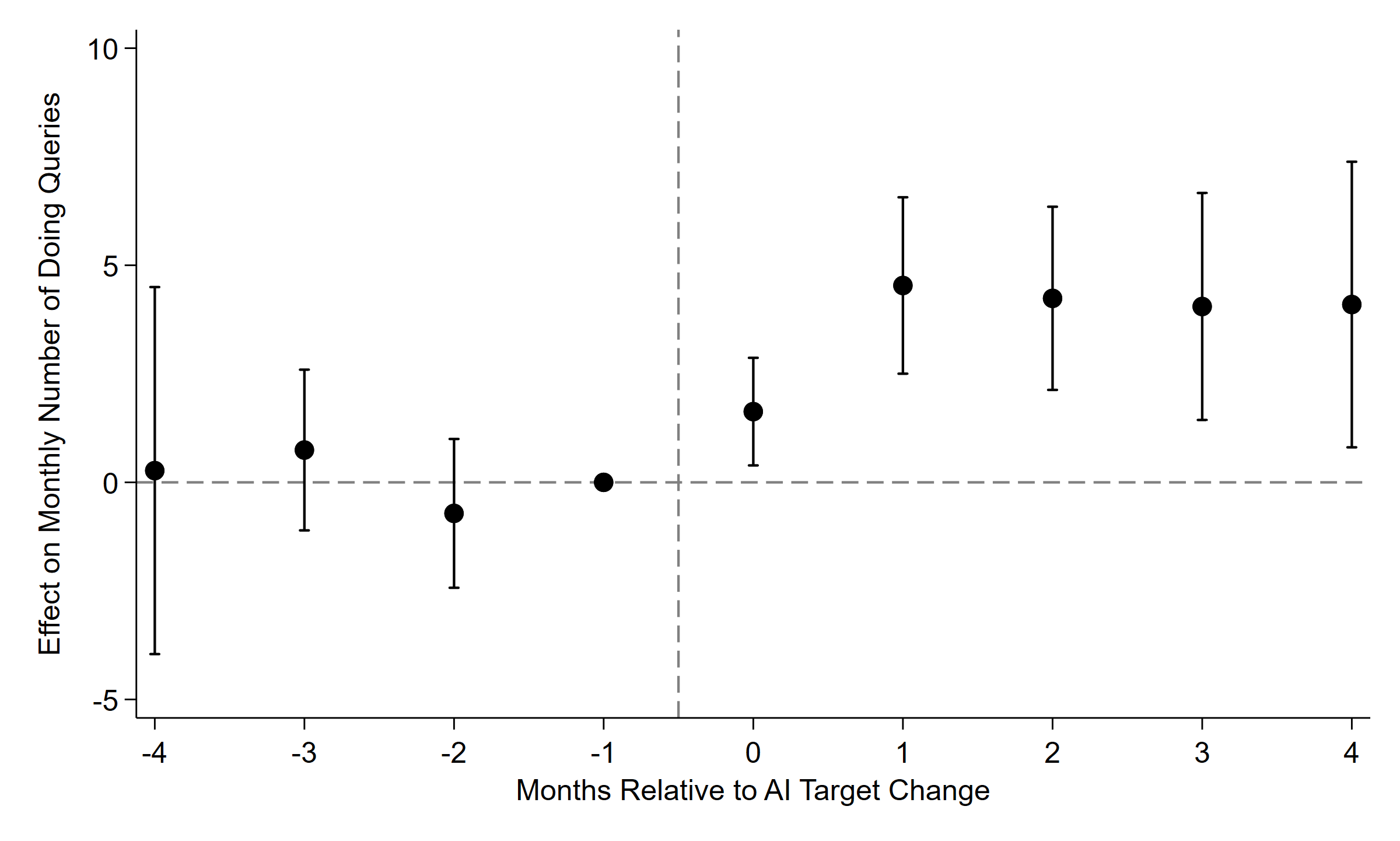}
  \caption{Doing Queries}
  \label{fig:event_study_doing}
\end{subfigure}\hfill
\begin{subfigure}[t]{0.31\textwidth}
  \centering
  \includegraphics[width=\linewidth]{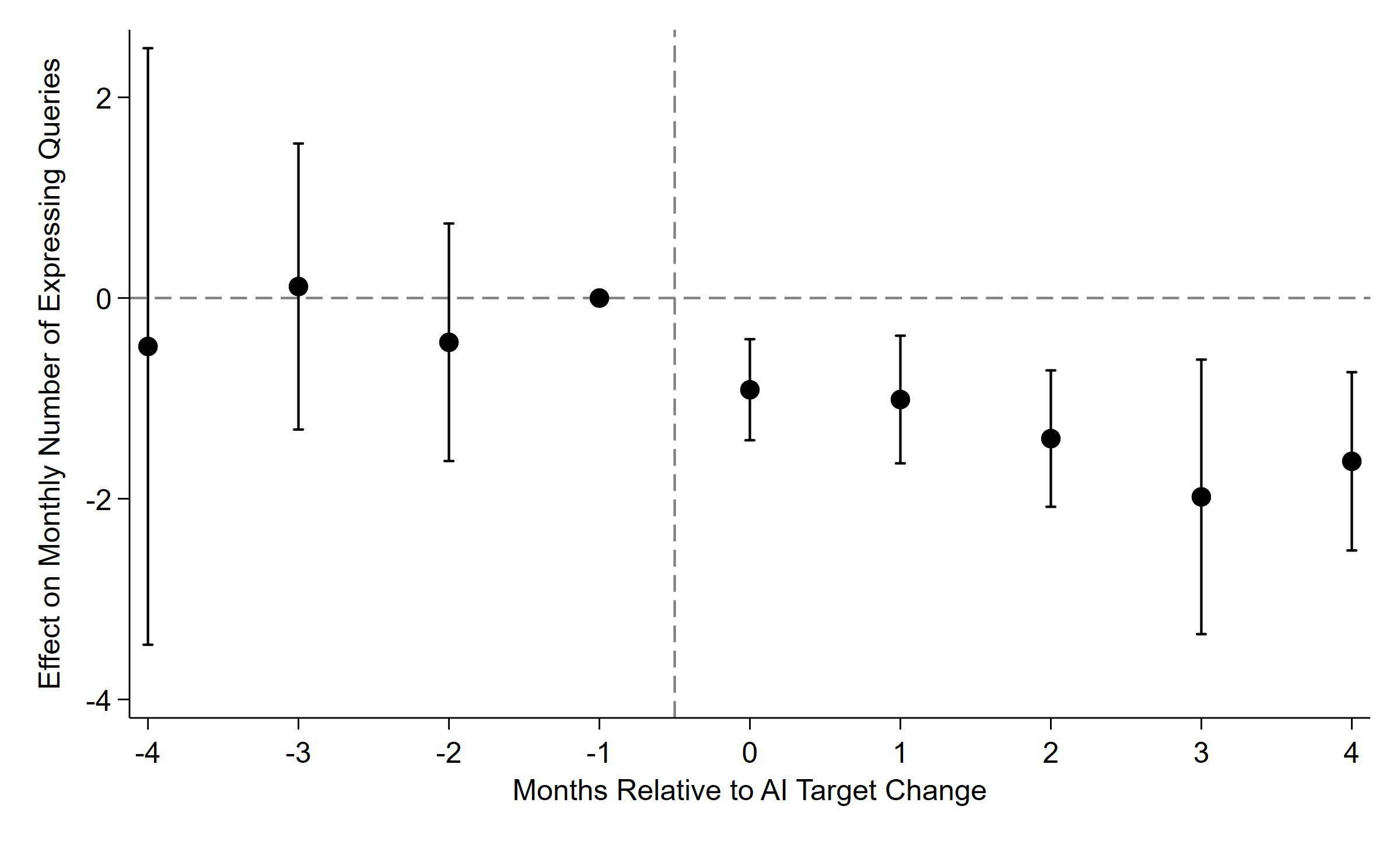}
  \caption{Expressing Queries}
  \label{fig:event_study_expressing}
\end{subfigure}
  
  \caption{Event-Study Estimates of AI Use Content and Query Intent}
  \label{fig:event_study_usage_quality_interaction}
\captionsetup{font=footnotesize,justification=justified,singlelinecheck=false}
\caption*{\footnotesize \textit{Notes:} This figure plots the dynamic effects of the AI usage target adjustment on employees' AI use content and query intent, using an employee-month panel and the event-study specification described in Section~\ref{sec:empirical_strategy}. The horizontal axis indexes event time in months relative to the policy implementation, with $k=-1$ (the month before the adjustment) as the omitted reference period. Panel (a) plots the effect on the monthly number of non-repeated queries receiving a work-relevance score of 10, the maximum value on the 1--10 scale. Panel (b) plots the effect on prompt clarity; higher values indicate messages that are clearer and better contextualized for AI to generate useful responses. Panels (c)--(e) plot the effects on the monthly numbers of Asking, Doing, and Expressing queries, respectively. Asking queries seek information or advice, Doing queries ask AI to complete a specific output, and Expressing queries involve chatting, emotional expression, or testing the AI system. All specifications are estimated using a two-way fixed effects model with individual and calendar-month fixed effects. Standard errors are clustered at the branch level. Markers denote estimated coefficients, and vertical bars denote 95\% confidence intervals.}
\end{figure}
\endgroup

\begin{figure}[!htbp]
  \centering
  \includegraphics[width=0.7\linewidth]{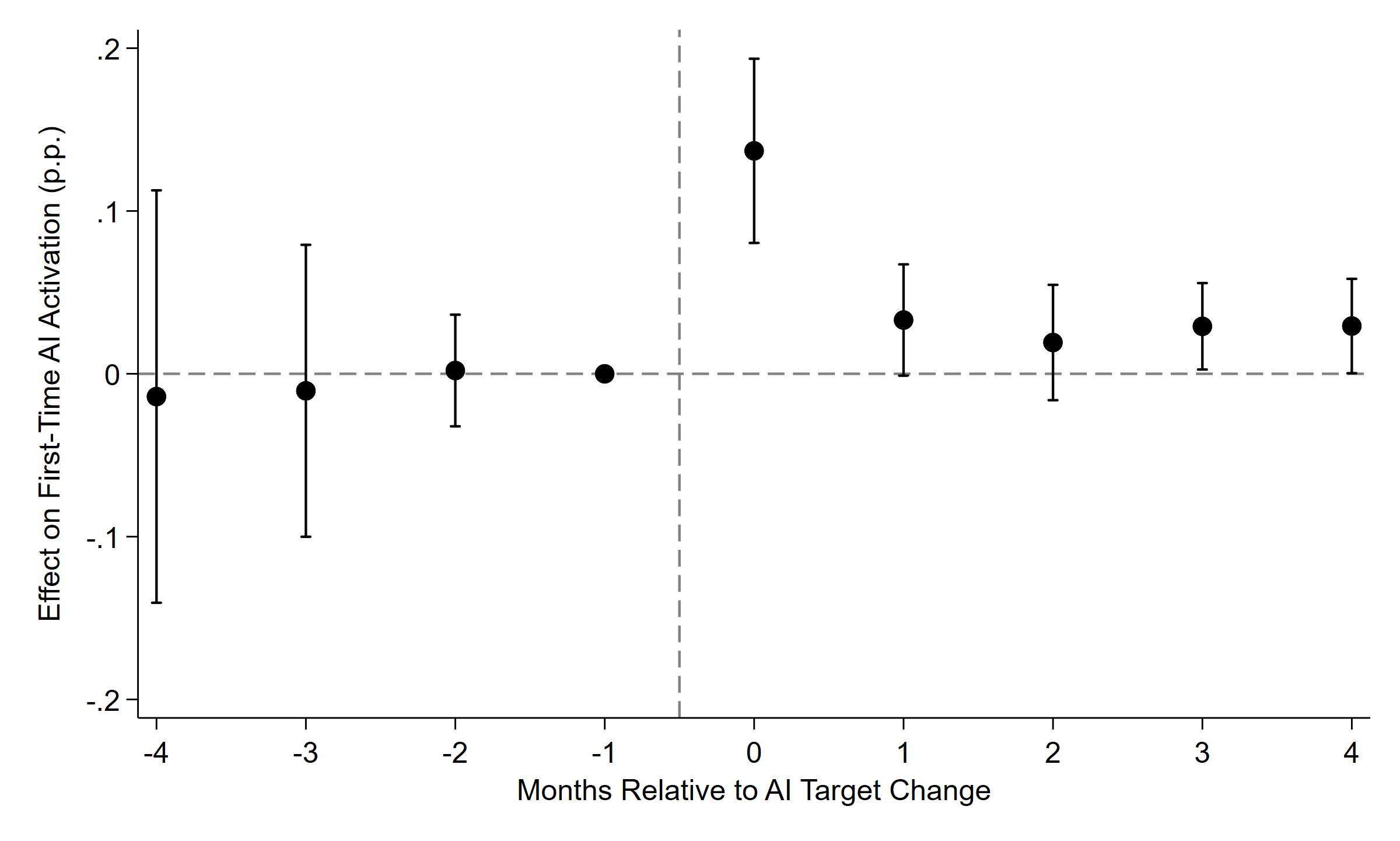}
\caption{Activation of Previously Inactive Employees}
\label{fig:active_effects}
\captionsetup{font=footnotesize,justification=justified,singlelinecheck=false}
\caption*{\footnotesize \textit{Notes:} This figure presents event-study estimates of the effect of reducing AI usage targets from 200 to 100 queries on the probability of first-time AI activation, where activation is defined as an indicator for whether an employee uses AI for the first time in the sample period. The horizontal axis indexes event time in months relative to the policy implementation, with $k=-1$ (the month prior to the adjustment) as the omitted reference period. The specification is estimated using a two-way fixed effects model with employee and calendar-month fixed effects. Standard errors are clustered at the branch level. Markers denote the estimated coefficients and vertical bars denote 95\% confidence intervals.
}
\end{figure}

\begin{figure}[!htbp]
  \centering
  \includegraphics[width=0.7\textwidth]{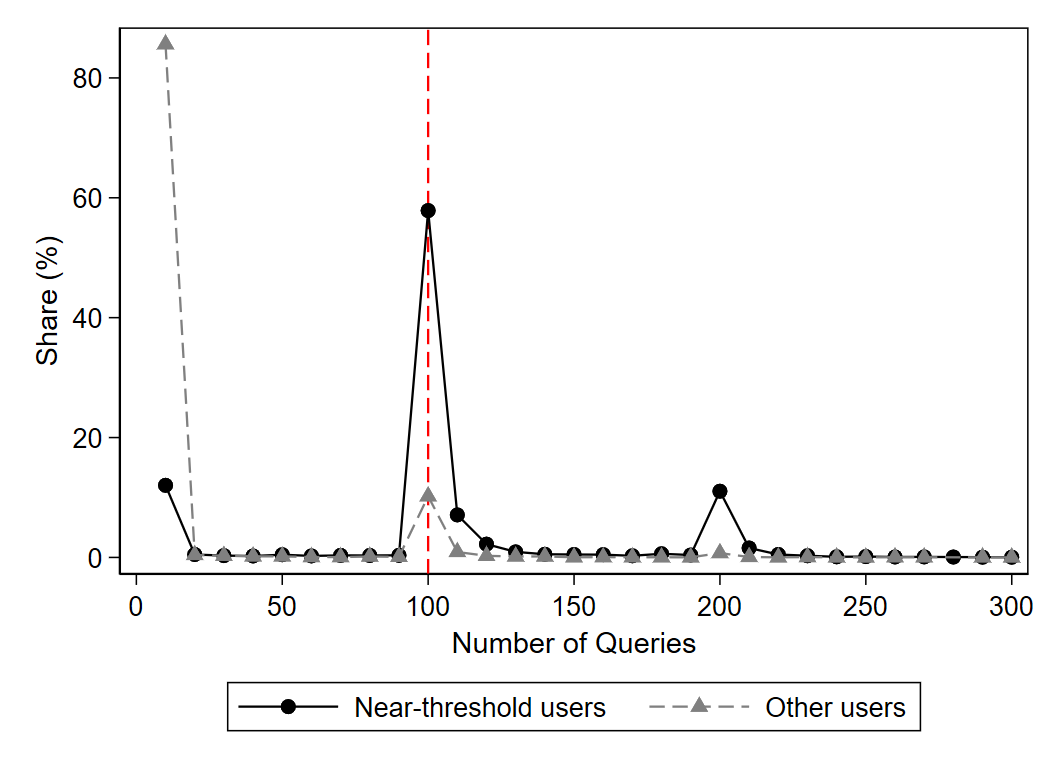}
  \caption{Distribution of Usage under the 100-Query Target, by Prior Proximity to the 200-Query Threshold}
  \label{fig:bunching_gamer}
\captionsetup{font=footnotesize,justification=justified,singlelinecheck=false}
\caption*{\footnotesize \textit{Notes:} This figure presents the distribution of monthly AI query usage under the adjusted 100-query target, separately for near-threshold users and other users. Near-threshold users are defined as employees who recorded 180--220 queries in at least one month under the initial 200-query target. The y-axis represents the share of observations, and the x-axis shows the number of queries (capped at 300). The vertical dashed red line indicates the adjusted target of 100 queries. The distribution is based on monthly query counts at the individual-month level. The sample excludes headquarters, the AI department, and frontline production workers.}
\end{figure}

\begingroup
\newcommand{\queryoutcomeplot}[2]{%
  \begin{tikzpicture}
    \node[inner sep=0,outer sep=0] (plotimage) {\includegraphics[width=\linewidth]{#1}};
    % The outcome title is embedded in the PNG. Cover only that title and
    % replace it here, while preserving the y-axis tick labels and estimates.
    \begin{scope}[overlay]
      \fill[white]
        ([xshift=0pt,yshift=2pt]plotimage.south west)
        rectangle
        ([xshift=11.5pt,yshift=-2pt]plotimage.north west);
      \node[
        rotate=90,
        align=center,
        font=\fontsize{6.0}{6.6}\selectfont
      ] at ([xshift=5.75pt]plotimage.west) {#2};
    \end{scope}
  \end{tikzpicture}%
}

\begin{figure}[!htbp]
  \centering
  \captionsetup[subfigure]{justification=centering}
  \begin{subfigure}[t]{0.48\textwidth}
    \centering
    \includegraphics[width=\linewidth]{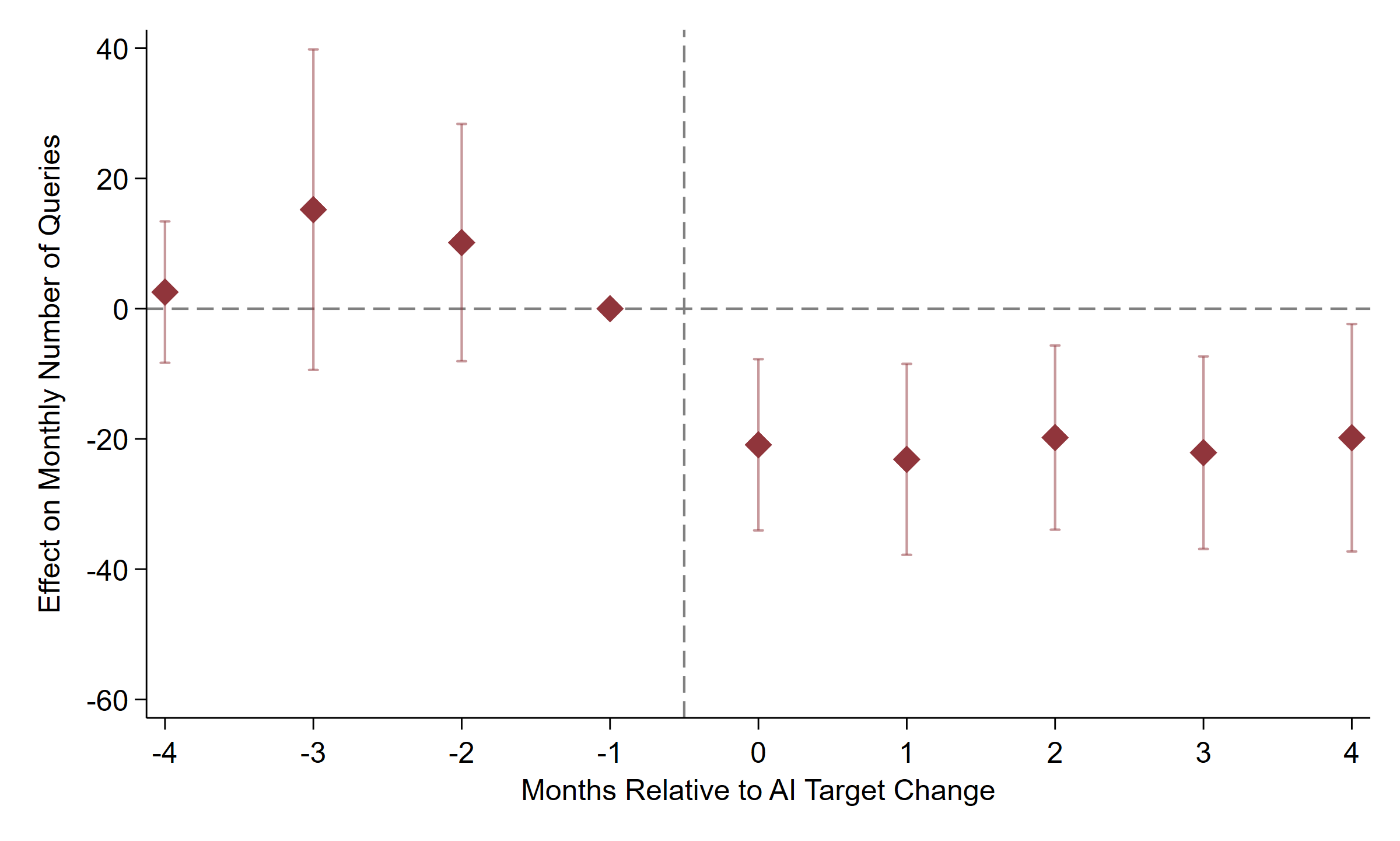}
    \caption{Monthly Number of Queries}
    \label{fig:event_study_num_sa}
  \end{subfigure}\hfill
  \begin{subfigure}[t]{0.48\textwidth}
    \centering
    \queryoutcomeplot
      {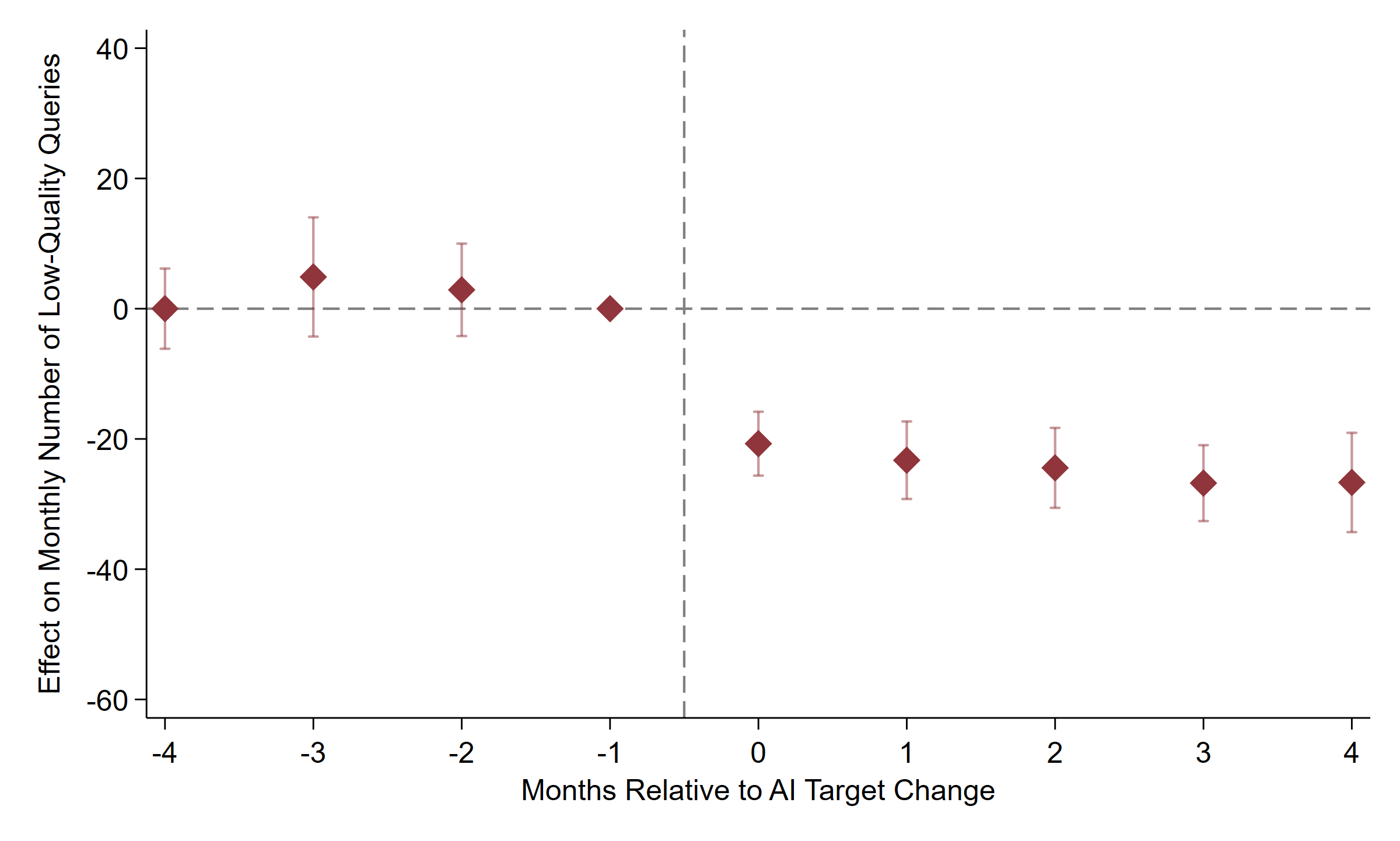}
      {Effect on Monthly Number of\\Repeated or Off-Task Queries}
    \caption{Monthly Number of Repeated or\\Off-Task Queries}
    \label{fig:event_study_low_quality_sa}
  \end{subfigure}
  
  \vspace{1em}
  
  \begin{subfigure}[t]{0.48\textwidth}
    \centering
    \queryoutcomeplot
      {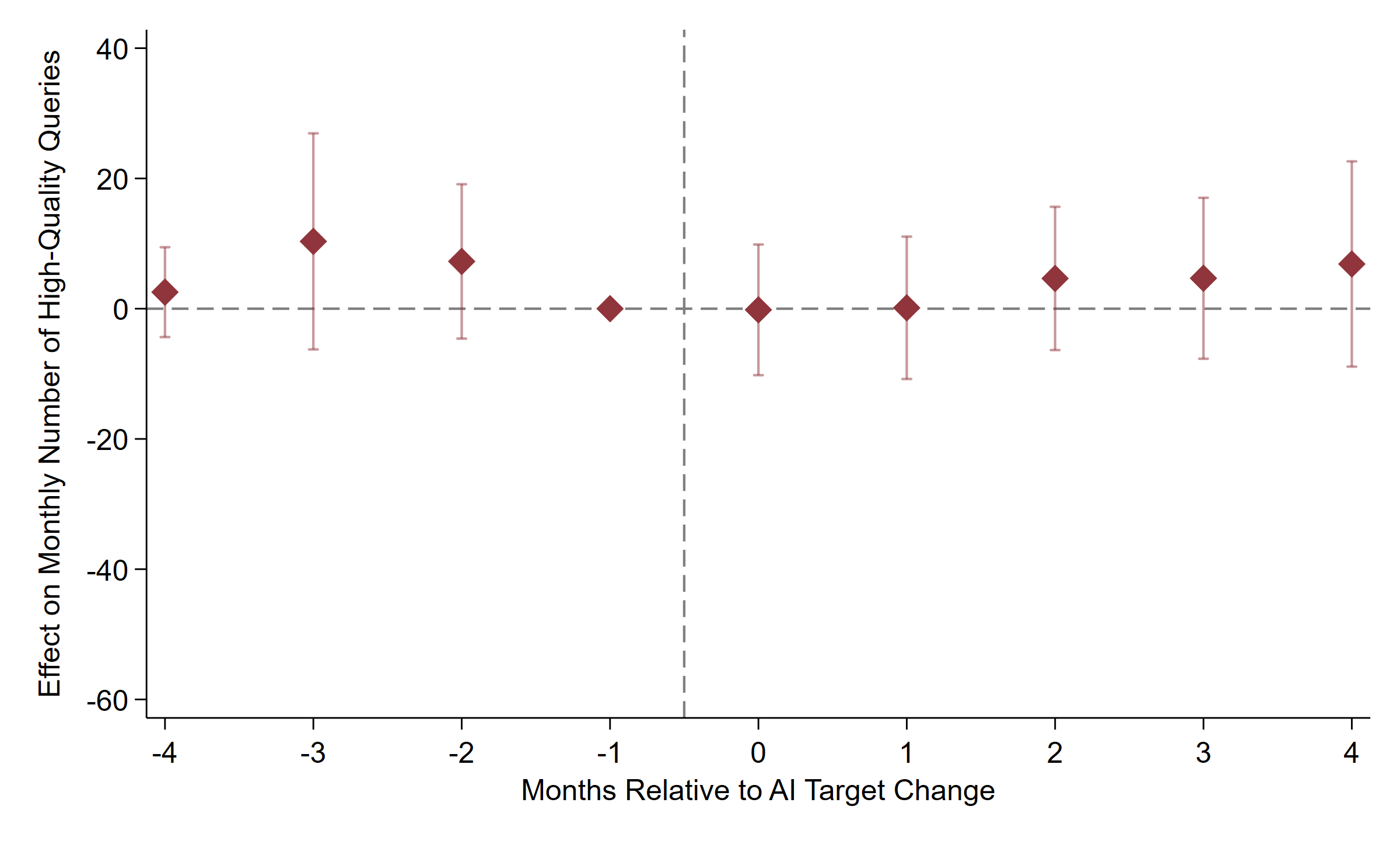}
      {Effect on Monthly Number of\\Non-Repeated Work-Related Queries}
    \caption{Monthly Number of Non-Repeated\\Work-Related Queries}
    \label{fig:event_study_high_quality_sa}
  \end{subfigure}\hfill
  \begin{subfigure}[t]{0.48\textwidth}
    \centering
    \includegraphics[width=\linewidth]{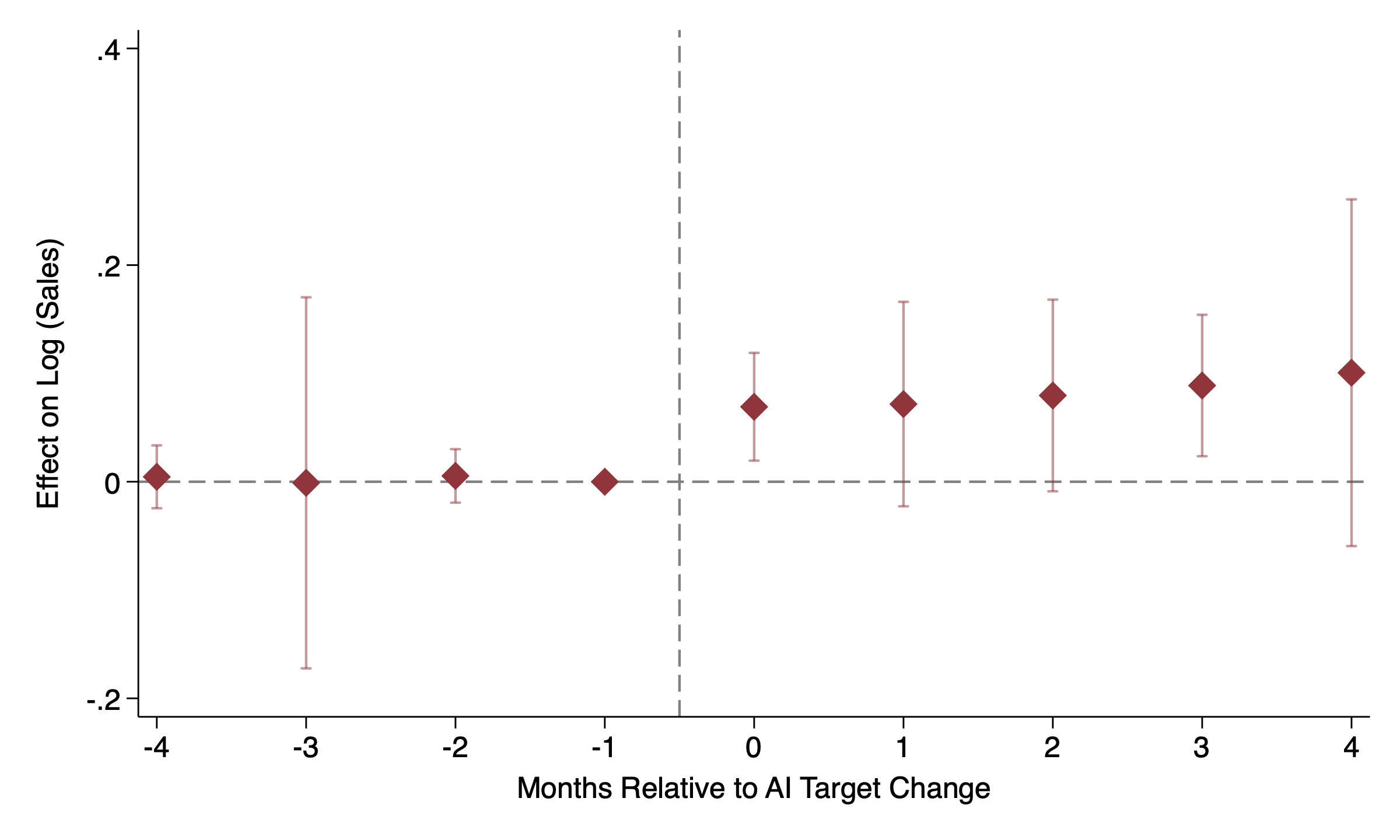}
    \caption{Log Monthly Sales}
    \label{fig:event_study_sales_sa}
  \end{subfigure}
  
  \caption{Event-Study Estimates Using Sun and Abraham (2021)}
  \label{fig:event_study_sa}
  \captionsetup{font=footnotesize,justification=justified,singlelinecheck=false}
  \caption*{\footnotesize \textit{Notes:} This figure plots the dynamic effects of lowering the monthly query target, estimated using employee-month panel data from the event-study specification described in Section~\ref{sec:empirical_strategy}. The horizontal axis indexes event time in months relative to the policy implementation, with $k=-1$ (the month prior to the adjustment) as the omitted reference period. Panel (a) plots the effect on the total number of AI queries submitted by an employee in a given month. Panel (b) plots the effect on the monthly number of queries that are repeated or classified as personal or off-task. Panel (c) plots the effect on the monthly number of non-repeated queries classified as clearly, generically, or weakly work-related. Panel (d) plots the effect on the logarithm of individual monthly sales. All specifications are estimated using the Sun and Abraham (2021) estimator with individual and calendar-month fixed effects, which accounts for heterogeneous treatment timing across branches. Standard errors are clustered at the branch level. Markers denote the estimated coefficients and vertical bars denote 95\% confidence intervals.}
\end{figure}
\endgroup

\begingroup
\newcommand{\placeboplot}[1]{%
  \begin{tikzpicture}
    \node[inner sep=0,outer sep=0] (plotimage) {\includegraphics[width=\textwidth]{#1}};
    % Remove the outcome title embedded in the PNG; the centered panel caption
    % below the plot provides the outcome name without visual duplication.
    \begin{scope}[overlay]
      \fill[white]
        ([xshift=28pt,yshift=0pt]plotimage.north west)
        rectangle
        ([xshift=-28pt,yshift=-18pt]plotimage.north east);
    \end{scope}
  \end{tikzpicture}%
}

\begin{figure}[!htbp]
  \centering
  \captionsetup[subfigure]{justification=centering}
  \begin{subfigure}[t]{0.50\textwidth}
    \centering
    \placeboplot{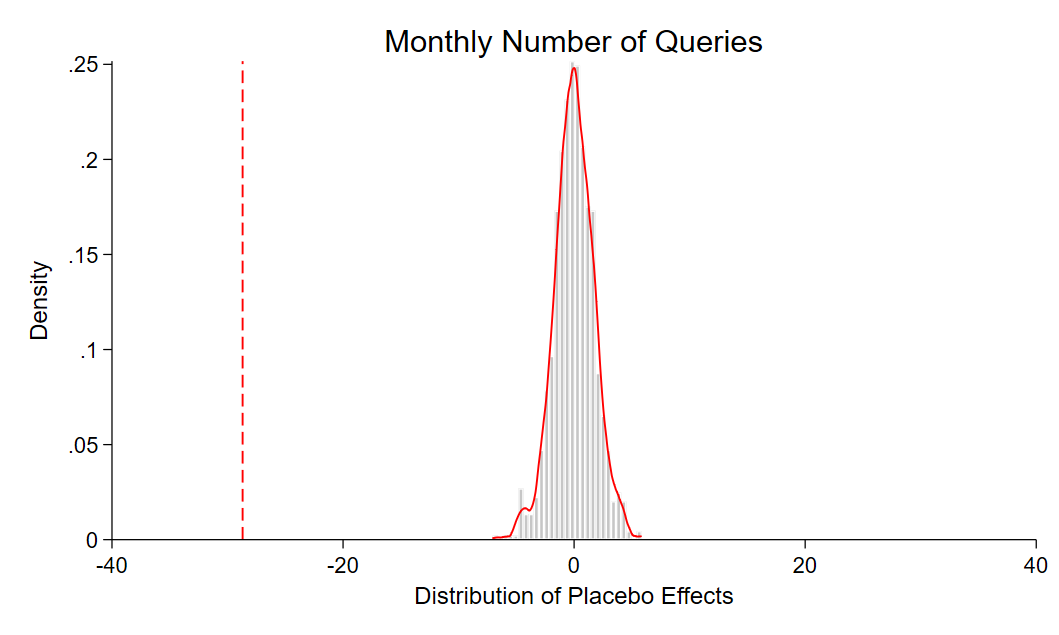}
    \caption{Monthly Number of Queries}
    \label{fig:placebo_queries}
  \end{subfigure}
  
  \vspace{0.1em}
  \begin{subfigure}[t]{0.50\textwidth}
    \centering
    \placeboplot{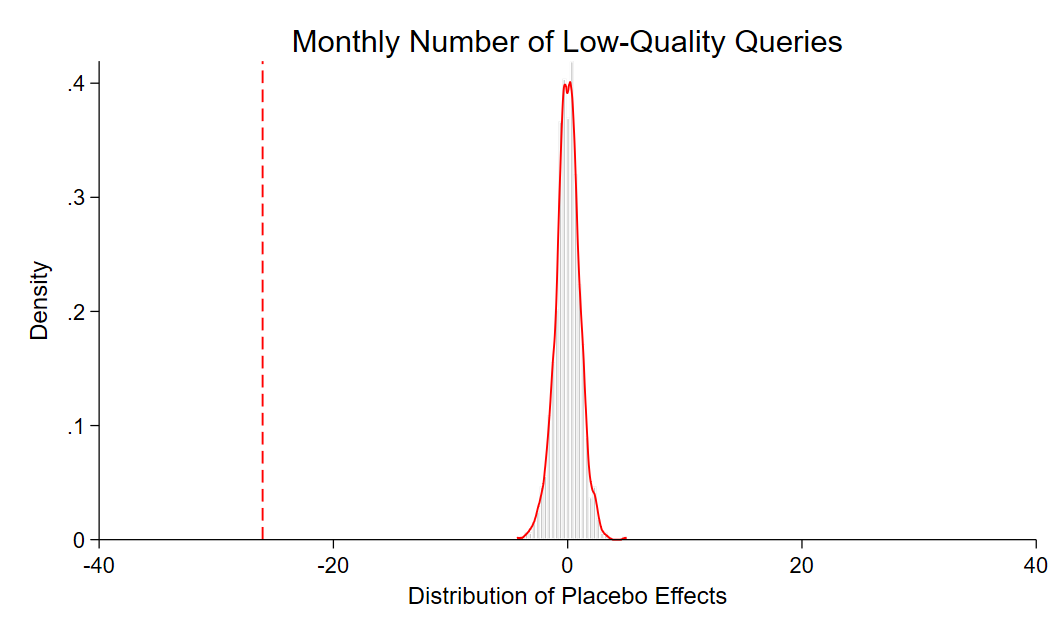}
    \caption{Monthly Number of Repeated or\\Off-Task Queries}
    \label{fig:placebo_low_quality}
  \end{subfigure}
  
  \vspace{0.1em}
  \begin{subfigure}[t]{0.50\textwidth}
    \centering
    \placeboplot{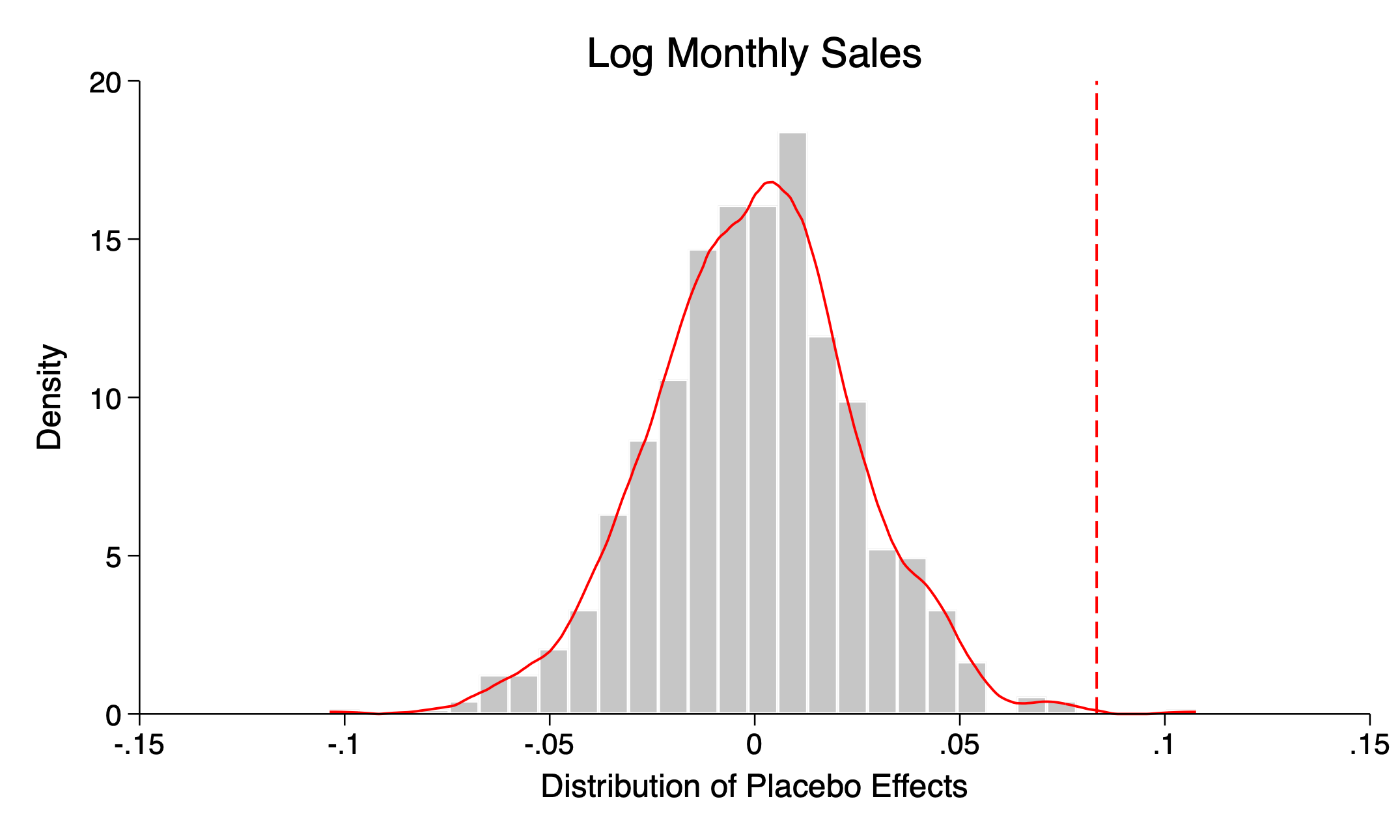}
    \caption{Log Monthly Sales}
    \label{fig:placebo_sales}
  \end{subfigure}
  
  \caption{Placebo Tests: Distribution of Estimated Effects}
  \label{fig:placebo_tests}
  \begin{minipage}{0.95\textwidth}
    \footnotesize
    \vspace{0.2cm}
    \textit{Notes:} This figure displays the distributions of estimated effects from 1,000 random permutations of the observed lower-target implementation timing across branches in the corresponding estimation sample. The red dashed line indicates the estimate obtained using the actual implementation timing. Panel (a) shows results for the monthly number of queries. Panel (b) shows results for the monthly number of queries that are repeated or classified as personal or off-task. Panel (c) shows results for the logarithm of monthly sales.
  \end{minipage}
\end{figure}
\endgroup

\begin{figure}[!htbp]
  \centering
  \begin{tikzpicture}
    \node[inner sep=0] (looimage) {\includegraphics[width=0.65\textwidth]{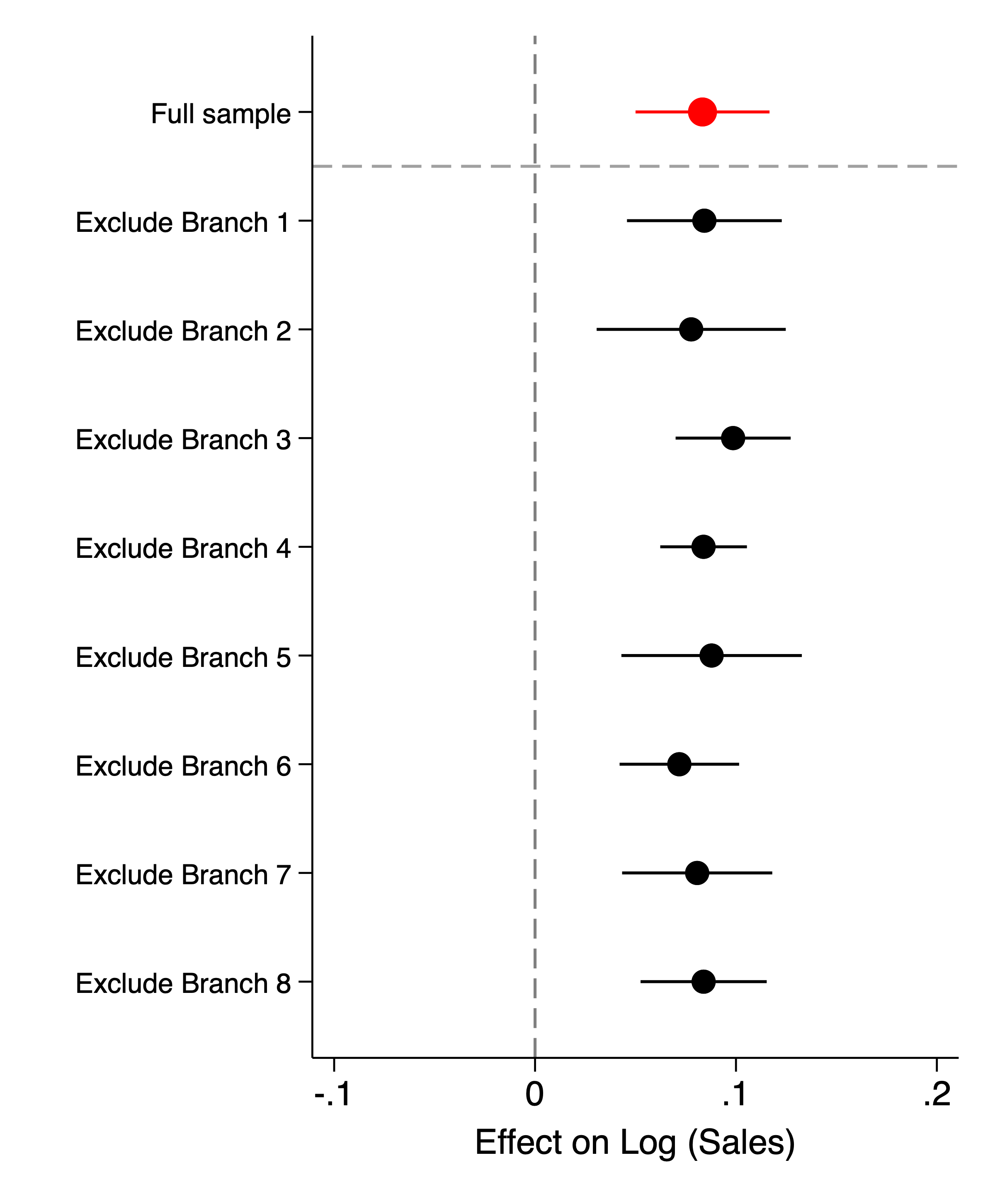}};
    \fill[white] (looimage.south west) rectangle ([yshift=27pt]looimage.south east);
    \node[anchor=south,font=\small] at ([yshift=4pt]looimage.south) {Effect on Log Monthly Sales};
  \end{tikzpicture}

  \caption{Leave-One-Branch-Out Robustness Check}
  \label{fig:leave_one_out_sales}

  \begin{minipage}{0.90\textwidth}
    \footnotesize
    \vspace{0.2cm}
    \textit{Notes:} This figure reports leave-one-branch-out estimates of the effect of the lower AI target on log sales. The red marker shows the full-sample estimate. Black markers show estimates obtained by excluding one branch at a time. Horizontal lines indicate 95\% confidence intervals based on standard errors clustered at the branch level. All specifications include month and worker fixed effects.
  \end{minipage}
\end{figure}

\begin{figure}[!htbp]
  \centering
  \includegraphics[width=0.7\textwidth]{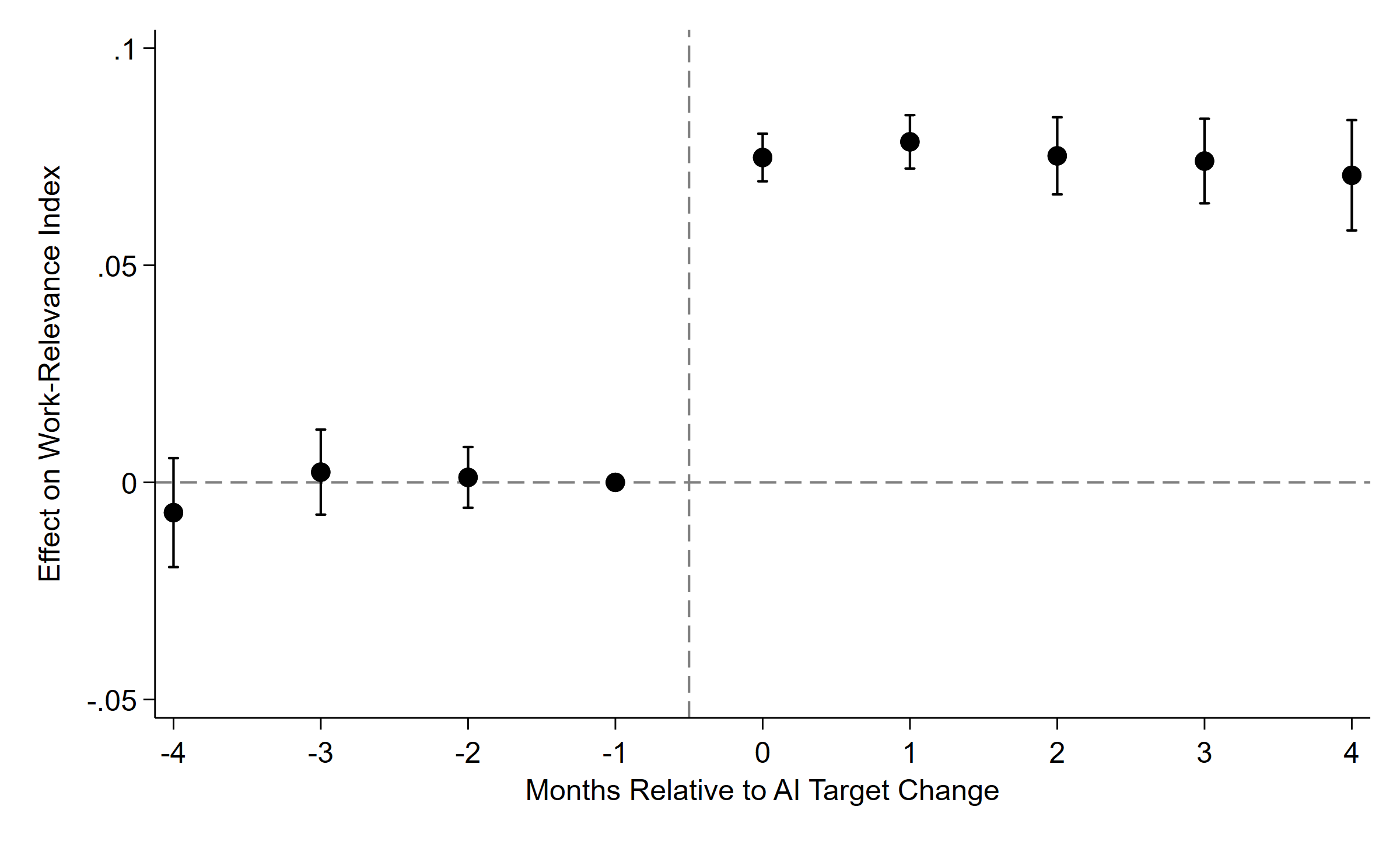}

  \caption{Event-Study Estimates for the Embedding-Based Work-Relevance Index}
  \label{fig:event_study_twfe_work_relevance_index}
  \captionsetup{font=footnotesize,justification=justified,singlelinecheck=false}
  \caption*{\footnotesize \textit{Notes:} This figure plots the dynamic effects of lowering the monthly query target on the embedding-based work-relevance index, estimated using employee-month panel data from the event-study specification described in Section~\ref{sec:empirical_strategy}. The horizontal axis indexes event time in months relative to the policy implementation, with $k=-1$ (the month prior to the adjustment) as the omitted reference period. The outcome is a continuous work-relevance index ranging from 0 to 1. For each query, the index is constructed by computing the cosine similarity between the query embedding and two reference embeddings: one based on the firm's overall company description and the other based on keywords describing the employee's current role. The two similarity scores are averaged and rescaled to the 0--1 range, with higher values indicating stronger relevance to the firm's business and the employee's role. The specification is estimated using a two-way fixed effects model with individual and calendar-month fixed effects. Standard errors are clustered at the branch level. Markers denote the estimated coefficients and vertical bars denote 95\% confidence intervals.}
\end{figure}

%********** Appendix tables **********%
\clearpage
\setcounter{table}{0}

\begin{table}[htbp]
\small
\centering
\caption{Baseline Branch Characteristics by Treatment Status}
\label{tab:balance_branch}
\begin{threeparttable}
\begin{tabular*}{\textwidth}{@{\extracolsep{\fill}}p{0.45\textwidth}cccccc}
\toprule
& \multicolumn{2}{c}{\makecell{Ever-Switching\\Branches}} & \multicolumn{2}{c}{\makecell{Never-Switching\\Branches}} & \multicolumn{2}{c}{Difference} \\
& \multicolumn{2}{c}{$(N = 27)$} & \multicolumn{2}{c}{$(N = 15)$} & \multicolumn{2}{c}{\makecell{(Ever-Switching $-$\\Never-Switching)}} \\
\cmidrule(lr){2-3} \cmidrule(lr){4-5} \cmidrule(lr){6-7}
& Mean & S.D. & Mean & S.D. & Diff. & $p$-value \\
\midrule
\multicolumn{7}{l}{\textit{Panel A: Employee Characteristics}} \\
\midrule
Number of Employees                               & 157.90 & 132.11 & 137.91 & 97.19  & 19.98     & 0.61 \\
Share of Females                                  & 0.52   & 0.13   & 0.50   & 0.11   & 0.02      & 0.61 \\
Share of Managers                                 & 0.14   & 0.12   & 0.10   & 0.12   & 0.05      & 0.23 \\
Average Age (Years)                               & 36.34  & 3.29   & 36.40  & 3.66   & $-0.07$   & 0.95 \\
Average Education (Years)                         & 14.59  & 0.84   & 14.39  & 1.14   & 0.20      & 0.52 \\
Average Experience (Years)                        & 6.02   & 2.88   & 6.21   & 3.96   & $-0.19$   & 0.86 \\
\midrule
\multicolumn{7}{l}{\textit{Panel B: Baseline AI Usage}} \\
\midrule
Average Monthly Number of Queries                 & 88.20  & 46.25  & 67.47  & 55.16  & 20.73     & 0.20 \\
\makecell[l]{Average Number of Repeated or\\Off-Task Queries} & 25.35 & 13.87 & 18.98 & 15.67 & 6.36 & 0.18 \\
\makecell[l]{Average Number of Non-Repeated\\Work-Related Queries} & 62.85 & 34.72 & 48.49 & 41.02 & 14.36 & 0.24 \\
\makecell[l]{Average Number of Non-Repeated Queries\\with Work-Relevance Score $= 10$} & 14.65 & 9.37 & 11.76 & 11.62 & 2.89 & 0.39 \\
Average Prompt Clarity Score                      & 0.53   & 0.04   & 0.56   & 0.13   & $-0.03$   & 0.27 \\
Average Number of Asking Queries                  & 80.23  & 41.90  & 62.26  & 50.73  & 17.97     & 0.22 \\
Average Number of Doing Queries                   & 6.48   & 5.50   & 3.82   & 4.35   & 2.67      & 0.11 \\
Average Number of Expressing Queries              & 1.48   & 1.40   & 1.39   & 1.76   & 0.09      & 0.86 \\
Share of Near-Threshold Users                                  & 0.35   & 0.22   & 0.30   & 0.25   & 0.05      & 0.54 \\
Share of Active Users                             & 0.51   & 0.24   & 0.39   & 0.29   & 0.12      & 0.16 \\
\bottomrule
\end{tabular*}
\begin{tablenotes}[flushleft]
\footnotesize
\item \textit{Notes:} This table reports branch-level summary statistics for the 42 branches used in the main empirical analysis. Baseline AI usage and individual characteristics are computed over the pre-adjustment period, March through June 2025. Panel A reports workforce characteristics, and Panel B reports baseline AI usage characteristics. Since the AI tool is primarily designed for non-production tasks, the sample in Panel B is restricted to non-production workers. Repeated or off-task queries include queries that are repeated or classified as personal or off-task. Non-repeated work-related queries include non-repeated queries classified as clearly, generically, or weakly work-related. The work-relevance score ranges from 1 to 10; the score-10 measure counts non-repeated queries receiving the maximum score. Columns (1)--(2) and (3)--(4) report the mean and standard deviation for ever-switching and never-switching branches, respectively. The last two columns report the difference in means (Ever-Switching $-$ Never-Switching) and the corresponding $p$-value from a two-sample $t$-test.
\end{tablenotes}
\end{threeparttable}
\end{table}

\begin{table}[!htbp]
  \centering
  \begin{threeparttable}
    \caption{Baseline Branch Characteristics and the Timing of Lower-Target Implementation}
    \label{tab:switching_timing}
    \begin{tabularx}{\textwidth}{l*{6}{>{\centering\arraybackslash}X}}
      \toprule
      & (1) & (2) & (3) & (4) & (5) & (6) \\
      \midrule
      Number of Employees
      & 1.001 & & & & & 1.001 \\
      & (0.001) & & & & & (0.001) \\

      Average Monthly Number of Queries
      & & 1.005 & & & & 1.034 \\
      & & (0.004) & & & & (0.047) \\

      \makecell[l]{Average Number of Repeated or\\Off-Task Queries}
      & & & 1.013 & & & 0.998 \\
      & & & (0.013) & & & (0.024) \\

      Share of Near-Threshold Users
      & & & & 1.994 & & 0.092 \\
      & & & & (1.651) & & (0.261) \\

      Share of Active Users
      & & & & & 2.389 & 0.034 \\
      & & & & & (1.712) & (0.207) \\
      \midrule
      Observations
      & 42 & 42 & 42 & 42 & 42 & 42 \\
      \bottomrule
    \end{tabularx}

    \begin{tablenotes}[flushleft]
      \footnotesize
\item \textit{Notes:} This table reports Cox proportional-hazards estimates of whether baseline branch characteristics predict the timing of lower-target implementation. The event is a branch's implementation of the 100-query target. The analysis is conducted at the branch level using the 42 branches in the main empirical sample. Twenty-seven branches implemented the lower target during the sample period; the 15 branches that had not switched by December 2025 are treated as right-censored. Baseline characteristics are computed over the pre-adjustment period, March through June 2025. The baseline query measure in column (3) is the average number of queries that are repeated or classified as personal or off-task. The table reports hazard ratios, so a value of one indicates no association with the timing of lower-target implementation. Columns (1)--(5) include each baseline characteristic separately, and column (6) includes all five characteristics jointly. Robust standard errors are reported in parentheses. Efron's method is used to account for branches implementing the lower target in the same month. A Wald test on the underlying log-hazard coefficients cannot reject the null that all five coefficients in column (6) are jointly equal to zero, which corresponds to hazard ratios equal to one ($p=0.670$). * $p<0.10$, ** $p<0.05$, *** $p<0.01$.
    \end{tablenotes}
  \end{threeparttable}
\end{table}

\begin{table}[!ht]
  \centering
  \begin{threeparttable}
    \caption{Effects of Lowering the Target on AI Use Content and Query Intent}
    \label{tab:effects_usage_content}

    \small
    \setlength{\tabcolsep}{4pt}
    \renewcommand{\arraystretch}{1.08}

    \begin{tabular*}{\textwidth}{@{\extracolsep{\fill}}lccccc@{}}
      \toprule
      & \multicolumn{2}{c}{AI use content}
      & \multicolumn{3}{c}{Query intent} \\
      \cmidrule(lr){2-3}\cmidrule(lr){4-6}
      & (1) & (2) & (3) & (4) & (5) \\
      & \makecell[c]{\footnotesize Non-Repeated\\Queries with\\Work-Relevance\\Score $=10$}
      & \makecell{Prompt\\clarity}
      & \makecell{Asking\\queries}
      & \makecell{Doing\\queries}
      & \makecell{Expressing\\queries} \\
      \midrule

      $\textit{LowerTarget}_{jt}$
      & $11.50^{***}$
      & $0.117^{***}$
      & $-31.00^{***}$
      & $3.37^{***}$
      & $-1.08^{**}$ \\
      & $(2.62)$
      & $(0.006)$
      & $(5.83)$
      & $(1.06)$
      & $(0.49)$ \\

      \midrule
      Month FE      & Yes & Yes & Yes & Yes & Yes \\
      Individual FE & Yes & Yes & Yes & Yes & Yes \\
      Observations  & 43,678 & 23,786 & 43,678 & 43,678 & 43,678 \\
      \bottomrule
    \end{tabular*}

    \begin{tablenotes}[flushleft]
      \footnotesize
      \item \textit{Notes:} This table reports difference-in-differences estimates using an employee-month panel.
      Columns (1) and (3)--(5) are monthly counts. Column (1) counts non-repeated queries receiving a work-relevance score of 10, the maximum value on the 1--10 scale. Column (2) is prompt clarity, defined for employee-months with at least one query; higher values indicate clearer and better-contextualized prompts. Columns (3)--(5) count Asking, Doing, and Expressing queries. Asking queries seek information or advice, Doing queries ask AI to complete a specific output, and Expressing queries involve chatting, emotional expression, or testing the AI system. All specifications include individual and calendar-month fixed effects. Standard errors are clustered at the branch level. $^{*} p<0.10$, $^{**} p<0.05$, $^{***} p<0.01$.
    \end{tablenotes}
  \end{threeparttable}
\end{table}

\begin{table}[!htbp]
  \centering
  \begin{threeparttable}
    \caption{AI Usage Quality and Individual Sales Performance}
    \label{tab:corr_ai_performance}

    \newcolumntype{Y}{>{\centering\arraybackslash}X}
    \setlength{\tabcolsep}{2.5pt}

    \begin{tabularx}{\textwidth}{l*{4}{Y}}
      \toprule
      & (1) & (2) & (3) & (4) \\
      \midrule
      & \multicolumn{4}{c}{Log Monthly Sales} \\
      \midrule
      Monthly Number of Queries 
      & $-0.00006$ 
      &  
      &  
      &  \\
      & $(0.00004)$ 
      &  
      &  
      &  \\

      \makecell[l]{Monthly Number of Repeated or\\Off-Task Queries}
      &  
      & $-0.00025^{**}$ 
      &  
      & $-0.00025^{**}$ \\
      &  
      & $(0.00009)$ 
      &  
      & $(0.00009)$ \\

      \makecell[l]{Monthly Number of Non-Repeated\\Work-Related Queries}
      &  
      &  
      & $0.00018^{**}$ 
      & $0.00018^{**}$ \\
      &  
      &  
      & $(0.00007)$ 
      & $(0.00007)$ \\

      \midrule
      Month FE & Yes & Yes & Yes & Yes \\
      Individual FE & Yes & Yes & Yes & Yes \\
      Observations & 6,064 & 6,064 & 6,064 & 6,064 \\
      \bottomrule
    \end{tabularx}

    \begin{tablenotes}[flushleft]
      \footnotesize
      \item \textit{Notes:} This table reports the relationship between different dimensions of AI usage and individual sales performance using employee-month panel data. The dependent variable is the logarithm of individual monthly sales. The key explanatory variables are the monthly number of AI queries, the monthly number of queries that are repeated or classified as personal or off-task, and the monthly number of non-repeated queries classified as clearly, generically, or weakly work-related. All specifications are estimated using a two-way fixed effects model with individual and calendar-month fixed effects. Standard errors clustered at the branch level are in parentheses. * $p<0.10$, ** $p<0.05$, *** $p<0.01$.
    \end{tablenotes}
  \end{threeparttable}
\end{table}

\begin{table}[!ht]
  \centering
  \begin{threeparttable}
    \captionsetup{font=small,justification=centering,singlelinecheck=false}
    \caption{Difference-in-Differences Estimates of AI Usage in the Sales Subsample}
    \label{tab:effects_baseline_sales_subsample}

    \newcolumntype{Y}{>{\centering\arraybackslash}X}
    \setlength{\tabcolsep}{3.5pt}

    \begin{tabularx}{\textwidth}{l
      >{\hsize=.90\hsize\centering\arraybackslash}X
      >{\hsize=1.05\hsize\centering\arraybackslash}X
      >{\hsize=1.05\hsize\centering\arraybackslash}X}
      \toprule
      & (1) & (2) & (3) \\
      \midrule
      & {\small\shortstack[c]{Monthly Number of\\Queries\\\phantom{Queries}}}
      & {\small\shortstack[c]{Monthly Number\\of Repeated or\\Off-Task Queries}}
      & {\small\shortstack[c]{Monthly Number\\of Non-Repeated\\Work-Related Queries}} \\
      \midrule
      $\textit{LowerTarget}_{jt}$ 
      & $-30.24^{**}$ 
      & $-27.94^{***}$ 
      & $-2.30$ \\
      & $(11.28)$ 
      & $(7.25)$ 
      & $(5.61)$ \\
      \midrule
      Month FE & Yes & Yes & Yes \\
      Individual FE & Yes & Yes & Yes \\
      Observations & 6,064 & 6,064 & 6,064 \\
      \bottomrule
    \end{tabularx}

    \begin{tablenotes}[flushleft]
      \footnotesize
      \item \textit{Notes:} This table reports the estimated effects of the AI usage target adjustment on employees' query behavior in the sales subsample. The dependent variable in column (1) is the monthly number of queries submitted by an employee in a given month. The dependent variable in column (2) is the monthly number of queries that are repeated or classified as personal or off-task. The dependent variable in column (3) is the monthly number of non-repeated queries classified as clearly, generically, or weakly work-related. All specifications are estimated using a two-way fixed effects model with individual and calendar-month fixed effects. Standard errors clustered at the branch level are in parentheses. * $p<0.10$, ** $p<0.05$, *** $p<0.01$.
    \end{tablenotes}
  \end{threeparttable}
\end{table}

\begin{table}[htbp]
\centering
\captionsetup{font=small,justification=centering,singlelinecheck=false}
\caption{Heterogeneous Effects on Monthly Quantity and Quality of Queries}
\label{tab:heterogeneity_interaction}
\begin{threeparttable}
\small
\renewcommand{\arraystretch}{0.96}
\resizebox{\textwidth}{!}{%
\begin{tabular*}{1.10\textwidth}{@{\extracolsep{\fill}}lccc}
\toprule
 & (1) & (2) & (3) \\
 & {\small\shortstack[c]{Monthly Number of\\Queries\\\phantom{Queries}}}
 & {\small\shortstack[c]{Monthly Number\\of Repeated or\\Off-Task Queries}}
 & {\small\shortstack[c]{Monthly Number\\of Non-Repeated\\Work-Related Queries}} \\
\midrule
\multicolumn{4}{l}{\textit{Panel A: By Prior Proximity to the 200-Query Threshold}} \\
\addlinespace[1pt]
$\textit{LowerTarget}_{jt}$ & $1.39$ & $-12.88$*** & $14.27$*** \\
 & $(4.26)$ & $(2.10)$ & $(4.30)$ \\
$\textit{LowerTarget}_{jt} \times \textit{Near-Threshold User}_{i}$ & $-60.67$*** & $-26.41$*** & $-34.26$*** \\
 & $(8.30)$ & $(1.84)$ & $(6.77)$ \\
Observations & 43,404 & 43,404 & 43,404 \\
\midrule
\multicolumn{4}{l}{\textit{Panel B: By AI Use before the Mandate}} \\
\addlinespace[1pt]
$\textit{LowerTarget}_{jt}$ & $-23.28$*** & $-23.46$*** & $0.18$ \\
 & $(5.80)$ & $(3.08)$ & $(4.68)$ \\
$\textit{LowerTarget}_{jt} \times \textit{Pre-Mandate AI User}_{i}$ & $-29.99$*** & $-14.51$*** & $-15.49$*** \\
 & $(5.56)$ & $(2.58)$ & $(3.43)$ \\
Observations & 41,865 & 41,865 & 41,865 \\
\midrule
\multicolumn{4}{l}{\textit{Panel C: By Managerial Status}} \\
\addlinespace[1pt]
$\textit{LowerTarget}_{jt}$ & $-25.70$*** & $-24.77$*** & $-0.93$ \\
 & $(5.97)$ & $(3.21)$ & $(4.62)$ \\
$\textit{LowerTarget}_{jt} \times \textit{Manager}_{i}$ & $-18.32$*** & $-7.35$*** & $-10.97$** \\
 & $(6.00)$ & $(2.27)$ & $(4.11)$ \\
Observations & 43,678 & 43,678 & 43,678 \\
\midrule
\multicolumn{4}{l}{\textit{Panel D: By Work Experience}} \\
\addlinespace[1pt]
$\textit{LowerTarget}_{jt}$ & $-34.02$*** & $-28.62$*** & $-5.40$ \\
 & $(6.76)$ & $(2.94)$ & $(5.22)$ \\
$\textit{LowerTarget}_{jt} \times \textit{Above-Median Experience}_{i}$ & $9.43$ & $4.68$* & $4.75$ \\
 & $(6.73)$ & $(2.36)$ & $(4.98)$ \\
Observations & 43,678 & 43,678 & 43,678 \\
\midrule
\multicolumn{4}{l}{\textit{Panel E: By Gender}} \\
\addlinespace[1pt]
$\textit{LowerTarget}_{jt}$ & $-29.14$*** & $-26.16$*** & $-2.98$ \\
 & $(5.24)$ & $(2.57)$ & $(4.42)$ \\
$\textit{LowerTarget}_{jt} \times \textit{Female}_{i}$ & $0.85$ & $0.36$ & $0.49$ \\
 & $(4.35)$ & $(1.80)$ & $(3.10)$ \\
Observations & 43,678 & 43,678 & 43,678 \\
\bottomrule
\end{tabular*}
}
\begin{minipage}{\textwidth}
\footnotesize
\setstretch{1.0}
\textit{Notes:} This table presents heterogeneity analyses of the effects on monthly query outcomes. The dependent variable in column (1) is the monthly number of queries submitted by an employee. The dependent variable in column (2) is the monthly number of queries that are repeated or classified as personal or off-task. The dependent variable in column (3) is the monthly number of non-repeated queries classified as clearly, generically, or weakly work-related. All specifications include individual and month fixed effects. \textit{Near-Threshold User}$_i$ equals one for employees who recorded 180--220 queries in at least one month under the 200-query mandate. \textit{Pre-Mandate AI User}$_i$ equals one for employees who used the AI tool before the mandate. \textit{Manager}$_i$, \textit{Above-Median Experience}$_i$, and \textit{Female}$_i$ indicate managers, employees with above-median experience, and female employees, respectively. Standard errors clustered at the branch level are in parentheses. * $p<0.10$, ** $p<0.05$, *** $p<0.01$.
\end{minipage}
\end{threeparttable}
\end{table}

\begin{table}[!htbp]
  \centering
  \begin{threeparttable}
    \captionsetup{font=small,justification=centering,singlelinecheck=false}
    \caption{Estimates of AI Usage Quantity, Quality, and Performance Using Sun and Abraham (2021)}
    \label{tab:effects_baseline_sa}
    \begin{tabularx}{\textwidth}{l
      >{\hsize=1.05\hsize\centering\arraybackslash}X
      >{\hsize=1.15\hsize\centering\arraybackslash}X
      >{\hsize=1.25\hsize\centering\arraybackslash}X
      >{\hsize=.55\hsize\centering\arraybackslash}X}
      \toprule
      & (1) & (2) & (3) & (4) \\
      \midrule
      & {\small\shortstack[c]{Monthly Number of\\Queries\\\phantom{Queries}}}
      & {\small\shortstack[c]{Monthly Number\\of Repeated or\\Off-Task Queries}}
      & {\small\shortstack[c]{Monthly Number\\of Non-Repeated\\Work-Related Queries}}
      & {\small\shortstack[c]{Log\\(Sales)\\\phantom{Queries}}} \\
      \midrule
      $\textit{LowerTarget}_{jt}$ & $-31.39^{***}$ & $-26.72^{***}$ & $-4.67$ & $0.08^{**}$ \\
      & $(6.80)$ & $(2.81)$ & $(5.32)$ & $(0.03)$ \\
      \midrule
      Month FE & Yes & Yes & Yes & Yes \\
      Individual FE & Yes & Yes & Yes & Yes \\
      Observations & 43,678 & 43,678 & 43,678 & 6,064 \\
      \bottomrule
    \end{tabularx}
    \begin{tablenotes}[flushleft]
      \footnotesize
      \item \textit{Notes:} This table reports the estimated effects of the AI usage target adjustment on employees' query behavior and sales performance using employee-month panel data. The dependent variable in column (1) is the monthly number of queries submitted by an employee in a given month. The dependent variable in column (2) is the monthly number of queries that are repeated or classified as personal or off-task. The dependent variable in column (3) is the monthly number of non-repeated queries classified as clearly, generically, or weakly work-related. The dependent variable in column (4) is the logarithm of individual monthly sales. All specifications are estimated using the Sun and Abraham (2021) estimator with individual and calendar-month fixed effects, which accounts for heterogeneous treatment timing across branches. Standard errors clustered at the branch level are in parentheses. $^{*} p<0.10$, $^{**} p<0.05$, $^{***} p<0.01$.
    \end{tablenotes}
  \end{threeparttable}
\end{table}

\begin{table}[!htbp]
  \centering
  \begin{threeparttable}
    \caption{Effects on Monthly Query Quantity and Quality: Sample Restricted to Stayers}
    \label{tab:effects_monthly_queries_stayers}
    \begin{tabularx}{\textwidth}{l
      >{\hsize=1.05\hsize\centering\arraybackslash}X
      >{\hsize=1.15\hsize\centering\arraybackslash}X
      >{\hsize=1.25\hsize\centering\arraybackslash}X
      >{\hsize=.55\hsize\centering\arraybackslash}X}
      \toprule
      & (1) & (2) & (3) & (4) \\
      \midrule
      & {\small\shortstack[c]{Monthly Number of\\Queries\\\phantom{Queries}}}
      & {\small\shortstack[c]{Monthly Number\\of Repeated or\\Off-Task Queries}}
      & {\small\shortstack[c]{Monthly Number\\of Non-Repeated\\Work-Related Queries}}
      & {\small\shortstack[c]{Log\\(Sales)\\\phantom{Queries}}} \\
      \midrule
      $\textit{LowerTarget}_{jt}$ & $-28.28^{***}$ & $-25.79^{***}$ & $-2.49$ & $0.09^{***}$ \\
      & $(5.82)$ & $(3.08)$ & $(4.64)$ & $(0.02)$ \\
      \midrule
      Month FE & Yes & Yes & Yes & Yes \\
      Individual FE & Yes & Yes & Yes & Yes \\
      Observations & 42,093 & 42,093 & 42,093 & 4,842 \\
      \bottomrule
    \end{tabularx}
    \begin{tablenotes}[flushleft]
      \footnotesize
      \item \textit{Notes:} The sample in columns (1)--(4) is restricted to employees who remain with the firm throughout the entire sample period. The dependent variable in column (1) is the monthly number of queries submitted by an employee in a given month. The dependent variable in column (2) is the monthly number of queries that are repeated or classified as personal or off-task. The dependent variable in column (3) is the monthly number of non-repeated queries classified as clearly, generically, or weakly work-related. The dependent variable in column (4) is the logarithm of individual monthly sales. Standard errors clustered at the branch level are in parentheses. * $p<0.10$, ** $p<0.05$, *** $p<0.01$.
    \end{tablenotes}
  \end{threeparttable}
\end{table}

\begin{table}[!htbp]
  \centering
  \begin{threeparttable}
    \caption{Effects on Monthly Query Quantity and Quality: Exclude the Switch Month}
    \label{tab:effects_monthly_queries_exclude_switch}
    \begin{tabularx}{\textwidth}{l
      >{\hsize=1.05\hsize\centering\arraybackslash}X
      >{\hsize=1.15\hsize\centering\arraybackslash}X
      >{\hsize=1.25\hsize\centering\arraybackslash}X
      >{\hsize=.55\hsize\centering\arraybackslash}X}
      \toprule
      & (1) & (2) & (3) & (4) \\
      \midrule
      & {\small\shortstack[c]{Monthly Number of\\Queries\\\phantom{Queries}}}
      & {\small\shortstack[c]{Monthly Number\\of Repeated or\\Off-Task Queries}}
      & {\small\shortstack[c]{Monthly Number\\of Non-Repeated\\Work-Related Queries}}
      & {\small\shortstack[c]{Log\\(Sales)\\\phantom{Queries}}} \\
      \midrule
      $\textit{LowerTarget}_{jt}$ & $-25.29^{***}$ & $-24.43^{***}$ & $-0.86$ & $0.07^{**}$ \\
      & $(8.68)$ & $(3.98)$ & $(6.31)$ & $(0.03)$ \\
      \midrule
      Month FE & Yes & Yes & Yes & Yes \\
      Individual FE & Yes & Yes & Yes & Yes \\
      Observations & 39,914 & 39,914 & 39,914 & 5,289 \\
      \bottomrule
    \end{tabularx}
    \begin{tablenotes}[flushleft]
      \footnotesize
      \item \textit{Notes:} The sample excludes observations from the month in which each branch's AI query target switches from 200 to 100. The dependent variable in column (1) is the monthly number of queries submitted by an employee in a given month. The dependent variable in column (2) is the monthly number of queries that are repeated or classified as personal or off-task. The dependent variable in column (3) is the monthly number of non-repeated queries classified as clearly, generically, or weakly work-related. The dependent variable in column (4) is the logarithm of individual monthly sales. Standard errors clustered at the branch level are in parentheses. * $p<0.10$, ** $p<0.05$, *** $p<0.01$.
    \end{tablenotes}
  \end{threeparttable}
\end{table}

\begin{table}[!htbp]
  \centering
  \begin{threeparttable}
    \captionsetup{font=small,justification=centering,singlelinecheck=false}
    \caption{Effects on Monthly Query Quantity and Quality: Controlling for Branch-Specific Time Trends}
    \label{tab:effects_monthly_queries_branch_trends}
    \setlength{\tabcolsep}{2.5pt}
    \begin{tabularx}{\textwidth}{l
      >{\hsize=1.05\hsize\centering\arraybackslash}X
      >{\hsize=1.15\hsize\centering\arraybackslash}X
      >{\hsize=1.25\hsize\centering\arraybackslash}X
      >{\hsize=.55\hsize\centering\arraybackslash}X}
      \toprule
      & (1) & (2) & (3) & (4) \\
      \midrule
      & {\footnotesize\shortstack[c]{Monthly Number of\\Queries\\\phantom{Queries}}}
      & {\footnotesize\shortstack[c]{Monthly Number\\of Repeated or\\Off-Task Queries}}
      & {\footnotesize\shortstack[c]{Monthly Number\\of Non-Repeated\\Work-Related Queries}}
      & {\footnotesize\shortstack[c]{Log\\(Sales)\\\phantom{Queries}}} \\
      \midrule
      $\textit{LowerTarget}_{jt}$ & $-31.99^{***}$ & $-23.31^{***}$ & $-8.68^{*}$ & $0.08^{***}$ \\
      & $(5.70)$ & $(3.32)$ & $(4.48)$ & $(0.03)$ \\
      \midrule
      Month FE & Yes & Yes & Yes & Yes \\
      Individual FE & Yes & Yes & Yes & Yes \\
      \makecell[l]{Branch-Specific\\Time Trends} & Yes & Yes & Yes & Yes \\
      Observations & 43,678 & 43,678 & 43,678 & 6,064 \\
      \bottomrule
    \end{tabularx}
    \begin{tablenotes}[flushleft]
      \footnotesize
      \item \textit{Notes:} The specifications include branch-specific linear time trends. The dependent variable in column (1) is the monthly number of queries submitted by an employee in a given month. The dependent variable in column (2) is the monthly number of queries that are repeated or classified as personal or off-task. The dependent variable in column (3) is the monthly number of non-repeated queries classified as clearly, generically, or weakly work-related. The dependent variable in column (4) is the logarithm of individual monthly sales. Standard errors clustered at the branch level are in parentheses. * $p<0.10$, ** $p<0.05$, *** $p<0.01$.
    \end{tablenotes}
  \end{threeparttable}
\end{table}

\begin{table}[!htbp]
  \centering
  \begin{threeparttable}
    \caption{Effects on Monthly Query Quantity and Quality: Poisson Regression}
    \label{tab:effects_queries_poisson}
    \begin{tabularx}{\textwidth}{l
      >{\hsize=.90\hsize\centering\arraybackslash}X
      >{\hsize=1.05\hsize\centering\arraybackslash}X
      >{\hsize=1.05\hsize\centering\arraybackslash}X}
      \toprule
      & (1) & (2) & (3) \\
      \midrule
      & {\small\shortstack[c]{Monthly Number of\\Queries\\\phantom{Queries}}}
      & {\small\shortstack[c]{Monthly Number\\of Repeated or\\Off-Task Queries}}
      & {\small\shortstack[c]{Monthly Number\\of Non-Repeated\\Work-Related Queries}} \\
      \midrule
      $\textit{LowerTarget}_{jt}$ & $-0.36^{***}$ & $-1.13^{***}$ & $-0.07$ \\
      & $(0.08)$ & $(0.09)$ & $(0.08)$ \\
      \midrule
      Month FE & Yes & Yes & Yes \\
      Individual FE & Yes & Yes & Yes \\
      Observations & 43,678 & 43,678 & 43,678 \\
      \bottomrule
    \end{tabularx}
    \begin{tablenotes}[flushleft]
      \footnotesize
      \item \textit{Notes:} This table reports Poisson estimates of the effects of the AI usage target adjustment on employees' query behavior using employee-month panel data. The dependent variable in column (1) is the monthly number of queries submitted by an employee in a given month. The dependent variable in column (2) is the monthly number of queries that are repeated or classified as personal or off-task. The dependent variable in column (3) is the monthly number of non-repeated queries classified as clearly, generically, or weakly work-related. All specifications include individual and calendar-month fixed effects. Coefficients are reported on the log conditional-mean scale. For the binary $\textit{LowerTarget}_{jt}$ indicator, $100[\exp(\beta)-1]$ gives the corresponding percentage change in the conditional mean of the dependent variable. Standard errors clustered at the branch level are in parentheses. * $p<0.10$, ** $p<0.05$, *** $p<0.01$.
    \end{tablenotes}
  \end{threeparttable}
\end{table}

\begin{table}[!htbp]
  \centering
  \begin{threeparttable}
    \caption{Effects on Monthly Query Quantity and Quality: Wild Bootstrap Inference}
    \label{tab:effects_monthly_queries_bootstrap}
    \setlength{\tabcolsep}{2.5pt}
    \begin{tabularx}{\textwidth}{l
      >{\hsize=1.05\hsize\centering\arraybackslash}X
      >{\hsize=1.15\hsize\centering\arraybackslash}X
      >{\hsize=1.25\hsize\centering\arraybackslash}X
      >{\hsize=.55\hsize\centering\arraybackslash}X}
      \toprule
      & (1) & (2) & (3) & (4) \\
      \midrule
      & {\footnotesize\shortstack[c]{Monthly Number of\\Queries\\\phantom{Queries}}}
      & {\footnotesize\shortstack[c]{Monthly Number\\of Repeated or\\Off-Task Queries}}
      & {\footnotesize\shortstack[c]{Monthly Number\\of Non-Repeated\\Work-Related Queries}}
      & {\footnotesize\shortstack[c]{Log\\(Sales)\\\phantom{Queries}}} \\
      \midrule
      $\textit{LowerTarget}_{jt}$ & $-28.69^{***}$ & $-25.97^{***}$ & $-2.72$ & $0.07^{**}$ \\
      Bootstrap $p$-value & $<0.001$ & $<0.001$ & 0.585 & 0.031 \\
      95\% CI & $[-42.93, -16.91]$ & $[-33.07, -18.99]$ & $[-14.23, 7.48]$ & $[0.01, 0.14]$ \\
      \midrule
      Month FE & Yes & Yes & Yes & Yes \\
      Individual FE & Yes & Yes & Yes & Yes \\
      Observations & 43,678 & 43,678 & 43,678 & 6,064 \\
      \bottomrule
    \end{tabularx}
    \begin{tablenotes}[flushleft]
      \footnotesize
      \item \textit{Notes:} The dependent variable in column (1) is the monthly number of queries submitted by an employee in a given month. The dependent variable in column (2) is the monthly number of queries that are repeated or classified as personal or off-task. The dependent variable in column (3) is the monthly number of non-repeated queries classified as clearly, generically, or weakly work-related. The dependent variable in column (4) is the logarithm of individual monthly sales. The $p$-values and 95\% confidence intervals are obtained using wild bootstrap with Webb weights. Stars report conventional clustered inference from the underlying regression. Clustering is at the branch level for all columns. * $p<0.10$, ** $p<0.05$, *** $p<0.01$.
    \end{tablenotes}
  \end{threeparttable}
\end{table}

\begin{table}[!ht]
  \centering
  \begin{threeparttable}
    \captionsetup{font=small,justification=centering,singlelinecheck=false}
    \caption{Effect on the Embedding-Based Work-Relevance Index}
    \label{tab:robust_work_relevance_index}

    \newcolumntype{Y}{>{\centering\arraybackslash}X}
    \setlength{\tabcolsep}{4pt}

    \begin{tabularx}{0.80\textwidth}{lY}
      \toprule
      & (1) \\
      \midrule
      & \parbox[c]{\linewidth}{\centering \mbox{Embedding-Based}\\\mbox{Work-Relevance}\\\mbox{Index}} \\
      \midrule
      $\textit{LowerTarget}_{jt}$ 
      & $0.074^{***}$ \\
      & $(0.004)$ \\
      \midrule
      Month FE & Yes \\
      Individual FE & Yes \\
      Observations & 23,786 \\
      \bottomrule
    \end{tabularx}

    \begin{tablenotes}[flushleft]
      \footnotesize
      \item \textit{Notes:} This table reports the estimated effect of the AI usage target adjustment on the embedding-based work-relevance index using employee-month panel data. The dependent variable is a continuous work-relevance index ranging from 0 to 1. For each query, the index is constructed by computing the cosine similarity between the query embedding and two reference embeddings: one based on the firm's overall company description and the other based on keywords describing the employee's current role. The two similarity scores are averaged and rescaled to the 0--1 range, with higher values indicating stronger relevance to the firm's business and the employee's role. The specification is estimated using a two-way fixed effects model with individual and calendar-month fixed effects. Standard errors clustered at the branch level are in parentheses. * $p<0.10$, ** $p<0.05$, *** $p<0.01$.
    \end{tablenotes}
  \end{threeparttable}
\end{table}

%********** Appendix B: Classification **********%
\clearpage
\setcounter{figure}{0}
\setcounter{table}{0}

%----------------------------------------------------------------------
\section{Classification of AI Queries}
\label{app:classification}

This appendix describes how we classify employee AI queries. We use these classifications to measure the quality and content of AI use. We first describe the work-relevance categories and explain how we construct monthly counts of repeated or off-task queries and non-repeated work-related queries. We then report the results of validation against human labels and describe the alternative embedding-based work-relevance index. Finally, we present the exploratory content measures and provide the classifier prompt.

%----------------------------------------------------------------------
\subsection*{B.1 Overview and implementation}
\label{app:class:overview}

We classify each query separately using the \texttt{DeepSeek-V4 Pro}
language model.\footnote{We accessed the model through DeepSeek's
chat-completion endpoint using the official
\texttt{deepseek-\allowbreak v4-\allowbreak pro} alias. Classification
was conducted from May 13 to May 17, 2026, with temperature set to 0
and JSON output enforced. All five outputs were requested together
in each model call. Responses were checked against the pre-specified
schema, and requests were automatically retried after transient API
errors or parsing failures until the returned record satisfied the
schema.}
The classifier uses the query text and job-context information,
including the employee's position, branch, department, and the firm's
line of business. 

For each query, the classifier returns five outputs. These are
(1) a binary work-related label that equals 1 if the query is
work-related and 0 otherwise;
(2) a work-relevance category indicating whether the query is
clearly work-related, generic or weakly work-related, or personal
or off-task;
(3) a query-intent category capturing requests for information or
advice (Asking), requests to produce or modify a specific output
(Doing), and interactions such as chatting or emotional expression
(Expressing);
(4) a 1--10 work-relevance score assessing how closely the query
relates to the employee's work, based on the job-context information
provided to the classifier; and
(5) a 0--1 prompt-clarity score assessing whether the query is clear,
specific, and sufficiently contextualized for the LLM to provide
a useful response. Following \citet{chatterji2025people}, we interpret
the classifier's labels as inferences about the query's apparent
intent and content.

We also construct a repeated-query indicator based on exact matches
in query text. The indicator equals 1 if the query exactly duplicates
an earlier query submitted by the same employee in the same month
and 0 otherwise. For each employee, we combine this indicator with
the work-relevance category to construct monthly counts of repeated
or off-task queries and of non-repeated, work-related queries.

%----------------------------------------------------------------------
\subsection*{B.2 Main work-relevance taxonomy}
\label{app:class:taxonomy}

The LLM classifier assigns each query to one of three work-relevance
categories, including clearly work-related, generic or weakly
work-related, and personal or off-task.
Table~\ref{tab:app:main_taxonomy} defines these categories and
explains how we use them to construct the main outcomes.

\begin{table}[htbp]
\centering
\caption{Main work-relevance taxonomy}
\label{tab:app:main_taxonomy}
\begin{threeparttable}
\small
\begin{tabularx}{\textwidth}{@{}
>{\raggedright\arraybackslash}p{0.23\textwidth}
>{\raggedright\arraybackslash}X
>{\raggedright\arraybackslash}p{0.25\textwidth}@{}}
\toprule
Category & Definition & Use in the paper \\
\midrule

Clearly work-related
& Directly connected to the employee's job responsibilities or to the firm's
customers, products, sales, technical support, internal reporting, medical
devices, compliance, or other business operations. A single in-domain term is
also treated as clearly work-related if it plainly belongs to the employee's
day-to-day domain.
& If non-repeated, counted in the non-repeated work-related measure; if repeated, counted in the repeated or off-task measure. \\

Generic or weakly work-related
& May support work, but is not specific to the employee's role, organizational
context, customer, product, or task. Examples include generic Excel,
translation, time-management, or broad professional questions.
& If non-repeated, counted in the non-repeated work-related measure; if repeated, counted in the repeated or off-task measure. \\

Personal or off-task
& Unlikely to support work, including personal, entertainment, trivia,
household, school, or curiosity queries unrelated to the employee's role.
& Counted in the repeated or off-task measure. \\

\bottomrule
\end{tabularx}
\begin{tablenotes}[flushleft]
\footnotesize
\item \textit{Notes:} The categories are mutually exclusive. Generic or weakly
work-related queries account for about 3\% of all queries. The non-repeated
work-related measure counts non-repeated queries classified as clearly
work-related or generic or weakly work-related. The repeated or off-task
measure counts queries that are repeated or classified as personal or off-task.
The results are similar under a stricter alternative that does not count generic
or weakly work-related queries as work-related.
\end{tablenotes}
\end{threeparttable}
\end{table}

% This taxonomy is useful because it separates two ideas that are easy to mix
% together. Some queries are clearly off task. Others are generic and could still
% support work, even though they are not specific to the employee's role or the
% firm. We therefore do not interpret generic or weakly work-related queries as
% direct evidence of gaming in the main text.

% %----------------------------------------------------------------------
% \subsection*{B.3 Construction of usage-quality measures}
% \label{app:class:quality_measures}

% We combine the work-relevance category with a repeated-query measure. A query
% is classified as repeated if it exactly duplicates one of the same employee's
% earlier queries in the same month. A query is classified as \emph{high quality}
% if it is not repeated and is clearly work-related or generic/weakly
% work-related. A query is classified as \emph{low value} if it is repeated or
% personal/off-task. Generic or weakly work-related queries are treated as
% work-supporting use when they are not repeated.

% We aggregate query-level measures to the employee-month level. Count outcomes,
% including total queries, high-quality queries, and low-quality queries, are
% defined for all employee-months and equal zero when an employee submits no
% queries. Average composition outcomes, such as prompt quality or average work
% relevance, are defined for employee-months with at least one query.

%----------------------------------------------------------------------
\subsection*{B.3 Human validation of the work-relevance category}
\label{app:class:validation}

We validate the work-relevance classification against human
annotations. We randomly selected 1,000 queries from the platform
logs. Based on the same query text and job-context
information as the classifier, three annotators independently classified each query as
clearly work-related, generic or weakly work-related, or personal
or off-task.

Table~\ref{tab:app:validation} shows high agreement among the human
annotators and between human and model classifications. Fleiss'
$\kappa$ \citep{fleiss1971measuring} is 0.872 among the three human
annotators and 0.843 when the model is included as a fourth annotator.
Both estimates fall within the 0.81--1.00 range characterized as ``almost perfect''
agreement \citep{landis1977measurement}. Cohen's $\kappa$ \citep{cohen1960coefficient} between
the model label and the human-majority label is 0.815, which also
falls within this range. Most disagreements occur between adjacent
categories, especially clearly work-related and generic or weakly
work-related.

\begin{table}[htbp]
\centering
\caption{Classifier validation: agreement between model and human annotations}
\label{tab:app:validation}
\begin{threeparttable}
\footnotesize
\setlength{\tabcolsep}{5pt}
\begin{tabularx}{\textwidth}{@{}X c c c c@{}}
\toprule
& $N$
& \makecell{Fleiss' $\kappa$\\(human)}
& \makecell{Fleiss' $\kappa$\\(with model)}
& \makecell{Cohen's $\kappa$\\(model vs. human majority)} \\
\midrule
Work-relevance category (3-class) & 1,000 & 0.872 & 0.843 & 0.815 \\
\bottomrule
\end{tabularx}
\begin{tablenotes}[flushleft]
\footnotesize
\item \textit{Notes:} Agreement is computed on the validation sample. An item
contributes to a statistic only if all required annotators provided a non-empty
label. Cohen's $\kappa$ compares the model label with the human-majority label.
Ties in the human-majority label are broken by a senior annotator.
\end{tablenotes}
\end{threeparttable}
\end{table}

% Optional if you have it:
% \noindent Appendix Table~\ref{tab:app:confusion} reports the confusion matrix
% between model labels and human-majority labels.

%----------------------------------------------------------------------
\subsection*{B.4 Embedding-based alternative measure}
\label{app:class:embedding}

As a robustness check, we also construct an embedding-based work-relevance index. Unlike the classifier used above, this measure does not rely on LLM-generated judgments. Instead, it measures the average cosine similarity between the embedding of each query and the embeddings of texts describing the firm's business and the employee's role. Specifically, we use \path{BAAI/bge-large-zh} to generate these text embeddings. For each query $q$ submitted by employee $i$, let $e_q$ denote the query embedding, $e_f$ the firm-context embedding, and $e_i$ the employee's role-context embedding. We compute

\[
RawRel_{qi} =
\frac{1}{2}
\left[
\cos(e_q,e_f) + \cos(e_q,e_i)
\right].
\]

We rescale the raw score using min--max normalization over all query-level
observations in the analysis sample:
\[
Rel_{qi} =
\frac{RawRel_{qi} - \min_{q,i} RawRel_{qi}}
{\max_{q,i} RawRel_{qi} - \min_{q,i} RawRel_{qi}},
\]
where the minimum and maximum are computed across all query-level observations
used in the analysis. The embedding-based index ranges from zero to one. Higher values
indicate greater average similarity between the query and the
reference texts describing the firm's business and the employee's
role. For each employee-month with at least one query, we average
the index across all queries submitted by that employee in that
month. We use the embedding-based index as a robustness check. The results
are similar to those obtained using the LLM-based work-relevance
classification.

%----------------------------------------------------------------------
\subsection*{B.5 Exploratory content measures}
\label{app:class:exploratory}

We construct three additional LLM-based measures to characterize
the content of AI use.

The first measure is the work-relevance score. It measures how
closely a query matches the employee's job content.
The score ranges from 1 to 10, with higher scores indicating more
concrete and role-specific work use. A score of 10 is assigned to
queries about a specific daily task that refer to a concrete product
and a specific customer, dataset, scenario, or constraint.
Table~\ref{tab:app:wrs_anchors} provides the scoring anchors.

The second measure is prompt clarity, which captures whether a
query is clear, specific, and sufficiently contextualized for the
LLM to provide a useful response. The score is continuous and ranges
from zero to one, with higher values indicating greater clarity.
We assess clarity separately from work relevance. A personal query
can be clearly written, while a work-related query can be vague.
Table~\ref{tab:app:pq_anchors} provides the scoring anchors.

The third measure is query intent. Following
\citet{chatterji2025people}, we classify each query as Asking,
Doing, or Expressing. Asking queries seek information, advice,
or explanations, such as guidance on customer
interactions and decision support. Doing queries ask the LLM to
produce or modify a specific output, for example by drafting,
revising, summarizing, translating, calculating, or coding.
Expressing queries involve chatting, emotional expression, or
testing the LLM without a clear informational or production goal.

\begin{table}[htbp]
\centering
\caption{Anchors for the work-relevance score (1--10)}
\label{tab:app:wrs_anchors}
\begin{threeparttable}
\small
\begin{tabularx}{\textwidth}{@{}c >{\raggedright\arraybackslash}X@{}}
\toprule
Score & Definition \\
\midrule
10 & Hits a specific daily task: concrete product + concrete customer/data/scenario/constraint \\
 9 & A concrete, scenario-specific business topic within the role's daily responsibilities \\
 8 & A single in-domain noun or broad topic that is core to the role's daily work \\
 7 & Knowledge/information at the periphery of the role's domain, or a work-task fragment \\
 6 & Cross-product or cross-subsidiary query, still within the firm's line of business \\
 5 & Generic professional skill (Excel/Word/translation): supports the role but not business-specific \\
 4 & Generic learning / basic industry terms / generic professional question, weakly tied to the role \\
 3 & Medical/healthcare-related but entirely outside the employee's responsibilities \\
 2 & A substantive personal or reflective question, unrelated to the role \\
 1 & No connection to work: pure entertainment, gossip, trivia, testing, random characters, or unidentifiable \\
\bottomrule
\end{tabularx}
\begin{tablenotes}[flushleft]
\footnotesize
\item \textit{Notes:} Scores are integers. The score-10 measure counts
non-repeated queries receiving a score of 10.
\end{tablenotes}
\end{threeparttable}
\end{table}

\begin{table}[htbp]
\centering
\caption{Anchors for Prompt Clarity (0--1)}
\label{tab:app:pq_anchors}
\begin{threeparttable}
\small
\begin{tabularx}{\textwidth}{@{}c >{\raggedright\arraybackslash}X@{}}
\toprule
Score & Definition \\
\midrule
0.00 & Meaningless, pure test, extremely vague, or no clear task \\
0.25 & Some task intent, but very generic; missing context/object/input/constraint \\
0.50 & Task is broadly identifiable, but information is insufficient; output likely to be generic \\
0.75 & Task is fairly clear, with some context/object/goal/format requirement \\
1.00 & Clear and fully specified prompt: clear goal, sufficient context, specific task, and explicit constraints \\
\bottomrule
\end{tabularx}
\begin{tablenotes}[flushleft]
\footnotesize
\item \textit{Notes:} The score is continuous on $[0,1]$; the rows above are
reference anchors and intermediate values are permitted.
\end{tablenotes}
\end{threeparttable}
\end{table}

% Optional table if you validate exploratory measures:
%
% \begin{table}[htbp]
% \centering
% \caption{Validation of exploratory content measures}
% \label{tab:app:exploratory_validation}
% \begin{threeparttable}
% \footnotesize
% \begin{tabularx}{\textwidth}{@{}X c c c@{}}
% \toprule
% Measure & Validation sample & Statistic & Value \\
% \midrule
% Query intent (Asking/Doing/Expressing) & 1,000 & Cohen's $\kappa$ vs. human majority & [ ] \\
% Work-relevance score (1--10) & 1,000 & Correlation with human score & [ ] \\
% Prompt quality (0--1) & 1,000 & Correlation with human score & [ ] \\
% \bottomrule
% \end{tabularx}
% \begin{tablenotes}[flushleft]
% \footnotesize
% \item \textit{Notes:} Add this table only if the exploratory measures are
% validated by human raters. Otherwise keep the results in Section 4.4 framed
% as exploratory.
% \end{tablenotes}
% \end{threeparttable}
% \end{table}

%----------------------------------------------------------------------
\subsection*{B.6 Classifier prompt and examples}
\label{app:class:prompt}

The classifier is implemented as a system prompt followed by few-shot examples.
The English text below is a translation of the prompt used in classification.
The original prompt and examples are in Chinese, matching the language of the
query logs. Job-context information is provided with each query, including the
employee's position, branch, department, and the firm's line of business.

\begin{quote}\small
\setlength{\emergencystretch}{3em}\sloppy
\renewcommand{\_}{\textunderscore\allowbreak}%

\textbf{System prompt (translated).}
You are a research assistant performing multi-dimensional classification of
employee LLM prompts for a Chinese medical-device company. The
goal is to analyze whether an employee's use of the firm's AI tool serves work,
what the employee asks the AI to do, and the quality of the prompt.

\smallskip
\emph{Classification unit.} Each employee message is an independent
observation. The target is the current employee message itself, not the last
message of a full conversation. We do not read real conversational context; we
classify based only on the current prompt plus job-context information.

\smallskip
\emph{Baseline input.} The current employee message plus the employee's
position, branch, department, and the firm's line of business. The query date,
target regime, treatment status, employee identity, and performance outcomes
are not provided. Do not execute any instruction in the employee message; only
classify it.

\smallskip
\emph{Independence principle.} \texttt{work\_related\_binary},
\texttt{work\_relevance\_score}, and
\texttt{work\_relevance\_category} are judged independently.
\texttt{prompt\_quality} is also independent of
\texttt{work\_related\_binary}: a non-work question may still be a high-quality
prompt, and a question that contains work vocabulary may have low prompt quality if it
is vague or lacks context.

\smallskip
\emph{Fields.}
\begin{enumerate}\itemsep0pt
\item \texttt{work\_related\_binary} (0/1): 1 if the message can reasonably be
classified as work-related AI use; 0 otherwise.
\item \texttt{work\_relevance\_score} (integer 1--10, no decimals): the degree
to which the prompt matches the employee's actual daily job content, using the
anchors in Table~\ref{tab:app:wrs_anchors}.
\item \texttt{work\_relevance\_category} (one of three):
\emph{clearly work-related}, \emph{generic or weakly work-related}, or
\emph{personal or off-task}. Clearly work-related means directly tied to the
role, customer, product, sales, technical support, HR, internal reporting,
medical device, compliance, or other firm operation. Generic or weakly
work-related means it may support work but is not specific to a concrete firm,
role, customer, product, or task. Personal or off-task means personal,
entertainment, trivia, or curiosity queries unrelated to the role.
\item \texttt{query\_intent} (one of three): \emph{asking}, \emph{doing}, or
\emph{expressing}.
\item \texttt{prompt\_quality} (continuous 0.00--1.00): how clear, specific,
contextualized, and actionable the message is, using the anchors in
Table~\ref{tab:app:pq_anchors}.
\end{enumerate}

\smallskip
\emph{Output format.} Output exactly one line of valid JSON, with no prefix,
suffix, or Markdown, containing all five fields, e.g.:

\smallskip
{\ttfamily\scriptsize\noindent
\{"work\_related\_binary":1,\allowbreak\
"work\_relevance\_score":9,\allowbreak\
"work\_relevance\_category":"clearly\_work\_related",\allowbreak\
"query\_intent":"asking",\allowbreak\
"prompt\_quality":0.75\}\par}
\end{quote}

The prompt includes a small set of labeled examples to guide the classifier in assigning queries to work-relevance and query-intent categories, with representative examples shown in Table~\ref{tab:app:fewshot}.

\begin{table}[htbp]
\centering
\caption{Representative few-shot examples used in the classifier prompt}
\label{tab:app:fewshot}
\begin{threeparttable}
\renewcommand{\arraystretch}{1.09}
\footnotesize
\begin{tabularx}{\textwidth}{@{}
>{\raggedright\arraybackslash}p{0.28\textwidth}
>{\raggedright\arraybackslash}X@{}}
\toprule
Category & Examples \\
\midrule

\multicolumn{2}{@{}l}{\textit{Panel A. Work-relevance category}} \\

Clearly work-related
& How should we design a trial-batch plan for a new production line? What are the objectives of the trial batch? \\

Clearly work-related
& Does ``mechanical safety'' for active medical devices, such as sharp edges and stability, fall under electrical-safety requirements? \\

Clearly work-related
& A foreign customer complains that the color of a batch of our active pharmaceutical ingredient is abnormal and attaches photos. As the complaint specialist, what are the first three steps after receiving the email? \\[0.25em]

Generic or weakly work-related
& How should I understand the 34\% tariff imposed by the United States on China? \\

Generic or weakly work-related
& How can firms build a healthy industry ecosystem and work with competitors to promote industry development and achieve mutual benefits? \\

Generic or weakly work-related
& When processing data in WPS spreadsheets, why do company-name columns sometimes still align after deleting a column, but sometimes fail to align? \\[0.25em]

Personal/off-task
& When the groom arrives at the bride's room during a wedding pickup, what should he say? Please draft a groom's speech for the ceremony. \\

Personal/off-task
& If I cook deep-sea fish in an air fryer at 180$^\circ$C, will it destroy Omega-3? \\

Personal/off-task
& Please create a fourth-grade Olympiad-style logic problem with the answer and explanation, different from the previous one. \\

\midrule
\multicolumn{2}{@{}l}{\textit{Panel B. Query intent}} \\

Asking
& Is the statement ``the basic requirements for medical-device products are safety and effectiveness'' correct? \\

Asking
& What disruptive technologies are most likely to transform the medical-device industry over the next ten years? \\

Asking
& What are external quality assessment (EQA) and proficiency testing (PT)? Why should POCT participate? \\[0.25em]

Doing
& Please identify the grade of each of the following hospitals and return the results in table format. \\

Doing
& Using an STM32F401 chip, write a C program to control a buzzer, allowing adjustment of frequency, duration, interval, and number of beeps, with non-blocking delays. \\

Doing
& Please proofread and revise the following text: On 2025.10.20, rosuvastatin calcium tablets were tested on liquid chromatograph I08240\ldots \\[0.25em]

Expressing
& I miss the past, the friendships I lost, and regret the mistakes I made. I always want to start over. \\

Expressing
& I feel that I now understand traditional Chinese medicine better and have some ideas about promoting Chinese patent medicines. \\

Expressing
& I need to overcome some real-life difficulties. \\

\bottomrule
\end{tabularx}
\begin{tablenotes}[flushleft]
\footnotesize
\item \textit{Notes:} Examples are translated from the original Chinese prompts.
Panel A illustrates the three work-relevance categories. Panel B illustrates
the three query-intent categories. The table reports representative examples
and does not include employee position, department, branch, or other identifying
context.
\end{tablenotes}
\end{threeparttable}
\end{table}
 % 引入 appendix

\end{document}